\documentclass[a4paper,11pt]{article}
\pdfoutput=1 

\usepackage{jheppub} 
\usepackage[bottom]{footmisc}
\usepackage{amssymb}
\usepackage{amsmath}
\usepackage{amsthm}
\usepackage[usenames,dvipsnames]{xcolor}
\usepackage{epsfig}
\usepackage{dcolumn}
\usepackage{tikz}
\usetikzlibrary{shapes.geometric, arrows}
\usepackage{upgreek}
\usepackage{setspace}
\usepackage{subfig}
\usepackage{array,multirow,bigdelim,arydshln}
\usepackage{appendix}
\usepackage{xparse}
\usepackage{stmaryrd}
\usepackage[T1]{fontenc} 
\usepackage{mathtools}
\usepackage{physics} 
\usepackage{adjustbox}
\usepackage{multirow}
\usepackage{graphicx} 
\usepackage{float} 
\graphicspath{{./images/}}
\usepackage[nottoc]{tocbibind}
\usepackage{hyperref}
\usepackage[utf8]{inputenc}
\usepackage{CJK}
\hypersetup{
	colorlinks,
	urlcolor=Maroon,
	linkcolor=Maroon,
	citecolor=Maroon
	}

\NewDocumentCommand{\binomial}{omm}
 {%
  \genfrac(){0pt}{}{#2}{#3}%
  \IfValueT{#1}{_{\!#1}}%
 }
\NewDocumentCommand{\eulerian}{omm}
 {%
  \genfrac<>{0pt}{}{#2}{#3}%
  \IfValueT{#1}{_{\!#1}}%
 }

\usepackage{latexsym}
\usepackage{tikz}
\newcommand*\diff{\mathop{}\!\mathrm{d}}

\theoremstyle{plain}

\theoremstyle{definition}

\def\bea#1\eea{\begin{eqnarray}#1\end{eqnarray}}
\def\be#1\ee{\begin{equation}#1\end{equation}}
\def\ba#1\ea{\begin{align}#1\end{align}}

\usepackage{amsmath}
\usepackage{multicol}
\usepackage{bbm}
\usepackage{enumerate}

\usepackage{amsthm}
\usepackage{mathrsfs}
\usepackage{upgreek}
\usepackage{amssymb}
\usepackage{bm}
\usepackage{setspace}
\usepackage{array,multirow,arydshln}
\usepackage{bigdelim}
\usepackage{scalerel}
\usepackage{diagbox}

\usepackage{tabularx}

\usepackage{tikz}

\usetikzlibrary{shapes.geometric,arrows,arrows.meta,decorations.pathmorphing,decorations.markings,patterns}

\def\<{\langle}
\def\>{\rangle}

\usepackage[percent]{overpic}
\usepackage{multirow} 
\usepackage{slashed}

\title{Hidden zeros for particle/string amplitudes and the unity of colored scalars, pions and gluons}

\author[a]{Nima Arkani-Hamed,}
\author[b,c]{Qu Cao (曹 趣),}
\author[b,d]{Jin Dong (董 晋),}
\author[e]{Carolina Figueiredo,}
\author[b,f,g]{Song He (何 颂)}

\affiliation[a]{School of Natural Sciences, Institute for Advanced Study, Princeton, NJ, 08540, USA}
\affiliation[b]{CAS Key Laboratory of Theoretical Physics, Institute of Theoretical Physics, Chinese Academy of Sciences, Beijing 100190, China}
\affiliation[c]{Zhejiang Institute of Modern Physics, Department of Physics, Zhejiang University, Hangzhou, 310027, China}
\affiliation[d]{School of Physical Sciences, University of Chinese Academy of Sciences, No.19A Yuquan Road, Beijing 100049, China}
\affiliation[e]{Jadwin Hall, Princeton University, Princeton, NJ 08540, USA}
\affiliation[f]{School of Fundamental Physics and Mathematical Sciences, Hangzhou Institute for Advanced Study \& ICTP-AP, UCAS, Hangzhou 310024, China}
\affiliation[g]{Peng Huanwu Center for Fundamental Theory, Hefei, Anhui 230026, P. R. China}

\emailAdd{arkani@ias.edu}
\emailAdd{qucao@zju.edu.cn}
\emailAdd{dongjin@itp.ac.cn}
\emailAdd{cfigueiredo@princeton.edu}
\emailAdd{songhe@itp.ac.cn}

\abstract{Recent years have seen the emergence of a new understanding of scattering amplitudes in the simplest theory of colored scalar particles--the Tr$(\phi^3)$  theory--based on combinatorial and geometric ideas in the kinematic space of scattering data. In this paper we report a surprise: far from the toy model it appears to be, the ``stringy'' Tr$(\phi^3)$ amplitudes secretly {\it contains} the scattering amplitudes for pions, as well as non-supersymmetric gluons, in any number of dimensions. The amplitudes for the different theories are given by one and the same function, related by a simple shift of the kinematics. This discovery was spurred by another fundamental observation: the tree-level Tr$(\phi^3)$ field theory amplitudes have a hidden pattern of zeros when a special set of non-planar Mandelstam invariants is set to zero. These zeros are not manifest in Feynman diagrams but are made obvious by the connection of these amplitudes to the new understanding of associahedra arising from ``causal diamonds'' in kinematic space. Furthermore, near these zeros, the amplitudes simplify, by factoring into a non-trivial product of smaller amplitudes. Remarkably the amplitudes for pions and gluons are observed to {\it also} vanish in the same kinematical locus. These properties for Tr$(\phi^3)$ amplitudes hold and further generalize to the ``stringy'' Tr$(\phi^3)$ amplitudes. The ``kinematic causal diamond'' picture suggests a unique shift of the kinematic data that preserves the zeros, and this shift is precisely the one that unifies colored scalars, pions, and gluons into a single object. We will focus in this paper on explaining the hidden zeros and factorization properties and the connection between all the colored theories, working for simplicity at tree level. Subsequent works will describe this new formulation for the Non-linear Sigma Model and non-supersymmetric Yang-Mills theory, at all loop orders.}

\begin{document}

\begin{CJK*}{UTF8}{}
\CJKfamily{gbsn}
\maketitle
\end{CJK*}
\addtocontents{toc}{\protect\setcounter{tocdepth}{2}}

\numberwithin{equation}{section}

		\tikzset{
		particles/.style={dashed, postaction={decorate},
			decoration={markings,mark=at position .5 with {\arrow[scale=1.5]{>}}
		}}
	}
	\tikzset{
		particle/.style={draw=black, postaction={decorate},
			decoration={markings,mark=at position .5 with {\arrow[scale=1.1]{>}}
		}}
	}
	\def  \layersep {.6cm}

\section{From Toy Models to The Real World Via Numerators and Zeros}
\label{sec:Intro}
Over the last few decades, a rich combinatorial and geometric structure underlying scattering amplitudes has been revealed. These descriptions have been most successful in the context of theories, such as planar ${\cal N} = 4$ super-Yang-Mills (SYM)~\cite{Arkani-Hamed:2013jha} and the Tr$(\phi^3)$ theory for colored scalars~\cite{Arkani-Hamed:2017mur, Arkani-Hamed:2023lbd, Arkani-Hamed:2023mvg}, for which the amplitudes are relatively simple. Speaking most invariantly, these are theories for which the amplitudes can be entirely determined by their long-distance singularities, for instance, their factorization properties on massless poles at tree level. Loosely speaking, the combinatorics and geometry provide an alternate understanding of the rich and intricate pattern of {\bf denominators} of the amplitude, which turn out to non-trivially determine the entire amplitude as well. 

But as we turn towards describing much more interesting and physically relevant theories, such as the non-supersymmetric gauge interactions of the Standard Model, we must incorporate a qualitatively new feature present in the amplitudes. At the most basic and concrete level, there are {\bf numerator} factors associated with the more interesting interactions vertices of the realistic Lagrangians. Of course even ${\cal N} = 4$ SYM has such vertices, so the more precise and invariant statement is that more interesting and realistic theories have new ``poles at infinity'', not associated simply with massless factorizations, which must be incorporated in any new combinatorial/geometric formulation of this physics. Such poles are absent in planar ${\cal N}=4$ SYM (as one consequence of the famous hidden ``dual conformal invariance'' of the theory~\cite{Drummond:2006rz}), as well as in Tr$(\phi^3)$ theory (a consequence of a cousin hidden ``projective invariance'' of the amplitudes~\cite{Arkani-Hamed:2017mur}). Poles at infinity are naturally associated with non-trivial numerator factors, whose presence and purpose in life must be exposed in the next phase of the adventure of connecting combinatorial geometry to the real world. 

The most obvious place to search for new structure associated with numerators is to understand whether these give rise to interesting patterns of {\bf zeros} of the amplitude -- so this is where we start.  We will begin by studying the simplest theory of colored scalars, Tr$(\phi^3)$ theory, which has been much studied recently from the new perspectives of tropical geometry, $u$-variables and surfacehedra~\cite{Arkani-Hamed:2019mrd, Arkani-Hamed:2019plo, Arkani-Hamed:2023lbd, Arkani-Hamed:2023mvg}. This may appear to be an odd starting point since the Tr$(\phi^3)$ theory is precisely the most ``overly simple'' theory with {\bf no} numerator factors in its amplitude! And yet as we will see, even this seemingly boring theory {\it does} have a surprising and rich pattern of amplitude zeroes, and what is more, this pattern extends to much more interesting and realistic theories of pions and gluons, revealing a striking hidden unity between these three theories, which as we will show are in a precise sense ``contained'' in each other. It is well-known that the traditional double copy relations~\cite{Bern:2008qj,Bern:2010ue,Cachazo:2013gna,Cachazo:2013hca,Cachazo:2013iea,Cachazo:2014xea} (see also the recent reviews~\cite{Bern:2019prr,Bern:2022wqg,Adamo:2022dcm} and the references therein) have established a broad web of interconnected theories. However, our results offer a new prescription for understanding the relationships between these theories.

The presence of these hidden zeroes, at least for the Tr$(\phi^3)$ theory, is not at all manifest in the diagrammatic expansion for the amplitude, but {\it is} made obvious by the understanding of the Tr$(\phi^3)$ amplitudes as the so-called ``canonical form'' of the so-called ABHY associahedron polytope, which we review in section \ref{sec:ABHY}. 
The zeros are connected with the fact that ABHY associahedron is built out of simpler lower-dimensional objects -- it is a Minkowski sum of simplices. As we will see in section \ref{sec:zeros1}, by turning off a sufficient number of such building blocks the polytope collapses and the amplitude vanishes. This geometric picture also tells us that in the last step before the polytope collapses, it takes the form of a ``sandwich'', with an interval separating opposite facets of the associahedron. This implies that the amplitude factorizes into lower point amplitudes, in a completely predictable fashion. It is fascinating to discover a new sort of factorization of amplitudes, which has nothing to do with the usual factorization on poles, but instead characterizes the behavior of amplitudes as we approach the hidden zeros. 

Of course the behavior of amplitudes near poles is perhaps the best studied aspect of the physics of particle scattering. By stark contrast, the kinematic locus where amplitudes vanish has hardly been explored, and a clear interpretation of our zeros in familiar physical terms is still lacking. Indeed both the zeros and factorization near zeros are properties of the whole amplitude, not features manifest from the Feynman diagram perspective; they are instead made manifest by the alternative geometric description of the amplitudes provided by the associahedron.  
 
Amazingly, we will find that exactly the same patterns of zeros and an avatar of the factorization near zeros generalizes to all interesting theories of colored particles: the Non-linear Sigma Model (NLSM) for pions, as well as gluons in Yang-Mills theory (YM). In sections \ref{sec:FTNLSM} and \ref{sec:FTYM}, we explain how this generalization works. The universality of these zeros seen in other colored theories is especially surprising given that no obvious associahedron-ic formulation for these theories is known.

However, as we will see, there {\it is} a beautiful reason for the universality of these zeros, which turn out to be deeply related to a surprising relation between these colored theories revealed upon understanding a unified  ``stringy'' descriptions of all these amplitudes. These stringy generalizations have all the zeros and factorization patterns of the field theory amplitudes and in fact generalize them to infinite new families of zero/factorization patterns. They also allow us to see that amplitudes for colored scalars, pions, and gluons are all given by a single function, expanded about different points in the kinematic space. This remarkable connection between Tr$(\phi^3)$ scalars, pions, and gluons will be explored at length in sections \ref{sec:StringPhi} and \ref{sec:StringDelta}. This single function is originally known as the $n$-point Koba-Nielsen string amplitudes~\cite{Koba:1969rw}, thus the zeros and factorization properties of all these theories at tree level ultimately come from (bosonic) string theory. 

Our goal in this paper is to explain the hidden zero/factorization patterns in the simplest possible setting and use this to motivate the new descriptions for amplitudes of pions and gluons arising from a simple kinematic shift of the Tr$(\phi^3)$ theory. To keep the discussion as simple as possible, for the story of zeros and factorization we will focus on tree-level amplitudes; this will already be enough to suggest the kinematic shift relating all the colored theories, that naturally generalizes to all loop orders. With this impetus as a starting point, we will take up a detailed description of both the Non-linear Sigma Model and Yang-Mills amplitudes from this point of view in upcoming works \cite{NLSM,Gluons}. 

~\\ 
\noindent 
\textbf{Note added:} After the first version appeared on arXiv, we were notified that some zeros of the dual resonant amplitudes have also been studied in the early days of string theory~\cite{DAdda:1971wcy}. Using the monodromy relations, it has been shown that the tachyon amplitudes vanish under the same kinematic condition we found for stringy Tr$(\phi^3)$ here. In addition, more zero conditions that involve multiparticle Mandlestam variables are considered, as well as the obvious extensions of the zeros of amplitudes for excited states, where the conditions depend solely on Mandelstam variables and not on Lorentz products containing polarization vectors, which differs from the results presented here.

\section{Tr$(\phi^{3})$ Theory and the Associahedron}
\label{sec:ABHY}
In this section we will review the associahedron construction presented in \cite{Arkani-Hamed:2017mur}, and explain in detail the pattern of zeros and factorizations for Tr$(\phi^3)$ theory that is made obvious by this construction. Henceforth we will focus on tree-level amplitudes. However, since there is an analogous Minkowski sum picture for polytopes describing loop integrands \cite{Arkani-Hamed:2023lbd, surfacehedron}, we expect these observations to generalize at loop level.  
 
The theory we are interested in is a theory of colored massless scalars interacting via a cubic interaction, described by the following Lagrangian:
\begin{equation}
    \mathcal{L}_{\mathrm{Tr}(\phi^3)}= \operatorname{Tr}(\partial \phi)^2+ g \operatorname{Tr}(\phi^3),
\end{equation}
where $\phi$ is an $N\times N$ matrix. 
 
Since this is a scalar theory, the amplitude is exclusively a function of the Lorentz invariant dot products of momenta: $p_i \cdot p_j$. There are $\frac{n(n-1)}{2}$ of these invariants, however momentum conservation gives us $n$ relations between them via $\sum_j p_i \cdot p_j = -p_i^2 = 0$, so we have $\frac{n(n-1)}{2} - n$ independent invariants \footnote{Note that our discussion holds for general arbitrarily large spacetime dimension $D$; in any fixed spacetime dimension there are further ``gram determinant'' constraints on the dot products $p_i \cdot p_j$ for $n>D$ particles.}. But there is no canonical way of imposing the momentum conservation constraints, and no cyclically invariant choice of the $p_i \cdot p_j$ that form a basis. This is perhaps not a surprise, since there is nothing invariantly special about dot products between pairs of momenta; many different linear combinations of these objects can also be considered, so it is no surprise that a canonical basis for the invariants does not exist ab initio; a better basis should reflect the exigencies of the physics we are trying to describe.  
\begin{figure}[t]
    \centering
    \includegraphics[width=0.4\textwidth]{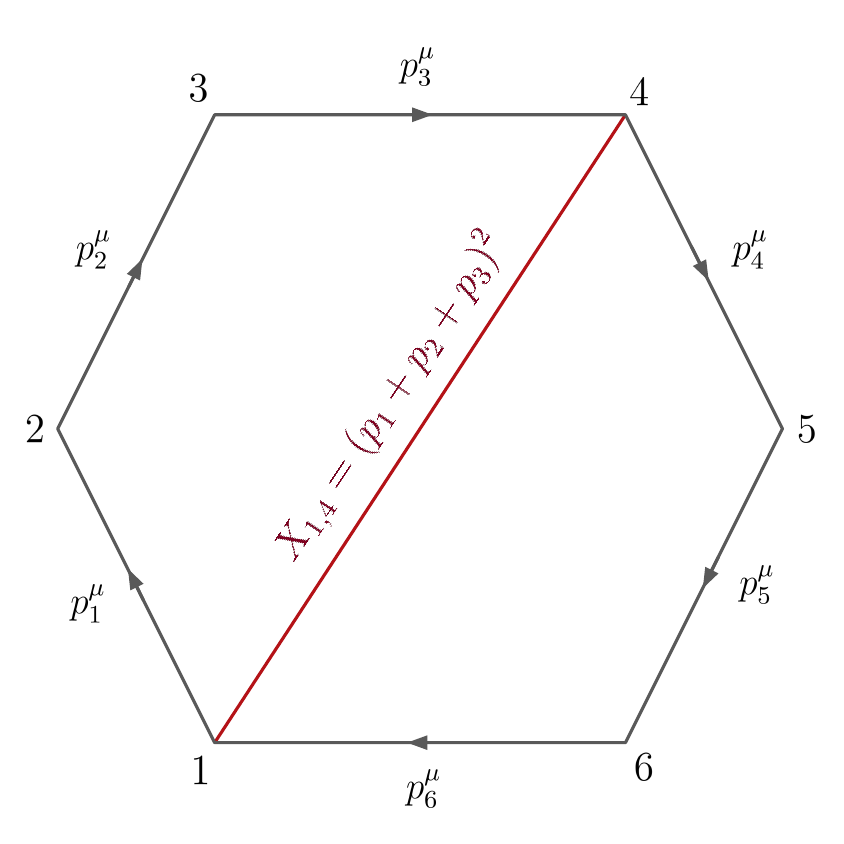}
    \caption{6-point momentum polygon.}
    \label{fig:6ptMomPoly}
\end{figure}

Indeed, there is a much nicer basis for the kinematic invariants that is directly tailored to our physical problem. Let us consider a fixed color ordering, which we can take to be without loss of generality $(1,2, \cdots, n)$.
We can keep track of the momenta of the particles in a familiar way by drawing each $p_i^{\mu}$ as an edge of the ``momentum polygon'' (see figure \ref{fig:6ptMomPoly} for the 6-point momentum polygon). We use the color ordering to order the momenta in the polygon one after the other in the same way, $(p_1^\mu, p_2^\mu, \cdots, p_n^\mu)$. The fact that the polygon closes reflects momentum conservation $\sum_i p_i^\mu = 0$. The vertices of this polygon can be associated with points $x_i^\mu$ so that $p_i^\mu = (x_{i+1}^\mu - x_i^\mu)$, which makes momentum conservation manifest. 

Now consider any chord in this polygon connecting the vertices $(i,j)$, and consider the squared $X_{i,j} = (x_i - x_j)^2$.  The $X_{ij}$ are naturally associated with the propagators that appear in Feynman diagrams for this color ordering: 
\begin{equation}
    X_{i,j} = (p_i + \dots +p_{j{-}1})^2.
\end{equation}

The tree amplitude is exclusively a function of these variables $\mathcal{A}^{\text{Tr}(\phi^3)}_n \equiv \mathcal{A}^{\text{Tr}(\phi^3)}_n(\{X_{i,j}\})$. 
Note that $X_{i,i+1} = p_i^2 = 0$ is not a dynamical variable, but all the rest of the $X_{i,j}$ are independent, and there are exactly $\frac{n(n-1)}{2} - n$ of them. Hence the $X_{i,j}$'s give a complete basis for all the kinematic invariants. This basis is much nicer than the one provided by dot products: the $X_{i,j}$ are what appears directly in the amplitudes, the basis respects the cyclic symmetry of the amplitudes,  and all the momentum conservation conditions are automatically taken into account: specifying an unconstrained set of $X_{i,j}$ fixes the kinematical data for our scattering process. 

Since the $X_{i,j}$ are a basis, we can express all the other kinematic invariants in terms of them, including the dot product we began with. If we define 
\begin{equation}
c_{i,j} = -2 p_i \cdot p_j,
\end{equation}
the relation is simply
\begin{equation}
    c_{i,j} = X_{i,j} + X_{i+1,j+1} - X_{i,j+1} - X_{i+1,j}.
    \label{eq:ceq}
\end{equation}

\subsection{The kinematic mesh}

One particularly nice way of encoding the kinematic data of the scattering process is using the \textbf{kinematic mesh}~\cite{Arkani-Hamed:2019vag}. As we will see, the mesh is not only useful for organizing the momentum invariants but will also be crucial for understanding the features of the amplitude we will be studying throughout this paper. 

\begin{figure}[t]
    \centering
    \includegraphics[width=\linewidth]{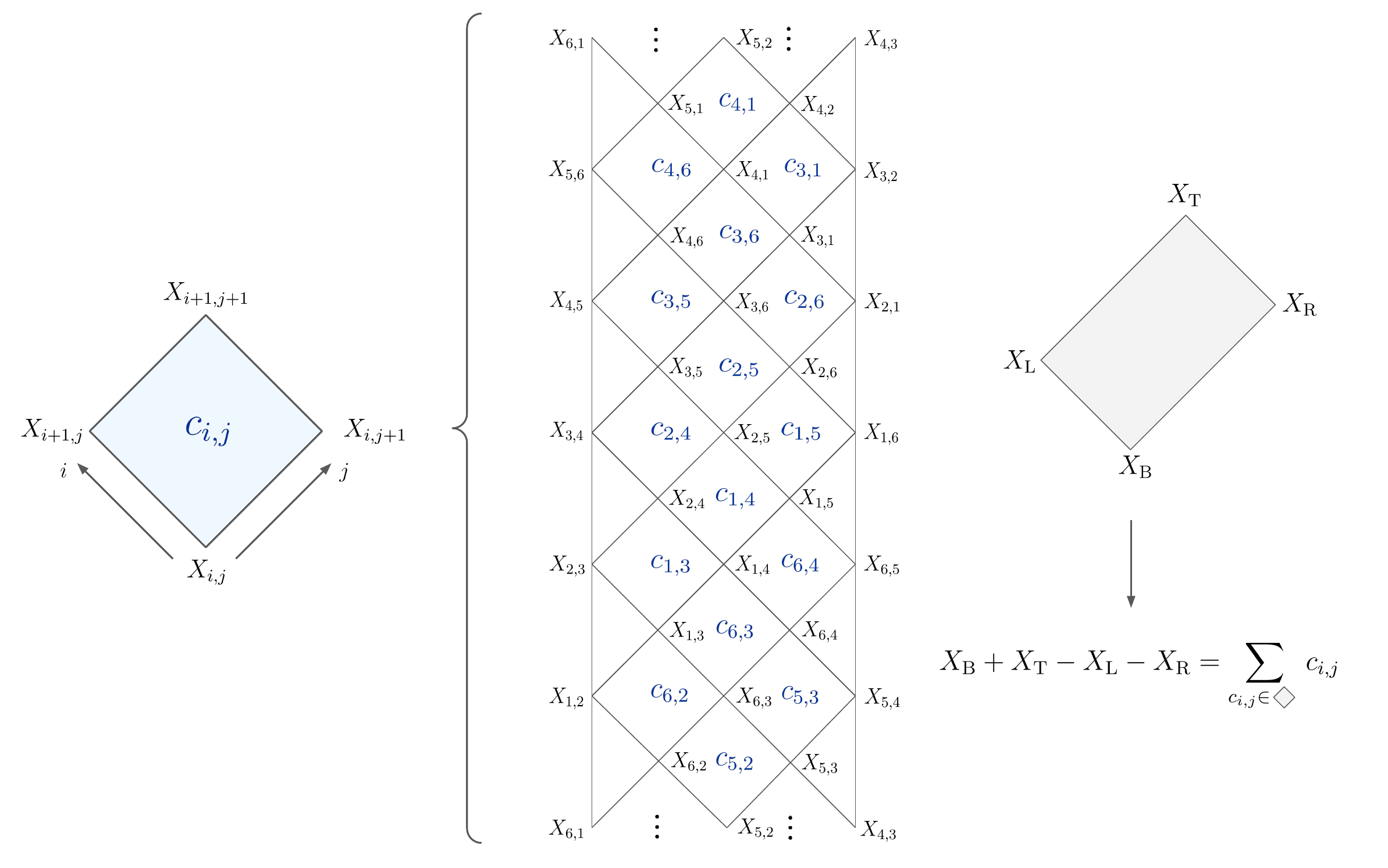}
    \caption{Kinematic mesh for 6-point.}
    \label{fig:6mesh}
\end{figure}

The guiding principle used to build the mesh is equation \eqref{eq:ceq}. We associate each $X_{i,j}$ in \eqref{eq:ceq} to the vertex of a square rotated by a $45^\circ$ angle (see figure \ref{fig:6mesh} on the left), and the corresponding $c_{i,j}$ to the square. By placing these squares together we form a square grid tilted $45^\circ$, that in the boundaries contains $X_{i,i+1} = p_i^2 =0$, since we are assuming massless particles. In figure \ref{fig:6mesh}, we present the mesh for the case of a 6-point process. We can see that all the planar variables, $X_{i,j}$, are associated with grid points. Meanwhile the {\it non-planar} dot product of momenta--which are the $c_{i,j}$ with $i, j$ non-adjacent indices-- correspond to the square tiles. The mesh extends for infinitely long but reflects the cyclic symmetry of the problem by an interesting ``Mobius'' symmetry, where we identify $X_{i,j} = X_{j,i}$ and $c_{i,j}=c_{j,i}$.

Due to the structure of \eqref{eq:ceq}, if we consider any causal diamond inside the mesh, the relation satisfied by the four $X$'s in the vertices of the causal diamond is exactly that of \eqref{eq:ceq}, where on the l.h.s. instead of $c_{i,j}$, we have the sum of all the $c_{i,j}$'s inside the causal diamond under consideration (see figure \ref{fig:6mesh} on the right). 

We have already seen that the $X_{i,j}$'s form a basis for all kinematic invariants. However, one can also build a basis in which we trade some of the planar variables for non-planar ones. In particular, we are interested in the case where we get a basis including the planar variables $X_{i^\star,j^\star}$ of a particular triangulation, $\mathcal{T}$ together with a set of non-planar Mandelstams, $c_{i^\star,j^\star}$. One way to find such a basis is by considering a minimal subregion of the mesh that includes all the planar variables, once and only once. All such subregions are in one-to-one correspondence with a triangulation of the $n$-gon, such that the basis we are interested in is formed by the $X_{i,j}$'s of this triangulation and the $c_{i,j}$'s inside this region. To obtain this subregion we do the following: start by picking a triangulation, $\mathcal{T}$, of the $n$-gon which is determining the set of $n-3$ chords, $X_{i,j}$ entering in the basis, with $(i,j)\in \mathcal{T}$. Now consider the ``rotated triangulation'', formed by chords $X_{i-1,j-1}$ with $(i,j) \in \mathcal{T}$. Then consider the region of the mesh that does \textbf{not} contain the meshes $c_{i-1,j-1}$ with $(ij) \in \mathcal{T}$. Since the mesh is infinite this will still be an infinite set of connected or disconnected regions, from which we further extract a finite subset that contains all the planar Mandelstams, once and only once -- this is the desired subregion. The respective kinematic basis we are interested in is constructed from the non-planar $c_{i,j}$'s inside the subregion together with $n{-}3$ $X_{i,j}$'s in the starting triangulation, $\mathcal{T}$. 

\begin{figure}
    \centering
    \includegraphics[width=\textwidth]{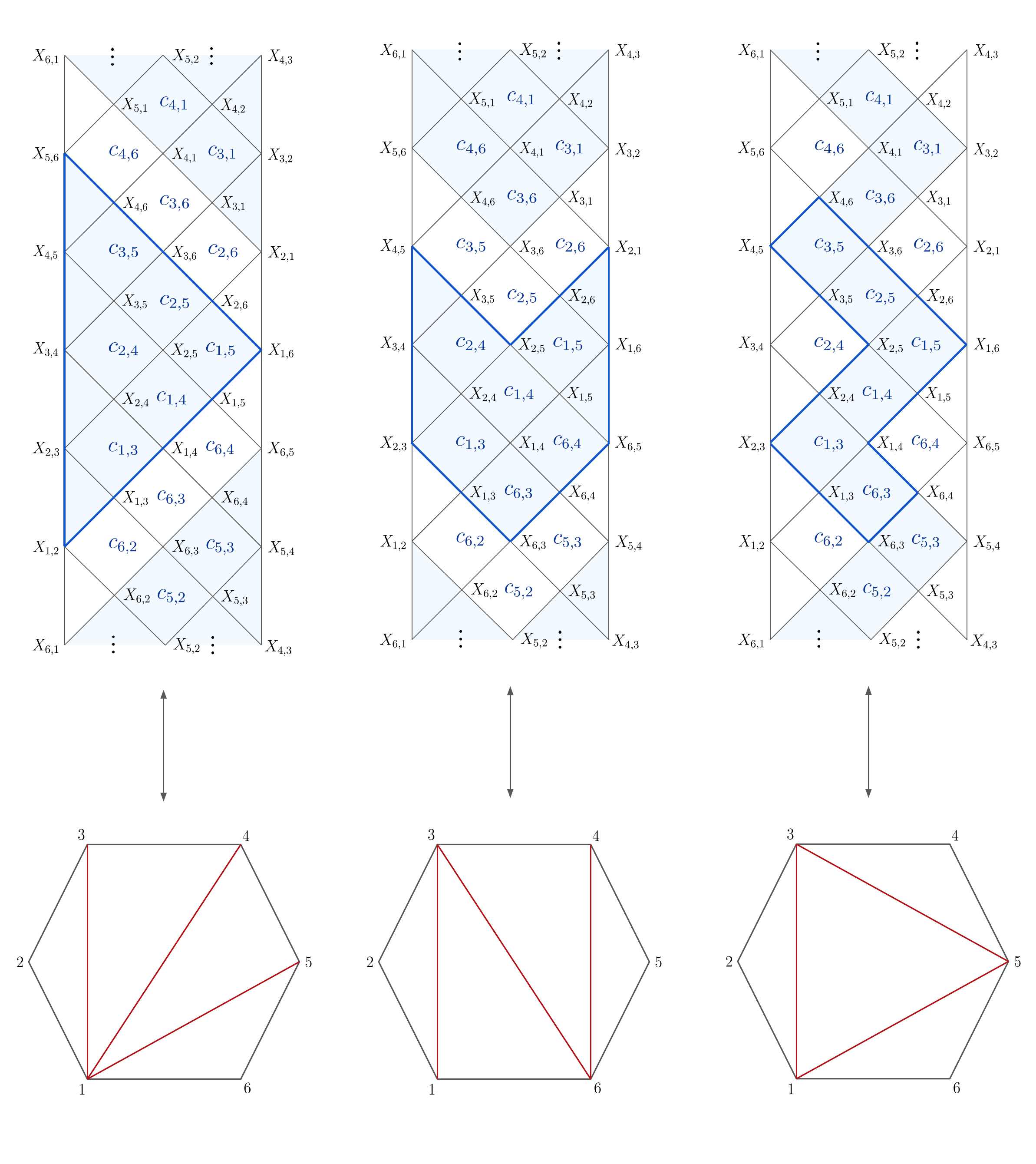}
    \caption{Triangulations and corresponding mesh subregions for 6-point kinematics.}
\label{fig:TriangMesh}
\end{figure}

In figure \ref{fig:TriangMesh} we present the regions of the mesh corresponding to the three different types of triangulation of the 6-point problem. All remaining triangulations correspond to cyclic shifts of the ones presented, and these cyclic shifts only translate the region in the mesh vertically, without altering its shape.  
 
We denote the triangulations like the one on the left of figure \ref{fig:TriangMesh} by \textbf{ray-like} triangulations. The kinematic basis associated with this region is $\{X_{1,3},X_{1,4},X_{1,5},c_{1,3},c_{1,4},c_{1,5},c_{2,4},$ $c_{2,5},c_{3,5}\}$. Such triangulations always lead to simple connected triangular regions in the mesh. As we will see, this choice of basis makes most of the interesting features manifest and therefore will be the preferred choice of kinematic basis henceforth.  
 
Now that we have properly defined and organized the kinematical data that the amplitude depends on, let us proceed to study the associahedron and understand how it is defined in kinematic space. 

\subsection{The ABHY associahedron}

As defined in \cite{Arkani-Hamed:2017mur}, the ABHY associahedron associated with an $n$-point amplitude is an $(n-3)$--dimensional polytope. Its embedding in kinematic space goes as follows:
\begin{enumerate}
    \item Define the region in kinematic space for which all the planar variables are \textbf{positive} -- $\Delta_+ = \{X_{i,j} >0\}$, for all $i<j \in \{1,\dots,n\}$.
    \item Pick a subregion of the mesh determining a basis of $n-3$ planar variables, $\tilde{X}_{i,j}$, and $(n-2)(n-3)/2$ non-planar, $\tilde{c}_{i,j}$, and solve for all the $X_{i,j}$'s in terms of this basis. 
    \item Impose that all $\tilde{c}_{i,j}>0$, so that, in this new basis, $\Delta_+$ defines a set of $n(n-3)/2$ inequalities in an $(n-3)$--dimensional space spanned by the $\tilde{X}_{i,j}$. The convex-hull of these inequalities is the ABHY associahedron. 
\end{enumerate}

From the previous procedure, we see that there are different ways of embedding this polytope in kinematic space, each of them corresponding to a different choice of basis. These different choices give rise to different realizations of the polytope, however, any statement about the amplitude should be realization-independent. 
 
Let us now see what the polytope looks like for a few simple examples. 

\subsubsection{4-point}
\label{sec:4ptAssoc}
At 4-point we only have two different planar variables, $X_{1,3}$ and $X_{2,4}$, corresponding, respectively, to the Madelstams $s$ and $t$. These are related to the non-planar variable, $c_{1,3} = -u$, in the following way:

\begin{equation}
    X_{1,3} + X_{2,4} = c_{1,3}.
\end{equation}

So picking the basis $\{X_{1,3},c_{1,3}\}$, the ABHY associahedron is:

\begin{equation}
    \{X_{1,3} >0 \, \wedge \, X_{2,4}>0 \Leftrightarrow X_{1,3}<c_{1,3}\} \Leftrightarrow 0<X_{1,3}<c_{1,3},
\end{equation}
which is simply a line segment -- a one-dimensional simplex. This is still a very small example, so to understand how it works in a less trivial case it is worth going to 5 points. 

\subsubsection{5-point}

At 5-point, there are five different planar variables. However, there is only one possible type of triangulation -- the ray-like triangulation. Let us pick the basis corresponding to the analogous of the 6-point ray-like triangulation discussed in the previous section, $i.e.$ $\{X_{1,3},X_{1,4},c_{1,3},c_{1,4},c_{2,4}\}$. Then the associahedron is defined in $(X_{1,3},X_{1,4})$ space as follows:
\begin{equation}
    \begin{cases}
        &X_{1,3}>0 \\
        &X_{1,4}>0 \\
        &X_{2,4}>0 \Leftrightarrow c_{1,3} -X_{1,3}+X_{1,4}>0 \\
        &X_{2,5}>0 \Leftrightarrow c_{1,3} +c_{1,4} -X_{1,3}>0 \\
        &X_{3,5}>0 \Leftrightarrow c_{1,4} +c_{2,4} -X_{1,4}>0 \\
    \end{cases}\quad ,
    \label{eq:5ptIneq}
\end{equation}
which is exactly given by the pentagon presented in figure \ref{fig:5ptMinkSum} (left). In section \ref{sec:Intro} we mentioned that the ABHY associahedron is given by the Minkowski sum of simplices, which is fulcral for the understanding of the zeros of the amplitude. At 4-point, the polytope is the one-dimensional simplex, however, at 5-point this decomposition is not so obvious. Let us now understand the decomposition of the polytope into its Minkowski summands. 

\subsection{The Minkowski summands and the mesh}
\label{sec:MeshMinkSum}
\begin{figure}[t]
    \centering
    \includegraphics[width=\textwidth]{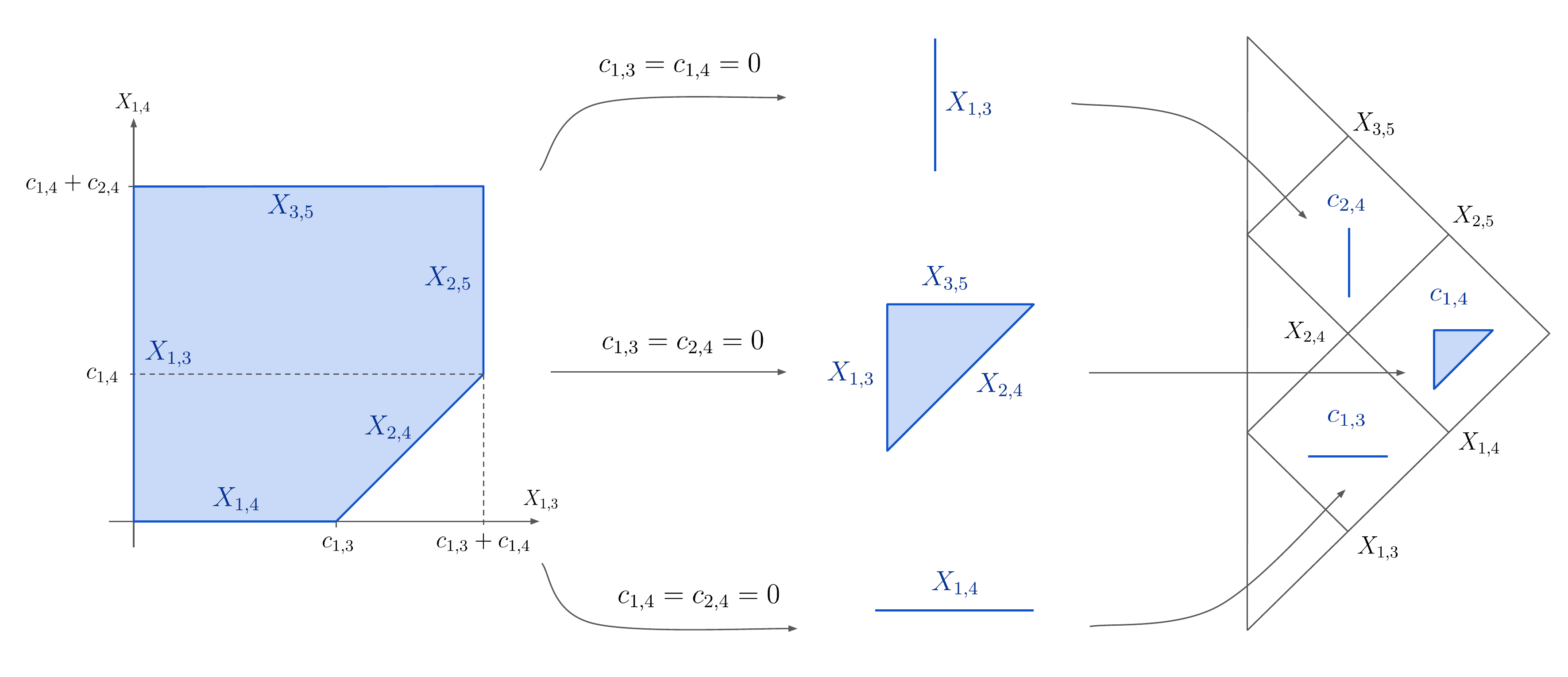}
    \caption{5-point ABHY associahedron and respective Minkowski summands.}
    \label{fig:5ptMinkSum}
\end{figure}
It turns out that each Minkowski summand is naturally associated with a mesh, $c_{i,j}$. Let us start by analyzing the 5-point example. The set of inequalities in \eqref{eq:5ptIneq} carves out a pentagon in $(X_{1,3},X_{1,4})$ space, in which each edge corresponds to an inequality $X_{i,j}>0$, and thus is uniquely associated to that planar Mandelstam  (see figure \ref{fig:5ptMinkSum}, left). In addition, we see that the location of the edges of the pentagon is set by the values of the non-planar variables $\{c_{1,3},c_{1,4},c_{2,4}\}$. Therefore we can study what happens when we set some of these non-planar variables to zero (see figure \ref{fig:5ptMinkSum}):
\begin{itemize}
    \item \underline{$c_{1,3}=c_{1,4} =0 $}: The pentagon collapses to a horizontal line segment, $i.e.$ a one-dimensional simplex. This is then the Minkowski summand corresponding to mesh $c_{2,4}$. 
    \item \underline{$c_{1,3}=c_{2,4} =0 $}: The pentagon collapses to triangle, $i.e.$ a two-dimensional simplex. This is then the Minkowski summand corresponding to mesh $c_{1,4}$. 
     \item \underline{$c_{1,4}=c_{2,4} =0 $}: The pentagon collapses to a horizontal line segment, again a one-dimensional simplex. This is then the Minkowski summand corresponding to mesh $c_{1,3}$. 
\end{itemize}

So Minkowski summing these three simplices associated with the three different non-planar variables, builds back the full pentagon. The fact that each Minkowski summand is associated with a non-planar variable, we can keep track of Minkowski summands using the mesh, just like it is shown in figure \ref{fig:5ptMinkSum} (right), where we present only the subregion of the mesh corresponding to the basis choice being used. We see that the lower-dimensional dimensional simplices are on the meshes on the left while the top-dimensional one is associated with the right-most corner. This pattern will persist at $n$-point as long as we are dealing with a basis associated with a ray-light triangulation: the meshes on the left-boundary will correspond to one-dimensional simplices and, as we move towards the right, the dimension of the simplices increases until it becomes top-dimensional in the right-most corner. 
 
This is a good point to highlight that, had we chosen a different basis, the pentagon would look different: it would be embedded in a different space, $(\tilde{X}_{i_1,j_1},\tilde{X}_{i_2,j_2})$, and the non-planar variables involved would be different. In particular, at higher points, different types of basis triangulation lead to different Minkowski summands.

\section{Zeros and Factorizations of Tr$(\phi^3)$ Tree Amplitudes}
\label{sec:zeros1}
\subsection{Zeros and factorizations -- two simple examples}

Now that we have understood how the ABHY associahedron is defined in terms of its Minkowski summands, let us proceed to the study of the zeros of the amplitude.
To do this we will start by studying in detail two simple examples: the 5-point and 6-point amplitudes. 
\subsubsection{5-point amplitude}
At 5-point, the amplitude is given by the sum of five different Feynman diagrams, corresponding to the five possible triangulations of the pentagon:
\begin{equation}
    \mathcal{A}_5 = \frac{1}{X_{1,3}X_{1,4}} +\frac{1}{X_{2,4}X_{2,5}} +\frac{1}{X_{1,3}X_{3,5}} 
    +\frac{1}{X_{1,4}X_{2,4}} 
    +\frac{1}{X_{2,5}X_{3,5}} .
    \label{eq:5ptAmpt}
\end{equation}

So to determine the kinematic locus where the amplitude vanishes, we can reduce \eqref{eq:5ptAmpt} to a common denominator and ask for the numerator to vanish. By doing this we get a cubic equation that obscures any possible simple zeros of the amplitude. 
 
However, we can recast the question about the zeros locus of the amplitude in polytopal language. As it is explained in \cite{Arkani-Hamed:2017mur}, the amplitude is the canonical form of the ABHY associahedron, and therefore if we are able to make the polytope collapse one in dimension, the amplitude will vanish. 
 
Looking back at the 5-point associahedron presented in figure \ref{fig:5ptMinkSum}, we see that by setting $c_{1,3}=c_{1,4}=0$ or $c_{1,4}=c_{2,4}=0$, the pentagon collapses into a line segment, which then means that the amplitude vanishes in this limit! Note that the same is no longer true for the case $c_{1,3}=c_{2,4}=0$, since in this limit we still get a top-dimensional object, and thus the amplitude does not vanish. 
 
Now by the cyclic invariance of the 5-point amplitude, any cyclic images of these conditions also make the amplitude vanish. Therefore, we get the simple family of zeros: pick an $i\in\{1,\dots,5\}$, the amplitude vanishes in the locus $c_{i,j}=0$, for all $j$ (non-adjacent to $i$). Even though the realization of the associahedron associated to the basis $\{X_{1,3},X_{1,4},c_{1,3},c_{1,4},c_{2,4}\}$ does \textbf{not} make manifest all these zeros, for each zero we can always find a realization in which it is manifest, as we will show in section \ref{sec:ZF_genStat}. 
 
Even though this example is still relatively simple it illustrates how the Minkowski sum picture of the associahedron justifies the presence of this simple class of zeros: Knowing that the non-planar variables are associated with individual summands that build up the full polytope then, by turning off enough of them, we can make the polytope collapse in dimension, and thus make the amplitude vanish. 

\begin{figure}[t]
    \centering
    \includegraphics[width=\textwidth]{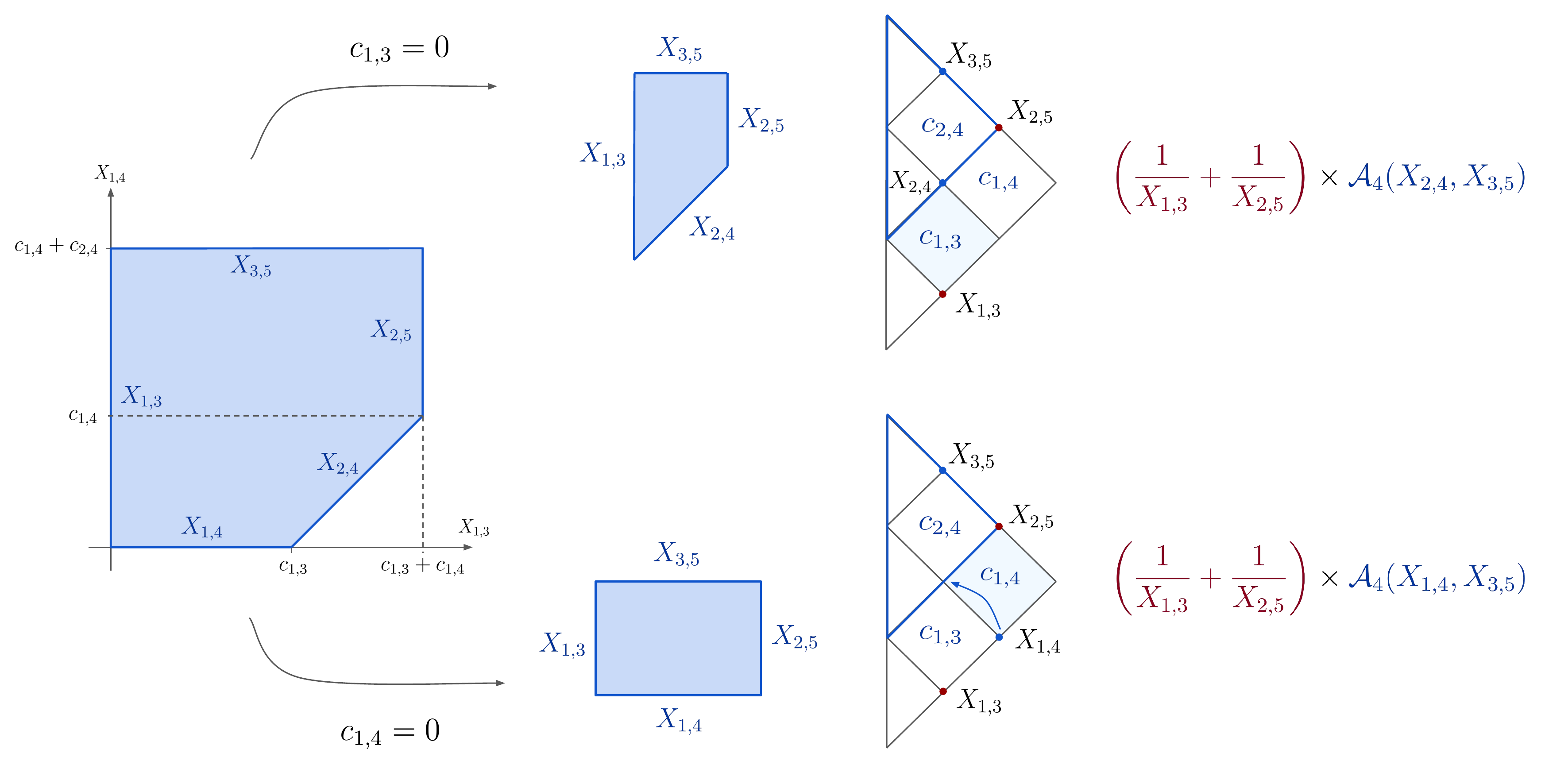}
    \caption{5-point factorization near zeros.}
    \label{fig:fact5pts}
\end{figure}
Let us now focus on the zero $c_{1,3}=c_{1,4}=0$, and understand what happens in the penultimate step before the polytope collapses, $i.e.$ when we \textbf{only} set $c_{1,3}=0$, or $c_{1,4}=0$. In the latter case, $c_{1,4}=0$, the two remaining Minkowski summands are two intervals, and so the polytope reduces to a square/rectangle (see figure \ref{fig:fact5pts}, bottom). In particular, starting with the full pentagon, we can see that by setting $c_{1,4}=0$, we lose the edge associated with $X_{2,4}$. One way to explain this fact is that since:
\begin{equation}
    X_{1,4}+X_{2,5}-X_{2,4} = c_{1,4} \xrightarrow{c_{1,4}=0} X_{2,4}= X_{1,4}+X_{2,5}.
\end{equation}

So the condition $X_{2,4}>0$, becomes redundant, since it is automatically satisfied for $X_{1,4}>0$ and $X_{2,5}>0$, and thus the corresponding facet disappears. Therefore, in this limit, the amplitude only depends on $\{X_{1,3},X_{1,4},X_{2,5},X_{3,5}\}$. In addition, from section \ref{sec:4ptAssoc}, we know that the 4-point ABHY associahedron is simply a line segment, so the fact that the geometry reduces to a product of two line segments hints that in this limit the amplitude turns into a product of two 4-point amplitudes. Indeed, by starting with the 5-point amplitude and imposing $c_{1,4}=0$, we obtain:
\begin{equation}  
\mathcal{A}_5(X_{1,3},X_{1,4},X_{2,4},X_{2,5},X_{3,5}) \xrightarrow{c_{1,4}=0}\left( \frac{1}{X_{1,3}} + \frac{1}{X_{2,5}} \right) \times \left( \frac{1}{X_{1,4}} + \frac{1}{X_{3,5}} \right) ,
\label{eq:factc14}
\end{equation}
which is indeed the product of two 4-point amplitudes with some interesting kinematics, $\mathcal{A}_4(X_{1,3},X_{2,5})$ and $\mathcal{A}_4(X_{1,4},X_{3,5})$. Let us now try to understand how we can read off this behavior from the kinematic mesh. We are currently exploiting the behavior near the zero associated with setting $c_{1,3}=c_{1,4}=0$, which corresponds to a $45^\circ$ titled rectangle in the bottom of the triangular mesh. The remaining part of the mesh, is exactly that of a 4-point problem. By setting only $c_{1,4}$ to zero, one of the 4-point factors we get exactly corresponds to this 4-point amplitude, with the kinematics entering the bottom diagonal depending on which $c_{i,j}$ we don't set to zero.  Before understanding the kinematic dependence, let us look at the other 4-point factor. This term is associated with the $X$'s at the bottom and the top of the causal diamond associated with the zero. Using the $c$-equation for the full causal diamond between $X_{1,3}$ and $X_{2,5}$, we have that $c_{1,3} = X_{1,3}+X_{2,5}$, and so we can rewrite \eqref{eq:factc14} as:
\begin{equation}
\mathcal{A}_5(X_{1,3},X_{1,4},X_{2,4},X_{2,5},X_{3,5}) \xrightarrow{c_{1,4}=0} \frac{c_{1,3}}{X_{1,3}X_{2,5}} \times \left( \frac{1}{X_{1,4}} + \frac{1}{X_{3,5}} \right), 
\end{equation}
so this factor is there to make manifest that the amplitude vanishes if we further set $c_{1,3}=0$. As we will see, for a generic $n$-point amplitude factorization, we always have such a factor that exactly ties to the zero we are exploiting.  
 
Let us now understand the kinematic dependence of the lower $4$-point amplitude. Say that, instead, we had set $c_{1,3}=0$. In this limit, the pentagon reduces to a trapezoid (see figure \ref{fig:fact5pts}, top), as we lose the edge corresponding to $X_{1,4}$. Similarly to the previous case, in this limit, the amplitude factorizes into a product of two 4-point amplitudes as follows:
\begin{equation}
\begin{aligned}
\mathcal{A}_5(X_{1,3},X_{1,4},X_{2,4},X_{2,5},X_{3,5}) \xrightarrow{c_{1,3}=0}&\left( \frac{1}{X_{1,3}} + \frac{1}{X_{2,5}} \right) \times \left( \frac{1}{X_{2,4}} + \frac{1}{X_{3,5}} \right) \\
=&\frac{c_{1,4}}{X_{1,3}X_{2,5}} \times \left( \frac{1}{X_{2,4}} + \frac{1}{X_{3,5}} \right),
\end{aligned} 
\end{equation}
\\ 
so the first factor remains the same, since we are still probing the same zero, while the second factor is now $\mathcal{A}_4(X_{2,4},X_{3,5})$. So the lower point amplitude no longer depends on $X_{1,4}$, as this edge of the polytope is now lost, and instead depends on $X_{2,4}$ which survives in this limit. 
 
In summary, we understand that by turning on different $c_{i,j}$ inside the zero causal diamond the factorization pattern does \textbf{not} change, however, the kinematic variables entering the lower point amplitudes \textbf{do} change. This is because by turning on different $c_{i,j}$ the facets of the polytope that survive in the limit are different, or, in other words, the set of inequalities $X_{i,j}>0$ that become redundant depends on the $c_{i,j}$ that we turn on. We will explain the general pattern in which the kinematics are inherited in the lower point amplitudes in section \ref{sec:ZF_genStat}. 
 
The discussion in this section is summarized pictorially in figure \ref{fig:fact5pts}.

\subsubsection{6-point amplitude}

The six-particle amplitude is a sum of 14 terms, corresponding to the Feynman diagrams in three cyclic classes: 
\begin{equation} 
{\cal A}_6 = \left(\frac{1}{X_{1,3} X_{1,4} X_{1,5}} + \frac{1}{X_{1,3} X_{3,6} X_{4,6}} + {\rm cyclic}\right) + \frac{1}{X_{1,3} X_{3,5} X_{1,5}} + \frac{1}{X_{2,4} X_{4,6} X_{2,6}}.
\end{equation}

Let us now see how the zeros and factorizations work, where we see the most generic behavior seen for all $n$, with cyclically inequivalent classes for patterns of zeros. 

\begin{figure}[t]
    \centering
    \includegraphics[width=\linewidth]{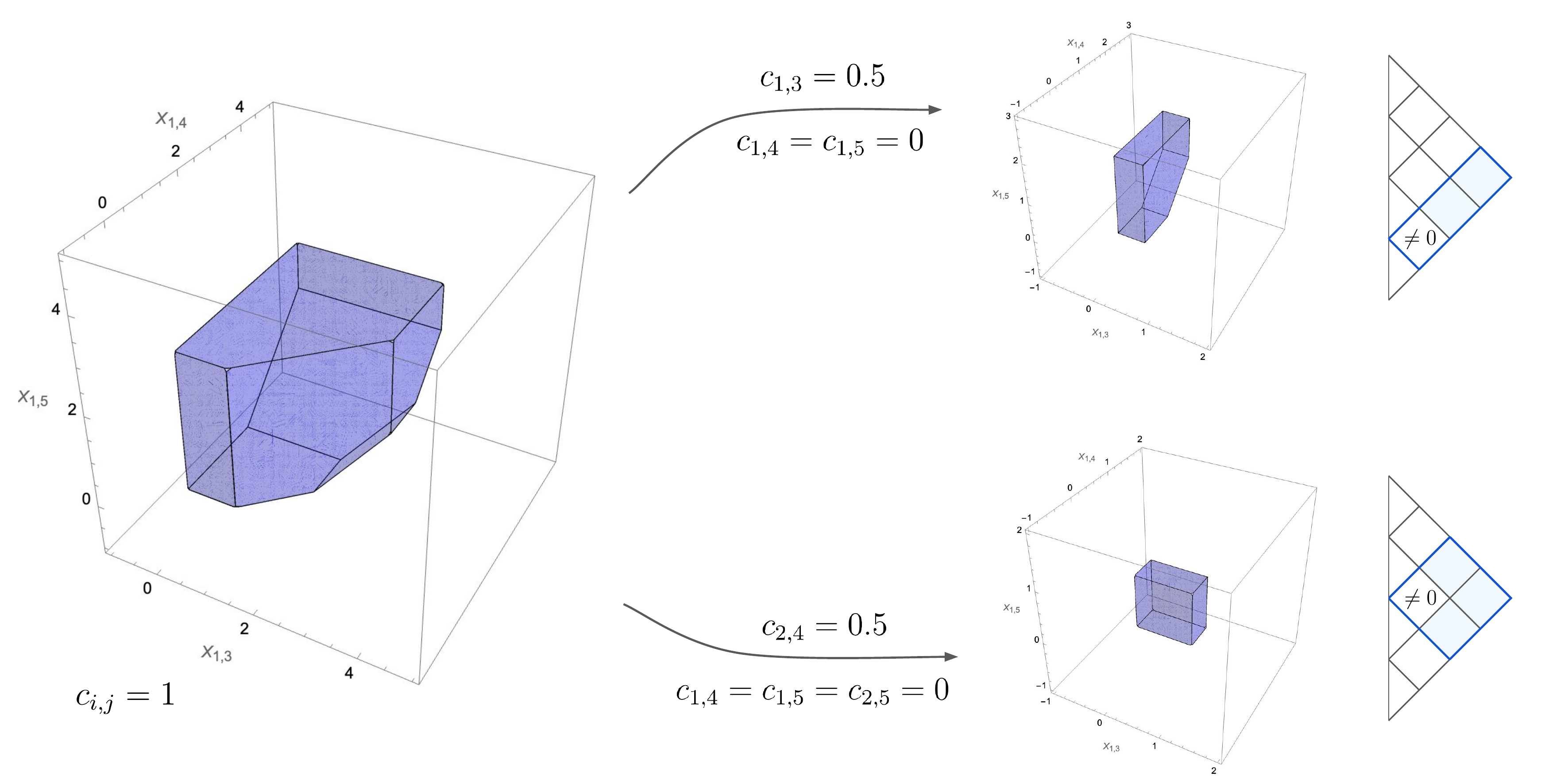}
    \caption{6-point factorization near zeros.}
   \label{fig:6ptfac}
\end{figure}

\begin{figure}
    \centering
    \includegraphics[width=\textwidth]{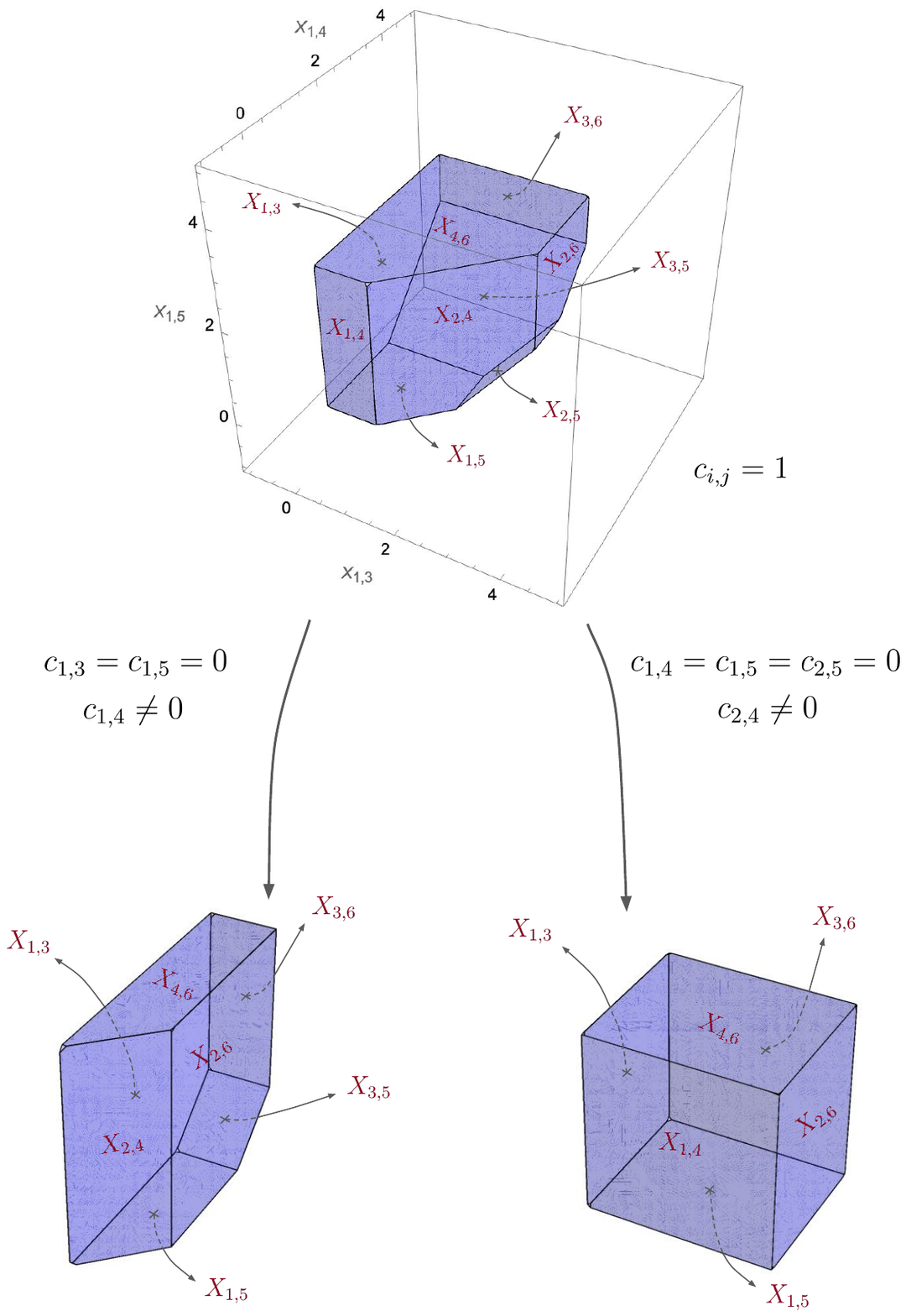}
    \caption{6-point factorization near zeros. Factorization into 5-point$\times$4-point (left) and factorization into 4-point$\times$4-point$\times$4-point (right).}
    \label{fig:6FactFull}
\end{figure}

We begin with the full ABHY associahedron, shown in the left panel of figure \ref{fig:6ptfac}. For some orientation, note that the facets lying on the co-ordinate planes $X_{1,3}, X_{1,4}, X_{1,5} \to 0$ are the pentagons ($X_{1,3} \to 0$ and $X_{1,5} \to 0$) and the square $(X_{1,4} \to 0$) as expected from the associated ${\cal A}_5 \times {\cal A}_3$ and ${\cal A}_4 \times {\cal A}_4$ factorizations. Note that we also have facets parallel to these, on the opposite side of the polytope. For instance, the facet parallel to the $X_{1,3}$ plane corresponds to $X_{2,6} \to 0$. This is obvious from the mesh picture since we have $X_{2,6} = C - X_{1,3}$ where $C=c_{1,3} + c_{1,4} + c_{1,5}$; this is the top corner of the maximal causal with $X_{1,3}$ on the bottom. In the same way the facet parallel to $X_{1,4}$ is $X_{3,6}$ and the one parallel to $X_{1,5}$ is $X_{4,6}$. This is a general feature of the ABHY associahedron: there are pairs of facets parallel to the coordinate planes that ``look the same'', corresponding to the poles associated with the bottom and top boundaries of the mesh picture. 
Let us begin with the analog of what we saw already at 5 points, the ``skinny rectangle'' zero. If we set $c_{1,3},c_{1,4},c_{1,5} \to 0$, then the three-dimensional associahedron collapses in the $X_{1,3}$ direction down to a two-dimensional pentagon. We can see this visually in figure \ref{fig:6ptfac} (a more detailed figure \ref{fig:6FactFull} appears at the end of the paper), where we have represented the penultimate step in shutting off $c$'s, where $c_{1,3},c_{1,5} \to 0$ but $c_{1,4}$ is still turned on. At this point, we have a 
``sandwich'', with the $X_{1,3}$ facet and its parallel cousin $X_{2,6}$ facet, separated by an interval. When we further shut off $c_{1,4}$ the interval shrinks to zero and we are left with the pentagon, showing that the amplitude vanishes in this limit.  We can also look at shutting off $c_{1,4},c_{1,5},c_{2,4},c_{2,5}$ where the associahedron collapses to a square. This is also shown in figure \ref{fig:6ptfac}, where we have again shown the penultimate step where we have turned $c_{2,4}$ back on. We again have a ``sandwich'' with $X_{1,4}$ and opposite $X_{3,6}$ facets, separated by an interval. The interval shrinks to zero when $c_{2,4} \to 0$; the associahedron collapses to a square and the amplitude vanishes. Just as at five points, in the penultimate step before the associahedron collapses, the amplitude factorizes, as we can see explicitly: 
\begin{eqnarray}
    {\cal A}_6 &\xrightarrow{c_{1,3}=0, c_{1,5}=0}&  \left(\frac{1}{X_{1,3}} + \frac{1}{X_{2,6}} \right) \times \left(\frac{1}{X_{2,4} X_{1,5}} + \frac{1}{X_{1,5} X_{3,5}} + \frac{1}{X_{3,5} X_{3,6}} + \frac{1}{X_{3,6} X_{4,6}} + \frac{1}{X_{4,6} X_{2,4}} \right) ,\nonumber \\ 
    {\cal A}_6 &\xrightarrow{c_{1,4}=0,c_{1,5}=0, c_{2,5}=0}&  \left(\frac{1}{X_{1,4}} + \frac{1}{X_{3,6}} \right) \times \left(\frac{1}{X_{1,3}} + \frac{1}{X_{2,6}} \right) \times \left(\frac{1}{X_{1,5}} + \frac{1}{X_{4,6}} \right).
\end{eqnarray}

\begin{figure}[t]
    \centering
    \includegraphics[width=\textwidth]{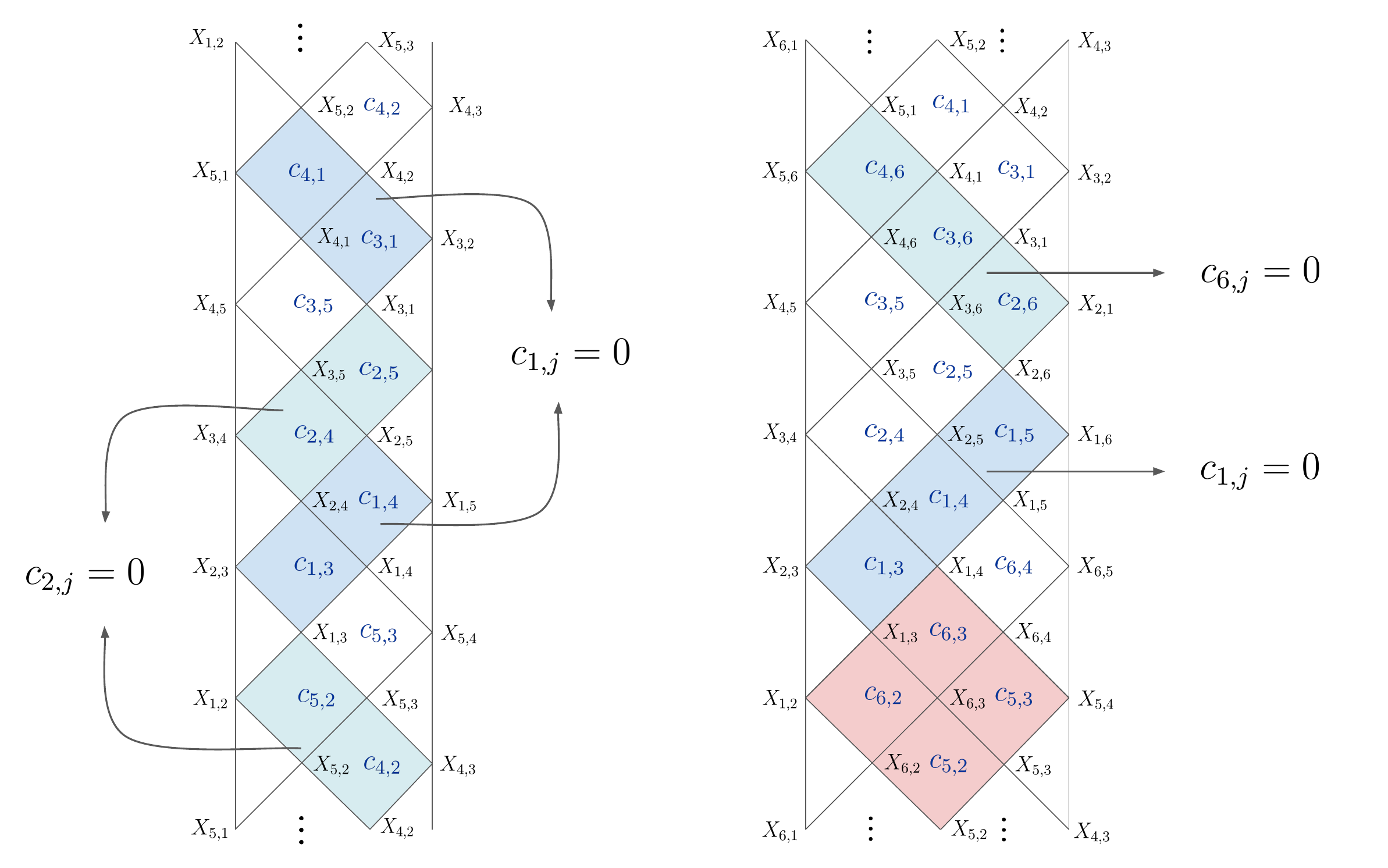}
    \caption{Some of the zeros for 5-point (left) and 6-point (right) amplitudes.}
    \label{fig:Zeros56}
\end{figure}
In the top line, we see the factor $(1/X_{1,3} + 1/X_{2,6})$, which vanishes when we further set $c_{1,4} \to 0$, since $X_{1,3} + X_{2,6} = c_{1,3} + c_{1,4} + c_{1,5} \to 0$ when we set $c_{1,4} \to 0$. It multiplies a five-particle amplitude, but with some interestingly redefined kinematic variables. If we look at the picture of the associahedron in this limit, the two bounding facets of the ``sandwich'' are precisely $X_{1,3}$ and $X_{2,6}$, while the facets keeping $X_{1,3}, X_{2,6}$ can be read off going around the pentagons as $X_{2,4}, X_{1,5},X_{3,5},X_{3,6},X_{4,6}$, which precisely defines the kinematics for an effective five-particle amplitude given in the second factor. The same story holds for the ``big square'' zero/factorization. The first factor $(1/X_{1,4} + 1/X_{3,6})$ vanishes when $c_{2,4} \to 0$ since $X_{1,3} + X_{2,6} = c_{1,4} + c_{1,5} + c_{2,4} + c_{2,5} \to 0$.  At finite $c_{2,4}$ the facets in between the $X_{1,4}$ and $X_{3,6}$ facets are $X_{1,3},X_{1,5},X_{2,6},X_{4,6}$. This is exactly the direct product of two effective 4-point problems with variables $(X_{1,3},X_{2,6})$ and $(X_{4,6},X_{1,5})$, which appear in the 4-point amplitude factors.   

In figure \ref{fig:Zeros56}, we summarize the patterns of zeros found for the 5-point and 6-point amplitudes using the kinematic mesh. We can see that all the zeros patterns are causal diamonds that extend to the boundaries of the mesh picture.

\subsection{Zeros and factorizations -- general statement}
\label{sec:ZF_genStat}

We now present the general statement about the zeros and factorization of $n$-point tree-level Tr$(\phi^3)$ amplitudes. 
As we will see the motivation and proof of these statements follow easily from simple properties of the associahedron. 
In later sections, we will give a different proof, beginning from the stringy integral representation of these amplitudes, that will generalize the statements beyond the field theory limit to full string amplitudes. 
\begin{figure}[t]
    \centering
    \includegraphics[width=0.8\textwidth]{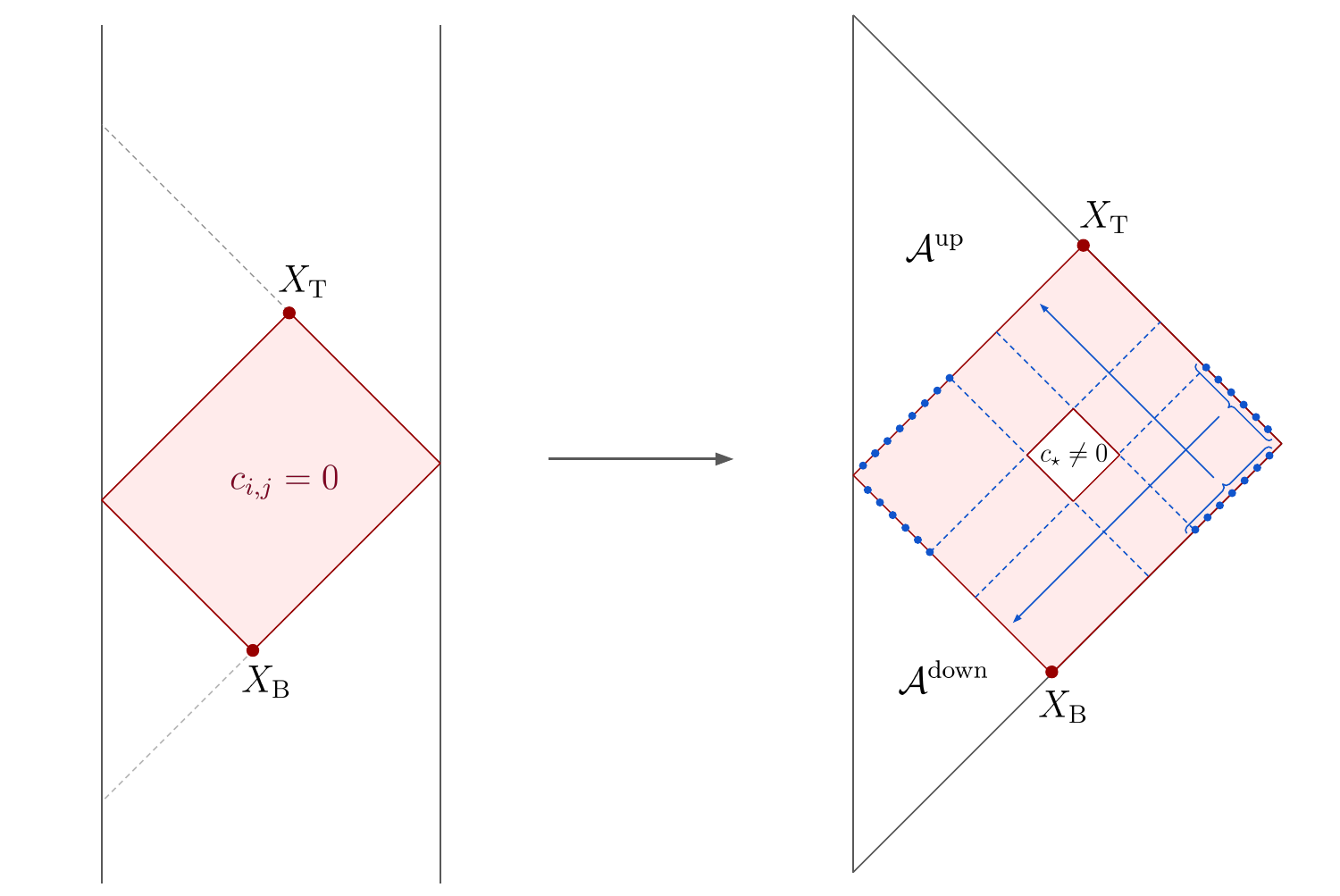}
    \caption{Zeros (left) and Factorizations with associated kinematic shifts (right).}
    \label{fig:ZerosFactGS}
\end{figure}
\paragraph{Zeros} Consider an $n$-point tree-level amplitude in Tr$(\phi^3)$ theory. Draw the corresponding $n$-point kinematic mesh. Pick a point in the mesh, $i.e.$ a planar variable $X_B$, and consider the causal diamond anchored in this variable: follow the two light rays starting at $X_B$, let them bounce in the boundaries of the mesh, and meet again in some other point, $X_T$. This encloses a region -- a causal diamond. Setting all the $c_{i,j}$ inside this causal diamond to zero will make the amplitude vanish (see figure \ref{fig:ZerosFactGS}). 

\paragraph{Factorizations} Let us consider turning back on one of the $c$'s inside the zero causal diamond, $c_\star \neq 0$. Then the amplitude factorizes into the product of lower point amplitudes in the following way (see figure \ref{fig:ZerosFactGS}):
\begin{equation}
    \mathcal{A}_n(c_\star \neq 0)= \left(\frac{1}{X_{\text{B}}}+ \frac{1}{X_{\text{T}}} \right) \times \mathcal{A}^{\text{down}}\times \mathcal{A}^{\text{up}}.
\end{equation}

Now the kinematic dependence of $\mathcal{A}^{\text{down}}$ and $\mathcal{A}^{\text{up}}$ can be read off from the kinematic mesh, and is summarized in figure \ref{fig:ZerosFactGS}. The question is what are the kinematic variables that enter in the upper/lower part of the down/up amplitudes. These are precisely those of the facets of the associahedron that are not lost in this kinematic limit. 
 
Let us start by looking at $\mathcal{A}^{\text{down}}$. We see that for any $X$ lying beneath the $c_\star$, we can build a rectangle with $X$ on the left vertex, $X_{i,i+1}=0$ in the boundary on the right vertex, $X_{B}$ on the bottom vertex, and $\tilde{X}$ on the top vertex, where $\tilde{X}$ lives on the upper boundary of the mesh. In this limit all the $c_{i,j}$ inside this rectangle are \textbf{zero}, and so using the $c$-equation we can write:
\begin{equation}
    X = X_B + \tilde{X}.
\end{equation}

Just like we saw in the 5-point example, this then means that the inequality $X>0$ becomes redundant, and so the facet of the polytope associated with $X$ disappears. Consequently, the variable that survives and enters the lower point amplitude is $\tilde{X}$. Exactly in the same way all the $X$'s above $c_\star$ for $\mathcal{A}^{\text{up}}$, disappear and instead the $\tilde{X}$ from the bottom boundary of the mesh are the ones that enter $\mathcal{A}^{\text{up}}$. 
 
Finally, for all the $X$'s lying above $c_\star$ in $\mathcal{A}^{\text{down}}$, if we try to build the same rectangle, now it will include $c_\star\neq 0$, and the argument does not hold anymore. However, we can now build a rectangle connecting $X$ to $\tilde{X}$, $X_{\text{T}}$ and a $X_{i,i+1}$ in the left boundary of the mesh. Inside this rectangle all the $c_{i,j}=0$, and so we get:

\begin{equation}
    \tilde{X} = X_{\text{T}} + X,
\end{equation}
which tells us that for this region the facets associated with $\tilde{X}$ disappear and the ones with $X$ survive. Exactly the same argument tells us that the $X$'s beneath $c_{\star}$ for $\mathcal{A}^{\text{up}}$ survive in this limit and thus appear in $\mathcal{A}^{\text{up}}$.
 
By now it should be clear that both the zeros and the factorization are properties of the amplitudes of Tr$(\phi^3)$ theory that are fundamentally hidden in the Feynman diagram formulation of these objects. It is instead the underlying geometry that gives us access to them and even suggests looking for them in the first place. 
 
As we will see shortly, these properties extend immediately to full string amplitudes. Before getting to that, we will first go through an even more magical fact -- that these generalize to the Non-linear Sigma Model and Yang-Mills theory. 
This generalization is surprising because there is no known geometrical formulation for these theories. Ultimately, the reason for the emergence of these properties will be manifest when we see that these theories are secretly simple deformations of each other. These deformations are defined at the level of stringy formulations of the amplitudes that we go over in section \ref{sec:StringDelta}.

\section{The Non-linear Sigma Model}
\label{sec:FTNLSM}

We were motivated to look for and predict zeros of the Tr$(\phi^3)$ amplitude from the Minkowski sum picture of the associahedron. But in the end, the zeros are associated with a simple locus in the space of non-planar Mandelstam invariants. As such it is natural to wonder whether other colored theories, which have the same notion of color-ordering and planarity/non-planarity, may also have such zeros. Clearly random theories will not have these zeros, for instance, the amplitude for Tr$(\phi^4)$ is simply a constant at four points and does not have our zero. But this is not an especially ``nice'' theory. Perhaps the most natural theory to examine is the Non-linear Sigma Model for pions, which already does have famous Adler zeros associated with soft limits. Of course on the surface the Tr$(\phi^3)$ theory and the NLSM could not appear more different: Tr$(\phi^3)$ theory has no derivatives in its interactions, and has non-vanishing amplitudes for all multiplicity; by contrast, pions are derivatively coupled and only have non-vanishing amplitudes for an even number of particles! Nonetheless, in this section we will see experimentally that the NLSM amplitudes have zeros in precisely the same locus of kinematics we uncovered for the Tr$(\phi^3)$ theory.

The scattering of massless pions can be described by the $\mathrm{U}(N)$ Non-Linear Sigma Model, and we record the Lagrangian in Cayley parametrization ({\it c.f.}~\cite{Kampf:2013vha}):
\begin{equation}
\mathcal{L}_{\mathrm{NLSM}}=\frac{1}{8 \lambda^2} \operatorname{Tr}\left(\partial_\mu \mathrm{U}^{\dagger} \partial^\mu \mathrm{U}\right), \quad \text { with } \quad \mathrm{U}=(\mathbb{I}+\lambda \Phi)(\mathbb{I}-\lambda \Phi)^{-1},
\end{equation}
where $\Phi=\phi_I T^I$, with $T^I$ the generators of $U(N)$ flavor group and $\lambda$ is the coupling constant. It is straightforward to derive color-ordered Feynman rules, {\it e.g.} for Tr$(1,2,\cdots, n)$, and the vertex with two derivatives for any even multiplicity $2m$ is given by
\begin{equation}
V_{2m}=-\frac{\lambda^{2m{-}2}}{2} \sum_{r=0}^{m-1} \sum_{a=1}^{2m} p_a \cdot p_{a+2 r+1}\,.
\end{equation}

NLSM tree amplitudes have been studied in {\it e.g.}~\cite{Cheung:2014dqa,Cachazo:2014xea}. It is well known that odd-point NLSM amplitudes vanish and even-point amplitudes have the Adler zero~\cite{Adler:1964um}, {\it i.e.} $A_{2n}^\mathrm{NLSM}\sim \mathcal{O}(\tau)$ when any external momentum becomes soft, $p_i^\mu=\tau \hat{p}_i^\mu$ with $\tau \to 0$. Hereafter, we will absorb the coupling constant $\lambda$ by defining: $\mathcal{A}_{2n}^\mathrm{NLSM}\equiv \lambda^{2-n} A_{2n}^\mathrm{NLSM}$, therefore, {\it e.g.} the 4-point amplitude reads $\mathcal{A}_{4}^\mathrm{NLSM}=X_{1,3}+X_{2,4}$. And the 6-point result is
\begin{equation}
\begin{aligned}   \mathcal{A}_6^{\text{NLSM}} =&  \frac{(X_{1,3}+X_{2,4})(X_{1,5}+X_{4,6})}{X_{1,4}} - X_{1,3} -X_{2,4} + (\text{cyclic},i\rightarrow i+2) .
\end{aligned}
\end{equation}
\subsection{Zeros and the soft limit}
Surprisingly, all the zeros that we described for Tr$(\phi^3)$ theory are also zeros of NLSM tree-level amplitudes. So starting with a mesh describing $2n$-kinematics, by picking any causal diamond like the one in figure \ref{fig:ZerosFactGS} and setting all the $c_{i,j}$'s inside it to zero, the NLSM amplitude vanishes. 
 
As explained previously, depending on the $X_{\text{B}}$ we pick to build the zero causal diamond, the shape of the diamond, and in particular, the codimension of the zero changes. The \textbf{smallest} codimension zeros are the \textbf{skinny rectangle} ones, where we pick an index $i$, and set all $c_{i,j}=0$, for all the $j$ not adjacent to $i$. It is well known that pion amplitudes have the Adler zero, $i.e.$ vanish when one particle is soft. Note however that the zero we are presenting is \textbf{not a soft limit}, instead it is stronger, in the sense that the fact that the amplitude vanishes in this zero implies that it has the Adler zero.
 
Let us look concretely at the 4-point amplitude. In this case, the zero would be $c_{1,3}=0$, and so $p_1 \cdot p_3 =0$, which, however does \textbf{not} require any particle to be soft. Indeed we have that the 4-point pion amplitude is:
\begin{equation}
    \mathcal{A}_4^{\text{NLSM}} = X_{1,3} + X_{2,4} = c_{1,3} \,\xrightarrow{c_{1,3}=0}\,\mathcal{A}_4^{\text{NLSM}} =0,
\end{equation}
and so vanishes as expected. Of course, for this zero to imply the Adler zero, it is crucial that the amplitude does not have poles when $p_i^\mu \rightarrow 0$, which is true for the NLSM since we always have even-point interactions. The same is not true for Tr$(\phi^3)$ theory, and this is why the zero in this context does not imply the vanishing in the soft limit. 
 
At last, it is worth highlighting that, despite the fact that the skinny rectangle zero is somehow related to the Adler zero, the same is not true for the other codimension zeros that we predict from the mesh picture. Just like in the Tr$(\phi^3)$ case, there is no clear physical explanation for the presence of these general families of zeros that are there for colored scalars and pions. 
\subsection{Factorizations}
Let us now understand how the statement about factorization near zeros generalizes to pions. For pion scattering, we always start with a $2n$ kinematical mesh and, for a particular codimension zero, $i.e.$ a particular zero causal diamond, we can ask what happens if we turn on one of the $c$'s inside the zero causal diamond, $c_\star \neq 0$.  
\begin{figure}[t]
    \centering
    \includegraphics[width=\linewidth]{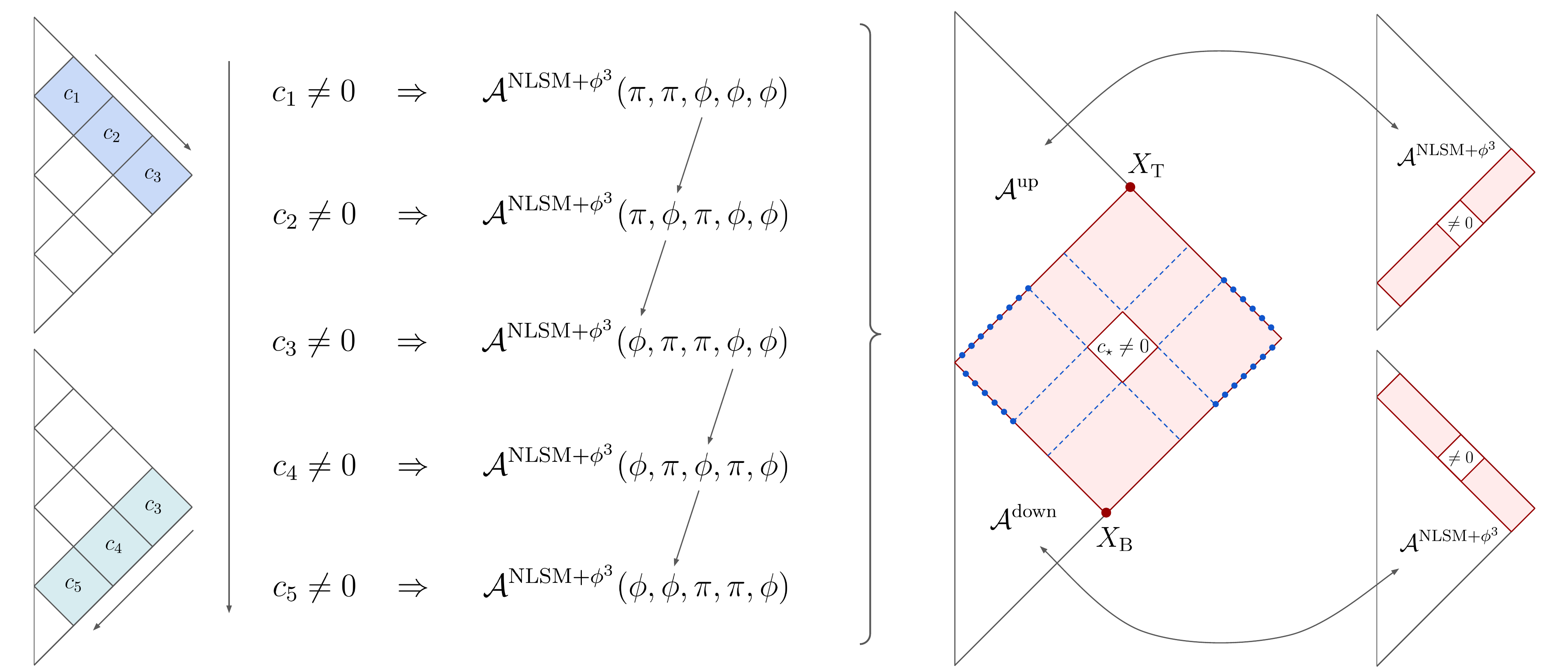}
    \caption{Factorization into mixed theory amplitudes, NLSM+$\phi^3$.}
    \label{fig:mixedFact}
\end{figure}
It turns out, that if the lower point amplitudes, $\mathcal{A}^{\text{down}}$ and $\mathcal{A}^{\text{up}}$, are both \textbf{even}-point amplitudes then the factorization holds exactly in the same way we described for Tr$(\phi^3)$:
\begin{equation}
     \mathcal{A}_{2n}^{\text{NLSM}}(c_\star \neq 0)= \left(\frac{1}{X_{\text{B}}}+ \frac{1}{X_{\text{T}}} \right) \times \mathcal{A}^{\text{down,NLSM}}\times \mathcal{A}^{\text{up,NLSM}},
     \label{eq:factpions}
\end{equation}
where the kinematic dependence of $\mathcal{A}^{\text{down,NLSM}}$ and $ \mathcal{A}^{\text{up,NLSM}}$ are determined exactly in the way we explained for Tr$(\phi^3)$ (according to figure \ref{fig:ZerosFactGS}). 
  
However, when the factorization pattern produces \textbf{odd}-point amplitudes we have something new. As we know there are no odd-point amplitudes for pions, so instead the amplitudes that enter in \eqref{eq:factpions} are amplitudes in the mixed theory of pions and scalars~\cite{Cachazo:2016njl}:
\begin{equation}
     \mathcal{A}_{2n}^{\text{NLSM}}(c_\star \neq 0)= \left(X_{\text{B}}+ X_{\text{T}} \right) \times \mathcal{A}^{\text{down,NLSM+$\phi^3$}}\times \mathcal{A}^{\text{up,NLSM+$\phi^3$}}.
     \label{eq:factpionsodd}
\end{equation}

In addition, we see that the prefactor involving $X_{\text{B}}$ and $X_{\text{T}}$ also changed: instead of the $4$-point scalar amplitude, we have the $4$-point NLSM amplitude. This change is due to the fact that mixed amplitudes have different units than NLSM amplitudes. But note that the prefactor is still such that makes manifest that the amplitude vanishes if we further set $c_\star=0$.
 
Now the kinematic dependence of $\mathcal{A}^{\text{down,NLSM+$\phi^3$}}$ and $\mathcal{A}^{\text{up,NLSM+$\phi^3$}}$ are once more determined by the kinematic shifts described for Tr$(\phi^3)$, however, we still need to specify the configuration of $\pi$'s and $\phi$'s entering these amplitudes. It turns out that this depends on the choice of $c_\star$ that is set to non-zero, but it will always be the case that for a $2n{-}1$ amplitude, we will always have 3 $\phi$'s and $2n{-}4$ $\pi$'s. 
 
To describe the rule let us start by considering a 6-point pion amplitude and consider the factorization that leads to a 5-point amplitude, corresponding to turning on a $c_{i,j}$ inside a skinny rectangle (see figure \ref{fig:mixedFact}, left). The skinny rectangle can either be on the top of the mesh or at the bottom. Let us start by looking at the case in which it is on the top, like the one highlighted in blue in figure \ref{fig:mixedFact}. Then by turning on the top $c$, we get the amplitude in which the three $\phi$'s are the last three particles, at five points this then means we have two $\pi$'s and three $\phi$'s. For a general $2n-1$ mixed amplitude, we would have $2n-4$ $\pi$'s followed by 3 $\phi$'s. Now as we go down the skinny rectangle turning on different $c_{i,j}$, the pattern is that the first $\phi$ next to the $\pi$'s starts moving past them, once at a time. Since there are $2n-3$ meshes in the skinny rectangle, once we reach the point to turn on the $c_{i,j}$ corresponding to the right-most mesh, the $\phi$ has now passed the full string of $\pi$'s so that we have $\mathcal{A}^{\text{NLSM+$\phi^3$}}(\phi, \pi, \pi, \phi, \phi)$, at $5$-point case (figure \ref{fig:mixedFact}, left), and $\mathcal{A}^{\text{NLSM+$\phi^3$}}(\phi, \pi,\dots,  \pi, \phi, \phi)$, for the general $2n-1$ mixed amplitude. 
 
Let us now continue by looking at the bottom skinny rectangle. Starting from the right-most mesh we have $\mathcal{A}^{\text{NLSM+$\phi^3$}}(\phi, \pi,\dots,  \pi, \phi, \phi)$, and going down is going to make the $\phi$ to the right of the string of $\pi$'s move through them, just like in the previous case. So that when we reach the left-most bottom mesh, we get $\mathcal{A}^{\text{NLSM+$\phi^3$}}(\phi, \phi, \pi,\dots,  \pi, \phi)$.
 
The way this generalizes for the factorization involving a general codimension zero is summarized in figure \ref{fig:mixedFact} (right). By picking some $c_\star \neq 0$ inside a given causal diamond, the $\phi$, $\pi$ configuration in $\mathcal{A}_{2n-1}^{\text{down,NLSM+$\phi^3$}}$ and $ \mathcal{A}_{2m-1}^{\text{up,NLSM+$\phi^3$}}$ is exactly the same as the ones we would have obtained considering a factorization from a skinny rectangle in a $2n$, $2m$ problem, respectively, where the $c_{\star}$ occupies the same relative position in these lower problems, as it does in the bigger one (see figure \ref{fig:mixedFact}, right).

\subsubsection{Examples}
Let us now see this factorization in action for the case of the 6-point pion amplitude. 
 
We start by looking at the factorization associated with the square causal diamond, for which the lower point amplitudes are 4-point amplitudes, and thus should be NLSM amplitudes. Consider the zero associated with setting $\{c_{1,4},c_{2,4},c_{1,5},c_{2,5}\}$ to zero, and let us look at the factorization we get by turning on $c_{2,4}$. From the kinematic shifts, we expect $X_{3,5}$ and $X_{2,4}$ to be fixed, and so the bottom 4-point amplitude will be a function of $(X_{1,3}$, $X_{2,6}$, $c_{1,3}$), while the top 4-point will depend on $(X_{1,5}$, $X_{4,6}$, $c_{3,5})$. In this kinematic limit, the amplitude becomes:
  \begin{equation}
   \begin{aligned}
  	   \mathcal{A}_6(c_{1,4}=c_{1,5}=c_{2,5}=0)&= \frac{c_{1,3} c_{2,4} c_{3,5}}{X_{1,4} (X_{1,4}-c_{2,4})} = \left(\frac{1}{X_{1,4}}+\frac{1}{X_{3,6}}\right)\cdot c_{1,3}\cdot c_{3,5} \\
  	 &= \left(\frac{1}{X_{1,4}}+\frac{1}{X_{3,6}}\right)\cdot \underbrace{\left(X_{1,3}+X_{2,6}\right)}_{\mathcal{A}_4^{\text{down,NLSM} }}\cdot \underbrace{\left(X_{1,5}+X_{4,6}\right)}_{\mathcal{A}_4^{\text{up,NLSM}}},
    \end{aligned}
  \end{equation}
which follows exactly the form predicted in \eqref{eq:factpions}.
 
Let us now consider the factorization into the 5-point mixed amplitude. The skinny rectangle we will consider will be the usual one containing $\{c_{1,3},c_{1,4},c_{1,5}\}$:

\begin{itemize}
	\item \underline{Turn on $c_{1,5}$}: 
	 
	According to the kinematic shifts, we should get a 5-point amplitude depending on ($X_{2,4}$, $X_{2,5}$, $X_{3,5}$, $X_{3,6}$, $X_{4,6}$, $c_{2,4}$, $c_{2,5}$, $c_{3,5}$). In this limit, the amplitude becomes:
	\begin{equation}
		\begin{aligned}
  	   \mathcal{A}_6(c_{1,3}=c_{1,4}=0)&= c_{1,5} \cdot \left(\frac{c_{3,5}}{X_{3,6}}+\frac{c_{2,4}}{X_{2,5}}+1\right)\\
  	 &= (X_{1,3}+X_{2,6}) \cdot \underbrace{\left(\frac{X_{3,5}+X_{4,6}}{X_{3,6}}+\frac{X_{2,4}+X_{3,5}}{X_{2,5}}-1\right)}_{\mathcal{A}_5^{\mathrm{NLSM} + \phi^3}(\phi,\pi,\pi,\phi,\phi)}.
  	\end{aligned}
	\end{equation}
	\item \underline{Turn on $c_{1,4}$}: 
	 
	According to the kinematic shifts, we should get a 5-point amplitude depending on ($X_{2,4}$, $X_{1,5}$, $X_{3,5}$, $X_{3,6}$, $X_{4,6}$, $c_{2,4}$, $c_{2,5}$, $c_{3,5}$). In this limit, the amplitude becomes:
	\begin{equation}
		\begin{aligned}
  	   \mathcal{A}_6(c_{1,3}=c_{1,5}=0)&= c_{1,4}\cdot \frac{c_{3,5}}{X_{3,6}}\\
  	 &= (X_{1,3}+X_{2,6}) \cdot \underbrace{\left(\frac{X_{3,5}+X_{4,6}}{X_{3,6}}-1\right)}_{\mathcal{A}_5^{\mathrm{NLSM} + \phi^3}(\phi,\pi,\phi,\pi,\phi)}.
  	\end{aligned}
	\end{equation}
	\item \underline{Turn on $c_{1,3}$}: 
	 
	According to the kinematic shifts, we should get a 5-point amplitude depending on ($X_{1,4}$, $X_{1,5}$, $X_{3,5}$, $X_{3,6}$, $X_{4,6}$, $c_{2,4}$, $c_{2,5}$, $c_{3,5}$). In this limit, the amplitude becomes:
	\begin{equation}
		\begin{aligned}
  	   \mathcal{A}_6(c_{1,4}=c_{1,5}=0)&= c_{1,3}\cdot \left(\frac{c_{3,5}}{X_{3,6}} +\frac{c_{3,5}+c_{2,5}}{X_{1,4}}\right)\\
  	 &= (X_{1,3}+X_{2,6}) \cdot \underbrace{\left(\frac{X_{3,5}+X_{4,6}}{X_{3,6}}+ \frac{X_{4,6}+X_{1,5}}{X_{1,4}}-1\right)}_{\mathcal{A}_5^{\mathrm{NLSM} +\phi^3}(\phi,\phi,\pi,\pi,\phi)}.
  	\end{aligned}
	\end{equation}
\end{itemize}

\section{Yang-Mills Theory}
\label{sec:FTYM}

Finally, let us look at the last and most important colored theory: pure Yang-Mills theory. In this case, we are describing massless spin 1 particles and therefore the amplitudes have a crucial new ingredient: the \textbf{polarizations} of the gluons, $\epsilon_i$. So we have that $\mathcal{A}^{\text{YM}} \equiv \mathcal{A}^{\text{YM}} (p_i \cdot p_j, \epsilon_i \cdot p_j, \epsilon_i \cdot \epsilon_j) $. 
 
Due to this new feature, the statement about the zeros requires a generalization to involve a statement about polarization vectors as well. The most natural extension is as follows: 

\paragraph{\bf Gluon Zeros} Consider an $n$-point tree-level amplitude in YM theory. Draw the corresponding $n$-point kinematic mesh. Draw a causal diamond just like the one described for Tr$(\phi^3)$.  Setting all the $c_{i,j}$ inside this causal diamond to zero as well as all $\epsilon_i \cdot p_j$, $\epsilon_j \cdot p_i$ and $\epsilon_i \cdot \epsilon_j$ will make the amplitude vanish. Note that setting $\epsilon_i \cdot \epsilon_j =0$ is not a gauge-invariant statement, unless we also have $\epsilon_i \cdot p_j = \epsilon_j \cdot p_i =0$. Therefore the zero condition is well-defined and physically meaningful. 

For example, if we set not only $c_{1,3}=-2 p_1\cdot p_3=0$ but also $\epsilon_1 \cdot \epsilon_3=\epsilon_1\cdot p_3=\epsilon_3 \cdot p_1=0$, the $4$-gluon amplitude vanishes. Similarly, we can have $4(n{-}3)$ zero conditions associated with ``skinny rectangle'' and other causal diamonds such as the $16$ conditions with $(i,j)=(1,4), (1,5), (2,4), (2,5)$ which makes the $6$-gluon amplitude vanishes.  
 
Now the question about factorization is more subtle. In particular there is only one inner product that we can turn back on that does not mess the gauge invariance of the other conditions: setting $c_{i,j} \neq 0$ makes $\epsilon \cdot p =0$ meaningless, $\epsilon_i \cdot p_j \neq 0$ makes $\epsilon_i \cdot \epsilon_j =0$ meaningless. So the only well-defined thing to do is set  $\epsilon_i \cdot \epsilon_j \neq 0$. Still doing so it is unclear the meaning of the factor multiplying $\epsilon_i\cdot \epsilon_j$ obtained in this limit. 
 
For this reason, at this stage, we do not have any generalization of the factorizations found for the previous theories. Such a generalization will only appear once we find a formulation of Yang-Mills theory that connects it to the other colored theories. Of course, while the connection between Tr$(\phi^3)$ and the NLSM is surprising, at least these are both theories of scalars, so it is even more surprising to connect Tr$(\phi^3)$ with Yang-Mills--for instance where do the polarization vectors come from? As we will see in section \ref{sec:StringDelta}, this new description of the $n$ gluon amplitudes will actually begin with a theory of $2n$ colored scalars, which will arise from a shift of the stringy Tr$(\phi^3)$ amplitudes for $2n$ scalars. Factorizing on poles where $n$ pairs of these scalars fuse to produce gluons then gives us general $n$-gluon amplitudes. This formulation will make the zeros and factorizations present in this theory manifest. 

\section{Stringy Tr$(\phi^3)$}
\label{sec:StringPhi}
In this section, we generalize the zeros and factorizations of Tr$(\phi^3)$ tree amplitude to the corresponding string amplitude: the so-called stringy integrals for ABHY associahedron~\cite{Arkani-Hamed:2019mrd}, which also represents a tree-level example of the so-called ``binary geometry''~\cite{Arkani-Hamed:2019plo}. 
These string amplitudes are in fact $n$-point generalizations~\cite{ Koba:1969rw} of the  Veneziano amplitude~\cite{Veneziano:1968yb}, known as dual resonance model in the early days of string theory (see~\cite{DiVecchia:2007vd} for a review), and more recently they arise as a natural basis for $n$-gluon tree amplitudes in type-I superstring theory (see ~\cite{Mafra:2011nw, Broedel:2013tta}). The $n$-point amplitude is given by an integral over the moduli-space of real points $z_1,\dots,z_n$ on the boundary of the disk:
\begin{equation}
	\mathcal{I}^{\mathrm{Tr}(\phi^3)}_n(1,2,...,n) = \int_{D(1\ldots n)} \frac{\mathrm{d} z_1 \ldots \mathrm{d} z_n}{\mathrm{vol~SL}(2, \mathbb{R})}\underbrace{\frac{1}{z_{1,2} z_{2,3}\ldots z_{n,1}}}_{\text{PT}(1,2,\cdots,n)}\times \underbrace{ \prod_{i<j}\,	z_{i,j}^{2 \alpha^\prime p_i \cdot p_j}}_{\text{Koba-Nielsen factor}},\label{eq:StringAmpZ}
\end{equation}
where the SL$(2, \mathbb{R})$ redundancy allows one to fix three punctures and the integration domain is the positive part of the real moduli space, ${\cal M}_{0,n}^+$, or $z_1<z_2<\dots<z_n$ (with $3$ of them fixed); here $z_{i,j}:=z_j -z_i >0$ for $i<j$, and the integrand is given by the Parke-Taylor factor, ${\rm PT}(1,2,\cdots, n)$ times the universal Koba-Nielsen factor (we have omitted the overall prefactor $\alpha^{\prime\, n-3}$). The low energy limit, where $\alpha^\prime X_{i,j} \ll 1$, yields the Tr$(\phi^3)$ tree amplitudes, and low-energy, $\alpha'$-expansion of these integrals have been extensively studied in the literature (the so-called $Z$-theory~\cite{Mafra:2016mcc, Carrasco:2016ldy}). 

To translate this to stringy integrals for binary geometries, we introduce $u$ variables (one for each chord $(i,j)$), which are $\text{SL(2,}\mathbb{R})$ invariant cross-ratios defined as follows:
\begin{equation}
    u_{i,j} = \frac{z_{i{-}1,j} z_{i,j{-}1}}{z_{i,j} z_{i{-}1,j{-}1}}.
	\label{eq:udef}
\end{equation}
In terms of $u$'s, the Koba-Nielsen factor becomes $\prod_{i,j} u_{i,j}^{\alpha' X_{i,j}}$ with planar variables $X_{i,j} = (p_i + p_{i{+}1} + ...+p_{j-1})^2$ (with $X_{i,i+1}=0$). Note that there are $n(n{-}3)/2$ $u$ variables, which satisfy $u$ equations~\cite{Arkani-Hamed:2017mur, Arkani-Hamed:2019vag} (see earlier works {\it e.g.}~\cite{Koba:1969rw,Brown:2009qja}):
\begin{equation}
u_{i,j}+ \prod_{(k,l)~{\rm cross}~(i,j)} u_{k,l}=1\,,
\end{equation}
such that for any $u_{i,j}\to 0$, all incompatible $u_{k,l}\to 1$ (the chord $(k,l)$ intersects with $(i,j)$), hence the name ``binary geometry". The ordering of $z_i$ is equivalent to requiring all $u$ variables to be positive (which implies that $0<u_{i,j}<1$), thus we have the $(n{-}3)$-dimensional positive binary geometry, $U_n^+\sim {\cal M}_{0,n}^+$, which has the shape of a (curvy) associahedron~\cite{Arkani-Hamed:2019plo}. As shown in~\cite{Arkani-Hamed:2017mur}, the Parke-Taylor factor ${\rm PT}(1,2,\cdots, n)$ with the measure is nothing but the canonical form of this space, which we denote as $\Omega(U_n^+)$, thus we have 
\begin{equation}
	\mathcal{I}^{\mathrm{Tr}(\phi^3)}_n(1,2,...,n) = 
 \int_{U_n^{+}}\Omega\left(U_n^{+}\right) \, \prod_{i<j} u_{i,j}^{\alpha^{\prime} X_{i,j}}.
	\label{eq:Trphi3U}
\end{equation}

To be more explicit, one can choose any {\it positive parametrization} of $U_n^+$, and a convenient one inspired by first fix the $\mathrm{SL}(2,\mathbb{R})$ by choosing a gauge fixing {\it e.g.} $z_{1}=0,z_{n-1}=1,z_{n}=\infty$) and change the remaining $z$ variables to the positive $y$ variables as
\begin{equation}
\begin{aligned}
&z_2=\frac{y_{1,3}\cdots y_{1,n{-}1}}{1+y_{1,n{-}1}+\cdots+y_{1,n{-}1} y_{1,n{-}2} \cdots y_{1,3}}\,, \, z_{3}{-}z_{2}=\frac{y_{1,4}\cdots y_{1,n{-}1}}{1+y_{1,n{-}1}+\cdots+y_{1,n{-}1} y_{1,n{-}2} \cdots y_{1,3}}\,, \\
&\, \cdots, \,1-z_{n{-}2}=\frac{1}{1+y_{1,n{-}1}+\cdots+y_{1,n{-}1} y_{1,n{-}2} \cdots y_{1,3}}\,.
\end{aligned}
\end{equation}

After the variable transformation, the integral becomes
\begin{equation}
	\mathcal{I}^{\mathrm{Tr}(\phi^3)}_n(1,2,...,n)=  \int_{\mathbb{R}_{>0}^{n-3}} \prod_{i=3}^{n{-}1} \frac{\diff y_{1,i}}{y_{1,i}} y_{1,i}^{\alpha' X_{1,i}} \prod_{1\leq i,j<n} F_{i,j} (y)^{-\alpha' c_{i,j}},
\end{equation}
where the {\bf F-polynomials} for our {\bf ray-like triangulation}, $F_{i,j}(y)$, are defined below in~\eqref{eq-F-poly}, the exponents $c_{i,j}$ are the non-planar variables (without $c_{i,n}$) satisfying \eqref{eq:ceq}.

More generally for any initial triangulation, the $n$-point tree-level amplitude can be written as 
\begin{equation}
\label{eq:stringyphi3}
\mathcal{I}_{n}^{\mathrm{Tr}(\phi^3)}=\int_{\mathbb{R}_{>0}^{n-3}} \prod_{I=1}^{n-3} \frac{\diff y_I}{y_I}~\prod_{(a, b)} u_{a,b}^{\alpha' X_{a,b}}(y)=\int_{\mathbb{R}_{>0}^{n-3}} \prod_{I=1}^{n{-}3} \frac{\diff y_I}{y_I} y_I^{\alpha' X_I} \prod_{i,j} F_{i,j} (y)^{-\alpha' c_{i,j}},
\end{equation}
where $\{y_I>0 \}$ for $I=1,\cdots, n{-}3$ specify a triangulation of the $n$-gon, which provides a positive parametrization of $U_n^+$ (thus the Parke-Taylor form becomes $\Omega(U_n^+)=\prod_I d\log y_I$; for each curve/chord, $(a,b)$, the $u$-variable $u_{a,b}$ is a nice rational function of $y_I$, which are discussed in detail in~\cite{Arkani-Hamed:2023lbd}, and the Koba-Nielsen factor becomes $\prod_I y_I^{\alpha' X_I}$ times the product of $\frac{(n{-}2)(n{-}3)}{2}$ $F_{i,j}(y)$, which are F polynomials of $y_I$'s for this triangulation~\cite{Arkani-Hamed:2023lbd}, with exponents $-\alpha' c_{i,j}$.

Of course \eqref{eq:stringyphi3} is just the tree-level/disk instance of stringy integrals associated with general surfaces ${\cal S}$, which correspond to ``stringy'' Tr$(\phi^3)$ amplitudes in the genus expansion (details can again be found in~\cite{Arkani-Hamed:2023lbd}):
\begin{equation}
{\cal I}_{\cal S}^{\mathrm{Tr}(\phi^3)}=\int_{\mathbb{R}_{>0}^{d}} \prod_{I=1}^d \frac{\diff y_I}{y_I} \prod u_\Gamma (y)^{\alpha' X_\Gamma} ,
\end{equation}
where given the triangulation of the surface ${\cal S}$, we have \textbf{positive coordinates} $y_I$ for $I=1,2, \cdots, d$ and for \textbf{every} curve on ${\cal S}$ denoted by $\Gamma$ we have a \textbf{$u$-variable} $u_\Gamma(y)$, which is a rational function of $y_I$, and the kinematic variable $X_\Gamma$ as its exponent. 
 
In section~\ref{sec:ABHY} we concluded that by setting a collection of $c_{i,j}=0$, the field theory Tr$(\phi^3)$ amplitude vanishes, and that by turning \textbf{one} $c_\star$ back on, the amplitude factorizes into three pieces, corresponding to lower point amplitudes. Both these properties heavily relied on the Minkowski sum picture for the ABHY associahedron encoding these amplitudes. This picture  emerges from \eqref{eq:stringyphi3} (for any triangulation) in the $\alpha'\to 0$ limit~\cite{Arkani-Hamed:2019mrd}. In this section, we study how the field-theory statements generalize to the stringy Tr$(\phi^3)$ integral \eqref{eq:stringyphi3}. 
 
Already in the field theory case, we concluded that we could access \textbf{all} zeros and factorizations by choosing the realization of the ABHY associahedron determined by \textbf{ray-like} triangulations (which produced a triangular region in the kinematic mesh). The same is true for string amplitudes and so we will be mostly considering positive parametrizations $\{y_I\}$ of \eqref{eq:stringyphi3} corresponding to \textbf{ray-like} triangulations~\cite{Arkani-Hamed:2019mrd}. In this case, the $F$-polynomials have a simple recursive structure, say at $n$-point with triangulation $\{y_{1,3}, y_{1,4}, \cdots, y_{1,n{-}1}\}$:
\begin{equation}\label{eq-F-poly}
    F_{i,j}=1+y_{1,j}+y_{1,j}y_{1,j-1}+\cdots+y_{1,j}\dots y_{1,i+2}.
\end{equation}

\begin{figure}[t]
    \centering
\includegraphics[width=\linewidth]{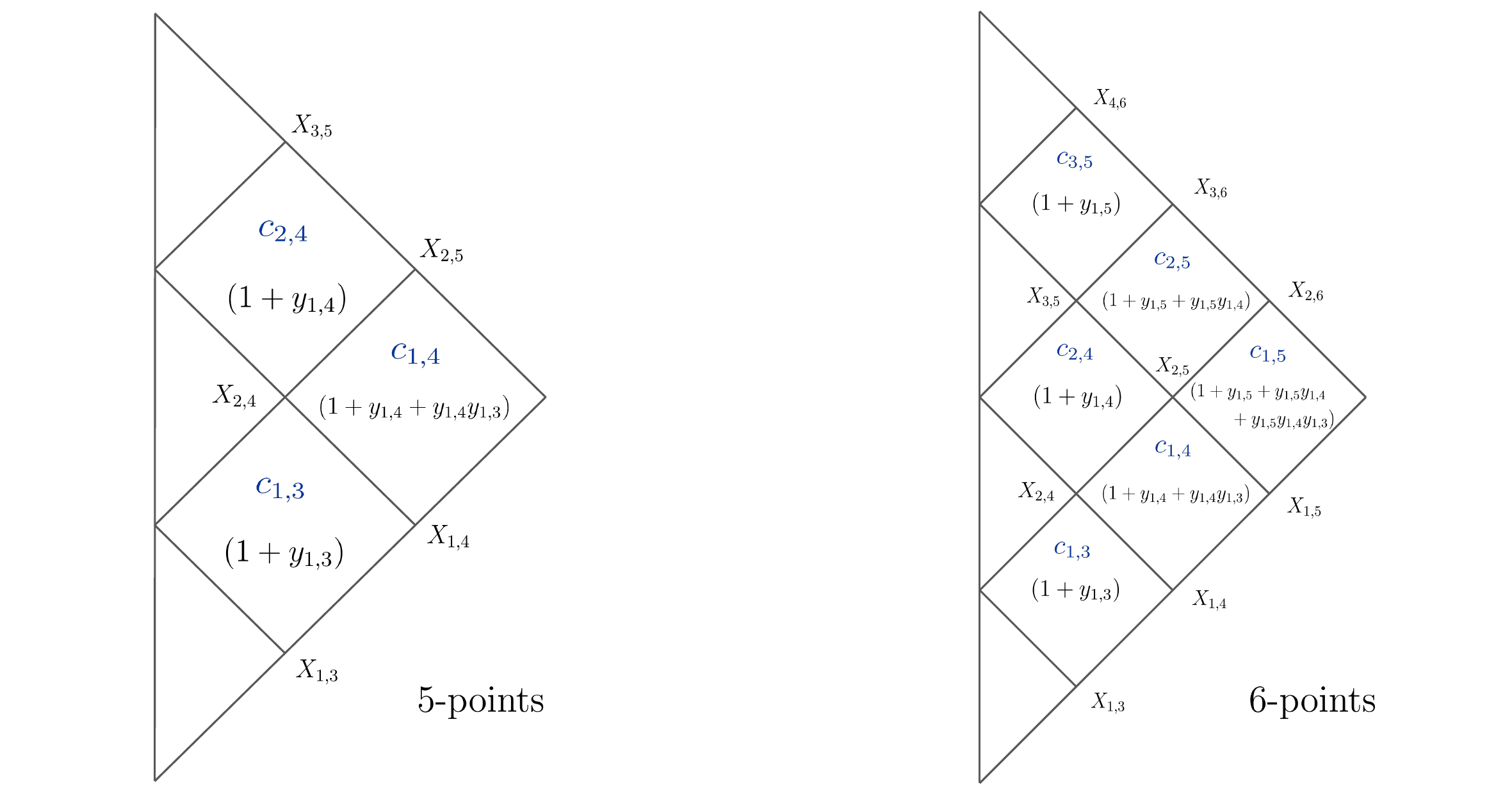}
    \caption{F-polynomials for ray-like triangulations, at 5 and 6 points.}
    \label{fig:FPol56}
\end{figure}
Since each $F_{i,j}$ is automatically associated with the non-planar Mandelstam, $c_{i,j}$, it is useful to organize them in the mesh picture. In figure \ref{fig:FPol56} we present the 5-point and 6-point kinematic mesh obtained when considering, respectively, triangulations $\{(1,3),(1,4)\}$ and $\{(1,3),(1,4),(1,5)\}$, as well as the corresponding $F$-polynomials.
 
Choosing a ray-like triangulation, the kinematic variables appearing in \eqref{eq:stringyphi3} once written in terms of $F_{i,j}$ are: $X_I$ and $c_{i,j}$ with $(i+1,j+1) \neq I$. This forms a kinematic basis naturally associated with the corresponding triangular region of the kinematic mesh. For example, at 6-point, for triangulation $\{y_{1,3},y_{1,4},y_{1,5}\}$ we have $\{X_{1,3},X_{1,4},X_{1,5}\}$ and $\{c_{1,3},c_{1,4},c_{1,5},c_{2,4},c_{2,5},c_{3,5}\}$.
  
It is well known that string amplitudes have poles when $X_{a,b}$ equals some non-positive integer (with $X_{a,b}=0$ corresponding to the pole of field-theory amplitudes)~\cite{Arkani-Hamed:2019mrd}. However, like in the field-theory case, the kinematical locus where they vanish has not been studied as extensively. Let us now understand how the field theory zeroes identified in section \ref{sec:ABHY} generalize to the string amplitudes.

\subsection{Zeros of Tr$(\phi^{3})$ string amplitudes}
\label{sec:stringZeros}
We begin with the simplest case, the 4-point amplitude. Considering the triangulation of the square containing chord $(1,3)$, we get the following string amplitude:
\begin{equation}
\mathcal{I}_{4}^{\mathrm{Tr}(\phi^3)} 
        =\int_{\mathbb{R}_{>0}}  \frac{\diff y_{1,3}}{y_{1,3}} y_{1,3}^{\alpha^{\prime}X_{1,3}}(1+y_{1,3})^{-\alpha^{\prime}c_{13}}=\frac{\Gamma[\alpha^{\prime}X_{1,3}]\Gamma[\alpha^{\prime}(c_{1,3}-X_{1,3})]}{\Gamma[\alpha^{\prime}c_{1,3}]}=\frac{\Gamma[\alpha^{\prime}X_{1,3}]\Gamma[\alpha^{\prime}X_{2,4}]}{\Gamma[\alpha^{\prime}c_{1,3}]},
        \label{eq:4ptString}
\end{equation}
which is exactly the Beta function $\mathrm{B}(\alpha^{\prime}X_{1,3},\alpha^{\prime}X_{2,4}$).  In the field-theory limit ($\alpha^{\prime}\to 0$), we get $\mathcal{I}_{4}^{\mathrm{Tr}(\phi^3)} \to c_{1,3}/(\alpha^{\prime}X_{1,3}X_{2,4})$, which vanishes for $c_{1,3}=0$, as  discussed previously. However, from the Beta function \eqref{eq:4ptString}, we see that, in addition to this zero, the full string amplitude vanishes whenever $\alpha^\prime c_{1,3}$ is a \textbf{non-positive integer}. 
 
From the integral representation, by setting $\alpha^\prime c_{1,3} = -n$, with $n\in \mathbb{N}_0$ we get:
\begin{equation}
    \mathcal{I}_{4}^{\mathrm{Tr}(\phi^3)} 
        \rightarrow \sum_{k=0}^n \binom{n}{k} \underbrace{\int_{\mathbb{R}_{>0}}  \frac{\diff y_{1,3}}{y_{1,3}} y_{1,3}^{\alpha^{\prime}X_{1,3}+k}}_{=0} =0.
    \label{eq:zeroInt}
\end{equation}

The integrals appearing in the sum are divergent, however, they all vanish by analytic continuation. The integral of the form $\int_{\mathbb{R}_{>0}} \frac{dy}{y}y^{\alpha^{\prime}X}$ bears resemblance to the concept of a \textbf{scaleless integral} within the context of Feynman integrals. The parameter $\alpha^{\prime}$ serves as a regulator factor, causing the integral to vanish. Specifically, we have:
\begin{equation}
\int_{\mathbb{R}_{>0}} \frac{dy}{y}y^{\alpha^{\prime}X}=\int_{0}^{1} \frac{dy}{y}y^{\alpha^{\prime}X}+\int_{1}^{\infty} \frac{dy}{y}y^{\alpha^{\prime}X}=-\frac{y^{\alpha^{\prime}X}}{\alpha^{\prime}X}\bigg \vert_{y=0}+\frac{y^{\alpha^{\prime}X}}{\alpha^{\prime}X}\bigg \vert_{y=\infty}.
\end{equation}

In the first part of the integral, $\alpha^{\prime}$ is analytically continued to $\alpha^{\prime}X> 0$, while in the second part of the integral, $\alpha^{\prime}$ is analytically continued to $\alpha^{\prime}X<0$. As a result, both sides of the integral evaluate to zero. As we will see shortly the vanishing of this class of one-dimensional integrals will be behind the general patterns of zeros we will describe in string amplitudes. 
 
Without loss of generality, let us choose the initial ray-triangulation to be $y_{1,i}$ for $i=3,\cdots,n-1$, then, from \eqref{eq-F-poly}, all the  $F$-polynomials depending on $y_{1,i}$ are those $F_{ab}$ with $1\leq a \leq i-2\,,i\leq b\leq n-1$. Looking at the mesh, this is saying that all the F-polynomials that depend on $y_{1,j}$ are contained \textbf{inside the causal diamond anchored at $X_{1,j}$} (see figure \ref{fig:FPol56}). So by setting all the $c_{a,b}=-n_{a,b}$, with $n_{a,b} \in \mathbb{N}_0$, inside this causal diamond, the stringy integral vanishes because the integration in $y_{1,i}$ reduces to \eqref{eq:zeroInt}: 
\begin{equation}
\begin{aligned}
\mathcal{I}_{n}^{\mathrm{Tr}(\phi^3)}\rightarrow \sum_{k_{a_1,b_1},\dots, k_{a_N,b_N}=0}^{n_{a_1,b_1},\dots,n_{a_N,b_N}}(\text{remaining integrals})\underbrace{\times\int_{\mathbb{R}_{>0}} \frac{\diff y_{1,i}}{y_{1,i}}y_{1,i}^{\alpha^{\prime}X_{1,i}+k_{a_1,b_1}+\dots + k_{a_N,b_N} }}_{=0}=0,
\end{aligned}
\end{equation}
where $N$ corresponds to the number of $c$'s inside the causal diamond considered. Note that the integral in $y_{1,i}$ can also be considered as the $4$-pt integral but setting $c$ to zero, which therefore vanishes.
\paragraph{Zeros} In field theory, we concluded that the zero locus corresponded to setting the $c$'s inside some causal diamond anchored in $X_{\text{B}}$ to zero. For string amplitudes the zero locus is now given by setting the same collection of $c$'s but to any non-positive integer, $i.e.$ for the mesh corresponding to triangulation $(1,3), \dots, (1,n-1)$, if we want the zero to come from the integration in $y_{1,i}$:
\begin{equation}\label{eq-zero}
	\alpha^{\prime}c_{ab}=-n_{a,b}\,,\quad \text{for}\quad 1\leq a \leq i-2\,,\quad i\leq b\leq n-1, \quad \text{and} \quad  n_{a,b}\in \mathbb{N}_0.
\end{equation}

Since each zero causal diamond is uniquely determined by the $X_{\text{B}}$ it is anchored on, the total number of zeros for the $\mathrm{Tr}(\phi^3)$ field theory amplitude is $\frac{n(n-3)}{2}$, which is equal to the number of $X_{i,j}$'s, and so the number of poles.
 
Next, we will study factorization near such zeros. Exactly like in the field theory case, we do this by relaxing one condition: so we set all but one $c_{a,b}$ inside the causal diamond to non-positive integers. 
 
\subsection{Factorization around the zeros}
\label{sec:stringFact} 
For simplicity, let us begin by understanding what happens when we set all the $c_{a,b}$ inside the causal diamond to \textbf{zero}, but for one of them $c_\star \neq 0$. In this case, the amplitude factorizes into three pieces, just like in the field theory case. If instead, we allow the $c$'s to be negative integers, then we get a sum of factorized terms, with interesting kinematic shifts, as we will see shortly. We have already understood the reason for factorization in the field theory limit from the perspective of the Minkowski-sum picture for the associahedron. We will now see a related but different derivation for factorization whose fundamental origin lies in certain separation properties of the $F$-polynomials appearing in the stringy integral. This will generalize our observations about factorization to the full stringy amplitude \footnote{While we will not dwell on this point, our previous derivation centered on Minkowski sums and the one we present now are closely connected, since the Minkowski summands are nothing other than the Newton polytopes of the $F$-polynomials.}. 

\subsubsection{Examples: $n=5,6$}

We begin by studying our two simple examples of $n=5,6$ string amplitudes. This will expose the basic mechanism for factorization arising from special properties of the $F$-polynomials, which motivate simple variable changes on the $y$ variables, giving rise both to factorization of the amplitude as well as precisely the same interesting kinematic shifts we encountered in our field-theoretical analysis. 

At $n=5$, the stringy integral reads:
\begin{equation}
\begin{aligned}
    \mathcal{I}_{5}^{\mathrm{Tr}(\phi^3)}&=\int_{0}^\infty \prod_{i=3}^{4} \frac{\diff y_{1,i}}{y_{1,i}} y_{1,i}^{\alpha^{\prime}X_{1,i}}\prod_{i<j} F_{i,j}(\textbf{y})^{-\alpha^{\prime}c_{i,j}}\\
    &= \int_{0}^\infty \frac{\diff y_{1,3}}{y_{1,3}} \frac{\diff y_{1,4}}{y_{1,4}}y_{1,3}^{\alpha^\prime  X_{1,3}} y_{1,4}^{\alpha^\prime  X_{1,4}}
    \left(1+y_{1,3}\right)^{-\alpha^\prime  c_{1,3}} \left(1+y_{1,4}\right)^{-\alpha^\prime  c_{2,4}} \left(1+y_{1,4}+y_{1,3} y_{1,4}\right)^{-\alpha^\prime  c_{1,4}}.
\end{aligned}
\end{equation}

Let us consider the zero associated with the skinny rectangle, $\{c_{1,3},c_{1,4}\}$. Setting $c_{1,3}=c_{1,4}=0$ the integral becomes:
\begin{equation}
  \mathcal{I}_{5}^{\mathrm{Tr}(\phi^3)} \to 
    \left(\int_{0}^\infty \frac{\diff y_{1,3}}{y_{1,3}} y_{1,3}^ {\alpha^\prime  X_{1,3}} \right)
    \int_{0}^\infty \frac{\diff y_{1,4}}{y_{1,4}} y_{1,4}^ {\alpha^\prime  X_{1,4}} \left(1+y_{1,4}\right)^{-\alpha^\prime  c_{2,4}}=0.
\end{equation}

Now letting $c_{1,4}\neq 0$, the answer factorizes as follows:
\begin{equation} \label{eq:skinnyfac5pt}
\begin{aligned}
    \mathcal{I}_{5}^{\mathrm{Tr}(\phi^3)} &\to 
   \int_{0}^\infty \frac{\diff y_{1,3}}{y_{1,3}} y_{1,3}^ {\alpha^\prime  X_{1,3}} 
   \int_{0}^\infty \frac{\diff y_{1,4}}{y_{1,4}} y_{1,4}^ {\alpha^\prime  X_{1,4}} \left(1+y_{1,4}\right)^{-\alpha^\prime  c_{2,4}} \left(1+y_{1,4}+y_{1,3}y_{1,4} \right)^{-\alpha^\prime  c_{1,4}}
    \\ 
    &=
   \int_{0}^\infty \frac{\diff y_{1,3}}{y_{1,3}} y_{1,3}^ {\alpha^\prime  X_{1,3}} 
   \int_{0}^\infty \frac{\diff y_{1,4}}{y_{1,4}} y_{1,4}^ {\alpha^\prime  X_{1,4}} \left(1+y_{1,4}\right)^{-\alpha^\prime  (c_{2,4}+c_{1,4})} \left(1+\frac{y_{1,3}y_{1,4}}{(1+y_{1,4})} \right)^{-\alpha^\prime  c_{1,4}}.
\end{aligned}
\end{equation}

Now changing variables to $\tilde{y}_{1,3} = y_{1,3}y_{1,4}/(1+y_{1,4})$, we get:
\begin{equation}
\begin{aligned}
    \mathcal{I}_{5}^{\mathrm{Tr}(\phi^3)} &\to \int_{0}^\infty \frac{\diff \tilde{y}_{1,3}}{\tilde{y}_{1,3}} \tilde{y}_{1,3}^ {\alpha^\prime  X_{1,3}} \left(1+\tilde{y}_{1,3} \right)^{-\alpha^\prime  c_{1,4}}\int_{0}^\infty \frac{\diff y_{1,4}}{y_{1,4}} y_{1,4}^ {\alpha^\prime  (X_{1,4}-X_{1,3})} \left(1+y_{1,4}\right)^{-\alpha^\prime  (c_{2,4}+c_{1,4} -X_{1,3})}  \\
    &=
    \mathcal{I}_{4}^{\mathrm{Tr}(\phi^3)}(\alpha^{\prime} X_{1,3},\alpha^{\prime}(c_{1,4}-X_{1,3}))\times \mathcal{I}_{4}^{\mathrm{up},\mathrm{Tr}(\phi^3)}(\alpha^{\prime} (X_{1,4}-X_{1,3}),\alpha^{\prime}(c_{2,4}+c_{1,4}-X_{1,4})) \\
    &= \mathcal{I}_{4}^{\mathrm{Tr}(\phi^3)}(\alpha^{\prime} X_{1,3},\alpha^{\prime}X_{2,5})\times \mathcal{I}_{4}^{\mathrm{up},\mathrm{Tr}(\phi^3)}(\alpha^{\prime} X_{2,4},\alpha^{\prime}X_{3,5}).
\end{aligned}
\end{equation} 

We see clearly a direct stringy generalization of the factorization we saw in the field theory limit. We have the 4-point amplitude pre-factor, now as a full stringy amplitude $\mathcal{I}_{4}^{\mathrm{Tr}(\phi^3)}(\alpha^{\prime} X_{1,3},\alpha^{\prime}X_{2,5})$, that vanishes when we further set $c_{1,4} \to 0$. Then we have a product of smaller stringy amplitudes, a (trivial) 3-point ``down'' amplitude and the ``up'' amplitude $\mathcal{I}_{4}^{\mathrm{up},\mathrm{Tr}(\phi^3)}(\alpha^{\prime} X_{2,4},\alpha^{\prime}X_{3,5})$, whose (interestingly redefined) kinematic variables $X_{2,4}, X_{3,5}$ precisely agree the same way as our previous analysis as summarized in figure \ref{fig:fact5pts}. The mechanism for factorization is easy to see as a consequence of a ``separation property'' of $F$-polynomials.  After setting $c_{1,3} \to 0$, all the non-trivial dependence on $y_{1,3}$ is in the $F$-polynomial $(1 + y_{1,4} + y_{1,4} y_{1,3})$ = $(A + y_{1,3} B)$ with $A = (1 + y_{1,4})$ and $B=y_{1,4}$ and quite nicely, both $A, B$ are polynomials that appear in smaller amplitudes. We can manifest the factorization by the change of variables $y_{1,3}B/A = \tilde{y}_{1,3}$, and this is what induces the interesting kinematic redefinitions in the smaller amplitude factors. 

At 6-point, choosing once more the triangulation $\{(1,3),(1,4),(1,5)\}$ for $n=6$, the stringy integral is given by:
\begin{equation}
    \begin{aligned}
    \mathcal{I}_{6}^{\mathrm{Tr}(\phi^3)}&=\int_{0}^\infty \prod_{i=3}^{5} \frac{\diff y_{1,i}}{y_{1,i}} y_{1,i}^{\alpha^{\prime}X_{1,i}}\prod_{i<j} F_{i,j}(\textbf{y})^{-\alpha^{\prime}c_{i,j}}\\
      &= \int_{0}^\infty \prod_{i=3}^{5} \frac{\diff y_{1,i}}{y_{1,i}} y_{1,i}^{\alpha^{\prime}X_{1,i}}(1+y_{1,3})^{-\alpha^{\prime}c_{1,3}} (1+y_{1,4})^{-\alpha^{\prime}c_{2,4}}(1+y_{1,5})^{-\alpha^{\prime}c_{3,5}}\\
      &\times(1+y_{1,4}(1+y_{1,3}))^{-\alpha^{\prime}c_{1,4}}(1+y_{1,5}(1+y_{1,4}))^{-\alpha^{\prime}c_{2,5}}(1+y_{1,5}(1+y_{1,4}(1+y_{1,3})))^{-\alpha^{\prime}c_{1,5}}.
    \end{aligned}
\end{equation}

At 6-point we have our two kinds of zeros, the ``skinny rectangle'' and ``big square'' patterns. The near-zero factorizations for the skinny rectangles exactly mirror what we have already seen at 5 points so we will look at the square pattern, where setting $c_{14}=c_{24}=c_{15}=c_{25}=0$ makes the amplitude vanish. 
\begin{equation}
    \begin{aligned}
        \mathcal{I}_{6}^{\mathrm{Tr}(\phi^3)}&\to \int_{0}^\infty \prod_{i=3}^{5} \frac{\diff y_{1,i}}{y_{1,i}} y_{1,i}^{\alpha^{\prime}X_{1,i}}(1+y_{1,3})^{-\alpha^{\prime}c_{1,3}} (1+y_{1,5})^{-\alpha^{\prime}c_{3,5}}\\
 &= \int_{0}^\infty \prod_{i\neq 4}\frac{\diff y_{1,i}}{y_{1,i}} y_{1,i}^{\alpha^{\prime}X_{1,i}}(1+y_{1,3})^{-\alpha^{\prime}c_{1,3}} (1+y_{1,5})^{-\alpha^{\prime}c_{3,5}} \underbrace{\left(\int_{0}^\infty \frac{\diff y_{1,4}}{y_{1,4}} y_{1,4}^{\alpha^{\prime}X_{1,4}}\right)}_{=0}\\
 &=0.
 \end{aligned}
\end{equation}

By turning on $c_{1,5}$, we expect the stringy integral to factorize into two 4-point amplitudes (up and down) and the 4-point prefactor:
\begin{equation}
    \begin{aligned}
        \mathcal{I}_{6}^{\mathrm{Tr}(\phi^3)}&\to \int_{0}^\infty \prod_{i=3}^{5} \frac{\diff y_{1,i}}{y_{1,i}} y_{1,i}^{\alpha^{\prime}X_{1,i}}(1+y_{1,3})^{-\alpha^{\prime}c_{1,3}} (1+y_{1,5})^{-\alpha^{\prime}c_{3,5}}(1+y_{1,5}+y_{1,4}y_{1,5}+y_{1,3}y_{1,4}y_{1,5})^{-\alpha^{\prime}c_{1,5}}\\
        &=\int_{0}^\infty \prod_{i=3}^{5} \frac{\diff y_{1,i}}{y_{1,i}} y_{1,i}^{\alpha^{\prime}X_{1,i}}(1+y_{1,3})^{-\alpha^{\prime}c_{1,3}} (1+y_{1,5})^{-\alpha^{\prime}(c_{3,5}+c_{1,5})}\left(1+\frac{(1+y_{1,3})y_{1,4}y_{1,5}}{(1+y_{1,5})}\right)^{-\alpha^{\prime}c_{1,5}}.\\
        \end{aligned}
 \end{equation}
 
So changing variables to $\tilde{y}_{1,4} = (1+y_{1,3})y_{1,4}y_{1,5}/(1+y_{1,5})$, we get:   
\begin{equation}
\begin{aligned}
 \mathcal{I}_{6}^{\mathrm{Tr}(\phi^3)}&\to \int_{0}^\infty \frac{\diff y_{1,3}}{y_{1,3}} y_{1,3}^{\alpha^{\prime}X_{1,3}}(1+y_{1,3})^{-\alpha^{\prime}(c_{1,3}+X_{1,4})} \times \int_{0}^\infty \frac{\diff \tilde{y}_{1,4}}{\tilde{y}_{1,4}} \tilde{y}_{1,4}^{\alpha^\prime X_{1,4}} (1+\tilde{y}_{1,4}) ^{-\alpha^{\prime}c_{1,5}}  \\&\quad \quad \times \int_{0}^\infty \frac{\diff y_{1,5}}{y_{1,5}} y_{1,5}^{\alpha^\prime(X_{1,5}-X_{1,4})}(1+y_{1,5})^{-\alpha^{\prime}(c_{3,5}+c_{1,5}-X_{1,4})}\\&=
 \mathcal{I}_{4}^{\mathrm{down},\mathrm{Tr}(\phi^3)}(\alpha^{\prime} X_{1,3},\alpha^{\prime}(c_{1,3}+X_{1,4}-X_{1,3}))\times \mathcal{I}_{4}^{\mathrm{Tr}(\phi^3)}(\alpha^{\prime} X_{1,4},\alpha^{\prime}(c_{1,5}-X_{1,4}))\\
 & \quad \quad \times \mathcal{I}_{4}^{\mathrm{up},\mathrm{Tr}(\phi^3)}(\alpha^{\prime} (X_{1,5}-X_{1,4}),\alpha^{\prime}(c_{3,5}+c_{1,5}-X_{1,5})) \\
 &= \mathcal{I}_{4}^{\mathrm{Tr}(\phi^3)}(\alpha^{\prime} X_{1,4},\alpha^{\prime}X_{3,6})  \times \mathcal{I}_{4}^{\mathrm{down},\mathrm{Tr}(\phi^3)}(\alpha^{\prime} X_{1,3},\alpha^{\prime}X_{2,4})\times \mathcal{I}_{4}^{\mathrm{up},\mathrm{Tr}(\phi^3)}(\alpha^{\prime} X_{3,5},\alpha^{\prime}X_{4,6}). 
    \end{aligned}
\end{equation}

As in the 5-point example the mechanism for factorization is a separation property of the $F$ polynomials. Having set $c_{1,4},c_{2,4},c_{2,5} \to 0$, the only non-trivial dependence on $y_{1,4}$ is in the $F$-polynomial $1 + y_{1,5} + y_{1,4} y_{1,5} + y_{1,3} y_{1,4} y_{1,5} = A + y_{1,4} B$, where $A = (1 + y_{1,5}), B = y_{1,5}(1 + y_{1,3})$. Again nicely $A,B$ are (up to monomial factors) $F$-polynomials for smaller amplitudes. We then make the change of variable to $B y_{1,4}/A = \tilde{y}_{1,4}$, and this induces the interesting kinematic shifts for the ``up'' and ``down'' four-point amplitudes.

\subsubsection{General Proof}

The general proof of factorization works in exactly the same way as we just saw in our examples; apart from decoration with indices, the steps are exactly the same as we saw above. We consider the zero associated with maximal causal diamond anchored to $X_{1,i}$, and consider turning back on $c_\star=c_{k,m}\neq 0$. Then stringy integral factorizes as follows:
\begin{equation}\label{eq-fac1}
	\begin{aligned}
		\mathcal{I}_{n}^{\mathrm{Tr}(\phi^3)}&\to \int \prod_{l=3}^{i-1} \frac{\diff y_{1,l}}{y_{1,l}} y_{1,l}^{\alpha^{\prime} X_{1,l}} \int \prod_{j=i}^{n-1} \frac{\diff y_{1,j}}{y_{1,j}} y_{1,j}^{\alpha^{\prime} X_{1,j}} \prod_{1\leq a< b-1\leq i-1} F_{a,b}(\mathbf{y})^{-\alpha^{\prime} c_{a,b}}\\
		&\times  F_{k,m}(\mathbf{y})^{-\alpha^{\prime} c_{k,m}}\times \prod_{i-1\leq e< f-1< n-1} F_{e,f}(\mathbf{y})^{-\alpha^{\prime} c_{e,f}}.
	\end{aligned}
\end{equation}

As in our examples, the key point is that the only $F-$polynomial depending on $y_{1,i}$ is $F_{k,m}(\mathbf{y})=1+y_{1,m}+\dots+y_{1,m}\cdots y_{1,k+2}\equiv A+y_{1,i} B$, with $A=F_{i-1,m}$ and $B=F_{k,i-1}\prod_{p=i+1}^{m}y_{1,p}$, which suggests the variable change $B y_{1,i}/A = \tilde{y}_{1,i}$. Following our noses this will give the factorization for the amplitude into smaller string amplitudes, with the same kinematical redefinition encountered in the field theory limit. Working everything out explicitly, we easily perform the integral over $y_{1,i}$:
\begin{equation}
	\begin{aligned}
			&\int \frac{\diff y_{1,i}}{y_{1,i}} y_{1,i}^{\alpha^{\prime} X_{1,i}}F_{k,m}(\mathbf{y})^{-\alpha^{\prime} c_{k,m}}=	\int \frac{\diff y_{1,i}}{y_{1,i}} y_{1,i}^{\alpha^{\prime} X_{1,i}} (A+y_{1,i} B)^{-\alpha^{\prime} c_{k,m}}\\
			&=A^{-\alpha^{\prime} c_{k,m}+\alpha^{\prime} X_{1,i}}B^{-\alpha^{\prime} X_{1,i}}\int \frac{\diff \tilde{y}_{1,i}}{\tilde{y}_{1,i}} \tilde{y}_{1,i}^{\alpha^{\prime} X_{1,i}} (1+\tilde{y}_{1,i} )^{-\alpha^{\prime} c_{k,m}}\\
			&=A^{-\alpha^{\prime} c_{k,m}+\alpha^{\prime} X_{1,i}}B^{-\alpha^{\prime} X_{1,i}}\times 	\underbrace{\mathcal{I}_{4}^{\mathrm{Tr}(\phi^3)}(\alpha^{\prime} X_{1,i},\alpha^{\prime}(c_{k,m}-X_{1,i}))}_{\mathcal{I}_{4}^{\mathrm{Tr}(\phi^3)}(\alpha^\prime X_{\text{B}},\alpha^\prime X_{\text{T}})},
	\end{aligned}
\end{equation}
where in the second line we used our change variables $\tilde{y}_{1,i} = B y_{1,i}/A$, which reduces the integral to the 4-point stringy integral. This 4-point factor is exactly analogous to the one we saw in the factorization of the field theory amplitudes, $(1/X_{\text{B}}+ 1/X_{\text{T}})$, except that now we get indeed the full 4-point string amplitude. Exactly in the same way as in the field theory, this factor makes manifest that if we further set $c_{k,m}$ to a non-positive integer the whole amplitude vanishes. 
Plugging this result back into~\eqref{eq-fac1}, 
	\begin{equation}
	\begin{aligned}
& \int \prod_{l=3}^{i-1} \frac{\diff y_{1,l}}{y_{1,l}} y_{1,l}^{\alpha^{\prime} X_{1,l}}  \prod_{1\leq a< b-1\leq i-1} F_{a b}(\mathbf{y})^{-\alpha^{\prime} c_{a b}}\int \prod_{j=i+1}^{n-1} \frac{\diff y_{1,j}}{y_{1,j}} y_{1,j}^{\alpha^{\prime} X_{1,j}} \prod_{i-1\leq e< f-1< n-1} F_{e,f}(\mathbf{y})^{-\alpha^{\prime} c_{e,f}}\\
		&\quad\times A^{-\alpha^{\prime} c_{k,m}+\alpha^{\prime} X_{1,i}}B^{-\alpha^{\prime} X_{1,i}}\times 	\mathcal{I}_{4}^{\mathrm{Tr}(\phi^3)}(\alpha^{\prime} X_{1,i},\alpha^{\prime}(c_{k,m}-X_{1,i}))\\
		&=\int \prod_{l=3}^{i-1} \frac{\diff y_{1,l}}{y_{1,l}} y_{1,l}^{\alpha^{\prime} X_{1,l}}  \prod_{1\leq a< b-1\leq i-1} F_{a,b}(\mathbf{y})^{-\alpha^{\prime} c_{a,b}^{\prime}}\int \prod_{j=i+1}^{n-1} \frac{\diff y_{1,j}}{y_{1,j}} y_{1,j}^{\alpha^{\prime} X_{1,j}^{\prime}} \prod_{i-1\leq e< f-1< n-1} F_{e,f}(\mathbf{y})^{-\alpha^{\prime} c_{e, f}^{\prime}}\\
		&\quad\times 	\mathcal{I}_{4}^{\mathrm{Tr}(\phi^3)}(\alpha^{\prime} X_{1,i},\alpha^{\prime}(c_{k,m}-X_{1,i}))\\
    &\equiv\mathcal{I}_{i}^{ \mathrm{down},\mathrm{Tr}(\phi^3) }\times \mathcal{I}_{n-i+2}^{\mathrm{up},\mathrm{Tr}(\phi^3)} \times \mathcal{I}_{4}^{\mathrm{Tr}(\phi^3)}(\alpha^{\prime} X_{1,i},\alpha^{\prime}(c_{k,m}-X_{1,i})),
	\end{aligned}
\end{equation}
where in the third line we replace $A(\vec{y})$ and $B(\vec{y})$, and define the shifted exponents $c^\prime$ and $X^\prime$. We have that: $c_{a,b}^{\prime}=c_{a,b}$ except for $c_{k,i-1}^{\prime}=c_{k,i-1}+X_{1,i}$, while $c_{e,f}^{\prime}=c_{e,f}$ except for $c_{i-1,m}^{\prime}=c_{i-1,m}+c_{k,m}-X_{1,i}$. Finally we also have $X_{1,j}^{\prime}=X_{1,j}-X_{1,i}$, for $j=i+1,\dots,m$. As we will see shortly, these shifts are exactly such that the up and down amplitudes depend on exactly the same kinematics as the respective up and down amplitudes in the field theory case, $i.e.$ the lower point string amplitudes appearing in the factorization also follow the kinematic shifts summarized in figure \ref{fig:ZerosFactGS}.
 
Looking at the kinematic mesh,  $\mathcal{I}_{i}^{\mathrm{down},\mathrm{Tr}(\phi^3)}$ denotes the string amplitude corresponding to the down triangular mesh, which is  $\mathcal{I}_{i}^{\mathrm{Tr}(\phi^3)}(1,\dots,i-1,I)$ with $p_{I}=-\sum_{j=1}^{i-1}p_{j}$, but now with shifted kinematic invariants:
\begin{equation}
\begin{aligned}
\mathcal{I}_{i}^{\mathrm{down},\mathrm{Tr}(\phi^3)}&=\int \prod_{l=3}^{i-1} \frac{\diff y_{1,l}}{y_{1,l}} y_{1,l}^{\alpha^{\prime} X_{1,l}}  \prod_{1\leq a< b-1\leq i-1} F_{a,b}(\mathbf{y})^{-\alpha^{\prime} c_{a,b}^{\prime}}\\
    &= \mathcal{I}_{i}^{\mathrm{Tr}(\phi^3)}(1,\dots,i-1,I)\bigg \vert_{X_{l,i}\to X_{l,i}+X_{1,i}=X_{l,n},  \text{ for } l=2,\dots,k.},
\end{aligned}
\end{equation}
where the shifted rules for the kinematic invariant are shifting the $X_{l,i}$ to $X_{l,i}+X_{1,i}$, which is equal to $X_{l,n}$ since the zero conditions, for $l=2,\dots,k$. These shifted rules are equivalent to the definition of $c_{a,b}^{\prime}$.
 
In turn, $\mathcal{I}_{n-i+2}^{\mathrm{up},\mathrm{Tr}(\phi^3)}$ denotes as the string amplitude corresponding to the upper triangular mesh, which is $\mathcal{I}_{n-i+2}^{\mathrm{Tr}(\phi^3)}(i,\dots,n,J)$ with $p_{J}=-\sum_{j=i}^{n}p_{j}$, but with shifted kinematic invariant:
\begin{equation}
\begin{aligned}
   \mathcal{I}_{n-i+2}^{\mathrm{up},\mathrm{Tr}(\phi^3)}&=\int \prod_{j=i+1}^{n-1} \frac{\diff y_{1,j}}{y_{1,j}} y_{1,j}^{\alpha^{\prime} X_{1,j}^{\prime}} \prod_{i-1\leq e< f-1< n-1} F_{e,f}(\mathbf{y})^{-\alpha^{\prime} c_{e,f}^{\prime}}\\
    &=\int \prod_{j=i+1}^{n-1} \frac{\diff y_{i-1,j}}{y_{i-1,j}} y_{i-1,j}^{\alpha^{\prime} X_{i-1,j}^{\prime}} \prod_{i-1\leq e< f-1< n-1} F_{e,f}(\mathbf{y})^{-\alpha^{\prime} c_{e, f}^{\prime}}\\
    &=\mathcal{I}_{n-i+2}^{\mathrm{Tr}(\phi^3)}(i,\dots,n,J)\vert_{X_{i-1,j}\to X_{i-1,j}-X_{i-1,n}=X_{1,j}, \text{ for } j=m+1,\dots,n-1.},
\end{aligned}
\end{equation}
where in the second line, we change the integration variables $y_{1,j}$ to $y_{i-1,j}$. The kinematic shifts send $X_{i-1,j}$ to $X_{i-1,j}-X_{i-1,n}$, which is equal to $X_{1,j}$ since the zero conditions, for $j=m+1,\dots,n-1$. This shifted rules are equivalent to the definition of $c_{e,f}^{\prime}$ and $X_{1,j}^{\prime}$.
 
In summary, by picking a zero causal diamond, $c_{ab}=0,\text{for}\, 1\leq a \leq i-2\,,i\leq b\leq n-1$, and letting $c_{k,m}\neq 0$, the stringy integral for $\mathrm{Tr}(\phi^3)$ factorizes into lower point amplitudes according to:
\begin{equation}\label{eq-3fac}
	\begin{aligned}
		\mathcal{I}_{n}^{\mathrm{Tr}(\phi^3)}&\to \mathcal{I}_{i}^{\mathrm{down},\mathrm{Tr}(\phi^3)}\times \mathcal{I}_{n-i+2}^{\mathrm{up},\mathrm{Tr}(\phi^3)}\times \mathcal{I}_{4}^{\mathrm{Tr}(\phi^3)}(\alpha^{\prime} X_{1,i},\alpha^{\prime}(c_{km}-X_{1,i})).
	\end{aligned}
\end{equation}

The shifted rules for up and down amplitudes are exactly those presented in figure \ref{fig:ZerosFactGS}, and can be summarized as follows:
\begin{align}\label{eq-shiftedrule}
&X_{l,i}\to X_{l,i}+X_{1,i}=X_{l,n}\,,\quad \text{for}\quad l=2,\dots,k.\\
&X_{i-1,j}\to X_{i-1,j}-X_{i-1,n}=X_{1,j}\,,\quad \text{for}\quad j=m+1,\dots,n-1.
\end{align}
\subsubsection{Factorization for general negative integers}
We have seen that when all the mesh constants but one in a maximal causal diamond are set to zero, the amplitude simplifies by factoring to a product of smaller amplitudes with non-trivially modified kinematics.  Just as the statement about zeros extends to the full string amplitude when the mesh constants are set more generally either to zero or negative integers, the near-zero factorizations also generalize to this case. We will see that instead of simply factoring into a product of smaller amplitudes, we get an interesting sum over products of smaller amplitudes with redefined kinematics. To see how this works let us consider as an example a skinny rectangle factorization where we set $c_{1,3}=-n_{1,3}, c_{1,5} = -n_{1,5}$ but we turn on $c_{1,4}$. The stringy integral becomes 
\begin{eqnarray}
{\cal I}_6 = \int_0^\infty \frac{\diff y_{1,3} \diff y_{1,4} \diff y_{1,5}}{y_{1,3} y_{1,4}y_{1,5}}  y_{1,3}^{X_{1,3}} y_{1,4}^{X_{1,4}} y_{1,5}^{X_{1,5}} \times (1 + y_{1,3})^{n_{1,3}} (1 + y_{1,5}(1 + y_{1,4}(1 + y_{1,3})))^{n_{1,5}} \times \nonumber \\
(1 + y_{1,4}(1 + y_{1,3}))^{-c_{1,4}}  (1 + y_{1,4})^{-c_{2,4}} (1 + y_{1,5}(1 + y_{1,4}))^{-c_{2,5}} (1 + y_{1,5})^{-c_{3,5}} .\nonumber
\end{eqnarray}

Relative to our usual expressions when mesh constants are set to zero, we have the extra factor on the first line, $(1 + y_{1,3})^{n_{1,3}} (1 + y_{1,5}(1 + y_{1,4}(1 + y_{1,3})))^{n_{1,5}}$. The important fact is that for $n_{1,3}, n_{1,5}$ positive integers, this factor is a {\it finite polynomial} in $y_{1,3},y_{1,4}, y_{1,5}$, we have 
\begin{equation}
    (1 + y_{1,3})^{n_{1,3}} (1 + y_{1,5}(1 + y_{1,4}(1 + y_{1,3})))^{n_{1,5}} = \sum_{k_{1,3},k_{1,4},k_{1,5}} C_{k_{1,3},k_{1,4},k_{1,5}}(n_{1,3},n_{1,5}) y_{1,3}^{k_{1,3}} y_{1,4}^{k_{1,4}} y_{1,5}^{k_{1,5}},
\end{equation}
where $C_{k_{1,3},k_{1,4},k_{1,5}}(n_{1,3},n_{1,5})$ are constants arising simply from performing the multinomial expansions of each term; while these can be trivially computed the detailed expressions are not important for us. But at this point, the expression is precisely the one we encountered previously when the mesh constants were set to zero, except as a weighted sum with weights $C_{k_{1,3},k_{1,4},k_{1,5}}$ of the previous amplitudes with kinematics shifted as $X_{1,3} \to X_{1,3} + k_{1,3}, \, X_{1,4} \to X_{1,4} + k_{1,4},\,  X_{1,5} \to X_{1,5} + k_{1,5}$. 

For instance if we set $n_{1,3}=1, n_{1,5} = 1$, we have $(1 + y_{1,3})^1 (1 + y_{1,5}(1 + y_{1,4}(1 + y_{1,3})))^1 = 1 + y_{1,3} + y_{1,5} + y_{1,3} y_{1,5} + y_{1,4} y_{1,5} + 2 y_{1,3} y_{1,4} y_{1,5} + y_{1,3}^2 y_{1,4} y_{1,5}$. 
Now recall that for factorization when mesh constants are set to zero, 
we have the factorization ${\cal I}_6 \to F_4(X) F_5(X)$ where $F_4(X) = \Gamma[X_{1,3}] \Gamma[X_{2,6}]/\Gamma[X_{1,3} + X_{2,6}]$ and $F_5(X) = {\cal I}_5(X_{2,4}, X_{2,5}\to X_{1,5}, X_{3,5},X_{3,6},X_{4,6})$. Then at $n_{1,3} = n_{1,5} = 1$ we instead have the factorization 
\begin{eqnarray}
{\cal I}_6 &\to & \left(F_4 F_5\right) + \left(F_4 F_5\right) (X_{1,3} \to X_{1,3} + 1) \nonumber \\ &+& \left(F_4 F_5\right)(X_{1,5} \to X_{1,5} + 1) + \left(F_4 F_5\right)(X_{1,3} \to X_{1,3} + 1, X_{1,5} \to X_{1,5} + 1) \nonumber  \\ &+&  \left(F_4 F_5\right)(X_{1,4} \to X_{1,4} + 1, X_{1,5} \to X_{1,5} + 1) \nonumber \\ &+&  2 \left(F_4 F_5\right)(X_{1,3} \to X_{1,3} + 1, X_{1,4} \to X_{1,4} + 1, X_{1,5} \to X_{1,5} + 1) \nonumber \\ &+& \left(F_4 F_5\right)(X_{1,3} \to X_{1,3} + 2, X_{1,4} \to X_{1,4} + 1, X_{1,5} \to X_{1,5} + 1).
\end{eqnarray}

The story for near-zero factorizations when mesh constants are set to general negative integers is the same. We always get a factor which is a polynomial $P$ in all the $y$'s associated with the $F$ polynomials raised to integer powers, which we can simply expand to get a big polynomial in the $y$'s. This then gives us a factorization of exactly the same form as with vanishing mesh constants, but as a sum over shifted kinematics determined by the polynomial $P$.

\section{Stringy $\delta$eformation}
\label{sec:StringDelta}

In this section, we propose a class of ``universal'' stringy models as a one-parameter deformation of the tree-level stringy Tr$(\phi^3)$ amplitude for an even number $2n$ of particles, which will be the basis of our unification of Tr$(\phi^3)$, pion and gluon amplitudes. 

This deformation amounts to inserting a factor $(\prod u_{e,e}/\prod u_{o,o})^{\alpha^\prime \delta}$ in the string integral \eqref{eq:stringyphi3}, where we take the product of all $u_{a,b}$ with $a,b$ both being even, over the product of $u_{a,b}$ both being odd, and $\delta$ is the deformation parameter:
\begin{equation}  \label{eq:stringyYMSNLSM}
\mathcal{I}_{2n}^{\delta}=\int_{\mathbb{R}_{>0}^{2n{-}3}} \prod_{I=1}^{2n{-}3} \frac{\diff y_I}{y_I}~\prod_{(a, b)} u_{a,b}^{\alpha' X_{a,b}} \left(\frac{\prod_{(e,e)}u_{e,e}}{\prod_{(o,o)}u_{o,o}}\right)^{\alpha^\prime \delta},
\end{equation}
expanding this ratio, we see that this factor is simply \textbf{shifting} the kinematics, $X_{a,b}$, of the un-deformed, stringy Tr$(\phi^3)$ amplitude:
\begin{equation}
{\cal I}_{2n}^{\delta}={\cal I}_{2n}^{\text{Tr}(\phi^3)} [\alpha' X_{e,e} \to \alpha'( X_{e,e} + \delta), \alpha' X_{o,o}\to \alpha' (X_{o,o}-\delta)]\,.    
\end{equation}

In addition, we claim that for different values of $\alpha^\prime \delta$ we get different colored theories:
\begin{enumerate}
    \item \underline{$\alpha^\prime \delta =0 $}: In this case, we get back the usual stringy Tr$(\phi^3)$ integral, so at low energies, we get the field theory amplitudes of the Tr$(\phi^3)$ Lagrangian:
   \begin{equation}
    \mathcal{L}_{\mathrm{Tr}(\phi^3)}= \operatorname{Tr}\,(\partial \phi)^2+ g \operatorname{Tr}(\phi^3).
      \end{equation} 
    \item \underline{$\alpha^\prime \delta \in (0,1) $}: In this case we claim that, at low energies, we get field theory amplitudes of the U(N) Non-linear Sigma Model ({\it c.f.}~\cite{Kampf:2013vha})
    :
    \begin{equation}
    \mathcal{L}_{\mathrm{NLSM}}=\frac{1}{8 \lambda^2} \operatorname{Tr}\left(\partial_\mu \mathrm{U}^{\dagger} \partial^\mu \mathrm{U}\right), \quad \text { with } \quad \mathrm{U}=(\mathbb{I}+\lambda \Phi)(\mathbb{I}-\lambda \Phi)^{-1}.
\end{equation}
\item \underline{$\alpha^\prime \delta =1 $}: In this case, we claim that, at low energies, we get Yang-Mills theory (YM). In fact, it is not simply YM, but instead gluons and adjoint scalars (YMS) (these YMS amplitudes have been studied in {\it e.g.}~\cite{Cachazo:2014xea}), with the following Lagrangian:
    
\begin{equation} \label{eq:LYMS}
    \mathcal{L}_{\mathrm{YMS}}=-\operatorname{Tr}\left(\frac{1}{4} F^{\mu \nu} F_{\mu \nu}+\frac{1}{2} D^\mu \phi^I D_\mu \phi^I-\frac{g_{\rm YM}^2}{4} \sum_{I \neq J}\left[\phi^I, \phi^J\right]^2\right).
\end{equation}
  
\end{enumerate}

In the rest of this section, we will study in detail how this deformation works. However, for a more complete explanation of each shift see \cite{NLSM,Gluons}.
 
There are multiple ways to motivate why this new factor is the correct deformation~\cite{NLSM,Gluons}. Notwithstanding, in the context of this note, what interests us the most is to understand how it explains the fact that the zeros and factorizations are present for these three colored theories. Indeed it turns out that this is the \textbf{only} kinematic shift that one can do on the $X_{i,j}'s$ that preserves the $c_{i,j}'s$! We will prove this shortly but let us start by understanding why this is the case and how it implies the generalization of the zeros/factorizations for any value of the deformation.
 
From equation \eqref{eq:ceq}, establishing the relation between the non-planar variables to the planar ones, we can easily see that by shifting $ X_{e,e} \to  X_{e,e} + \delta $ and $ X_{o,o} \to  X_{o,o}-\delta$, while keeping $X_{o,e}$ unchanged, the shift exactly cancels in \eqref{eq:ceq} thus preserving all $c_{i,j}$'s. This means that independent of the underlying triangulation we choose, $\mathcal{T}$, to parametrize $u_{i,j}[y_{\mathcal{T}}]$, the $c$'s appearing in the exponents of the $F$-polynomials in the string integral remain unchanged. Therefore the result of this shift can only change the exponents of $y_i$. Note that our derivation in sections ~\ref{sec:stringZeros} and ~\ref{sec:stringFact} of the zeros/factorizations of the stringy integral is \textbf{independent} of these exponents of $y_i$. Therefore, the fact that the $c_{i,j}'s$ remain unchanged under this shift implies that the zeros and factorizations are also true for ${\cal I}^\delta_n$!

We stress that the existence of zeros and factorizations for both field theory and string theory amplitudes is made obvious from the associahedron and the stringy $\delta$eformation we describe in this section. We expect that with these statements in hand, it should be possible to prove them from different starting points as well. For instance we have understood how the zeros and factorization of Tr $\phi^3$, NLSM, and YMS field-theory amplitudes can be proven starting from their CHY formulas~\cite{Cachazo:2013hca, Cachazo:2013iea, Cachazo:2014xea}, though the formalism doesn't make these facts obvious. We find it more natural and satisfying that these hidden properties of both string and field-theory amplitudes, together with the startling unity of all these colored theories, are manifested via stringy $\delta$eformation. 

\subsection{Uniqueness of kinematical shift}

\begin{figure}[t]
    \centering
\includegraphics[width=0.8\textwidth]{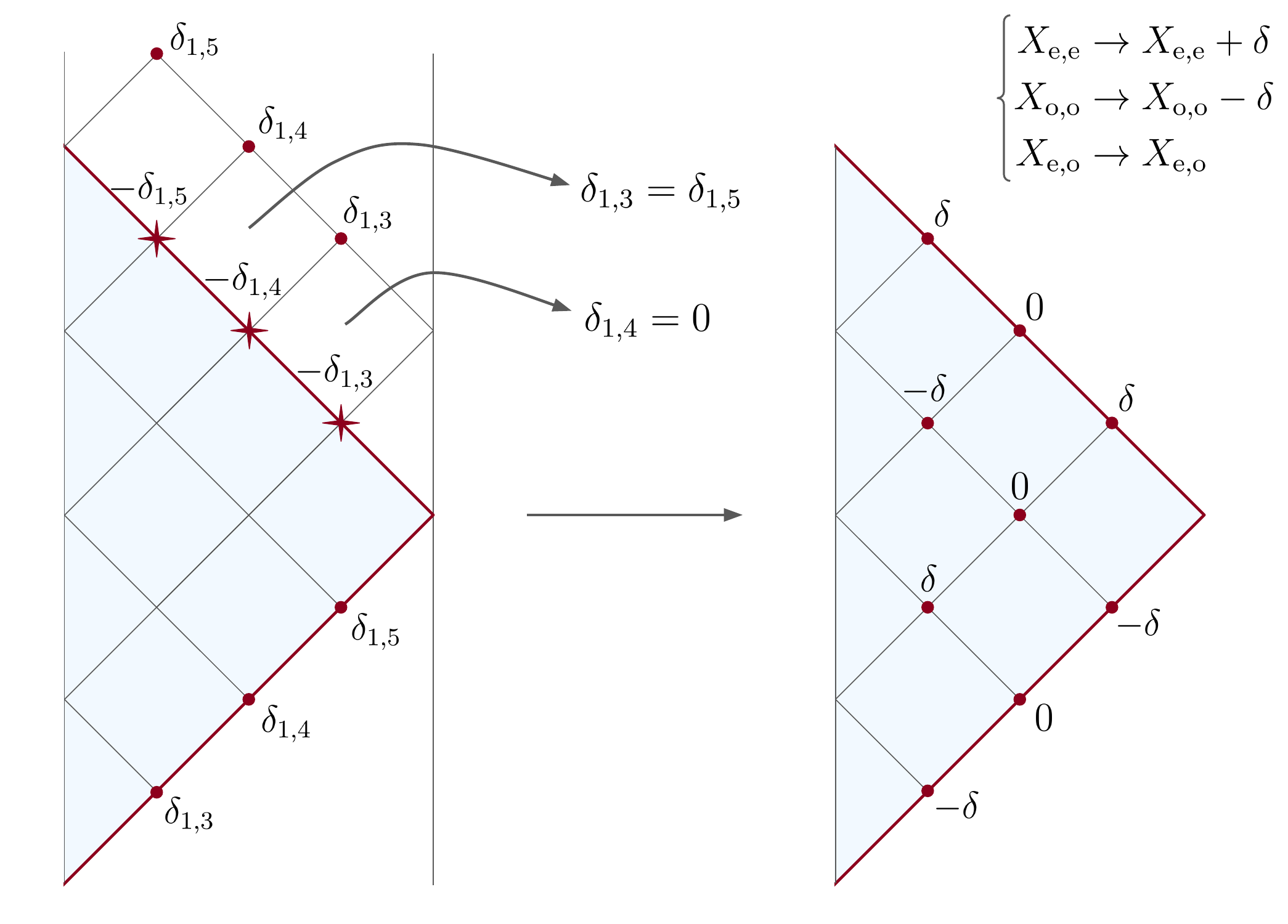}
    \caption{The $c$-preserving shift for $6$-point stringy integral.}
    \label{6pt-c-preserve}
\end{figure}

The striking fact is not only that this shift exactly preserves the non-planar variables, but it actually is the \textbf{unique} shift that does so. 
 
To prove this, we start by noticing that in order to preserve all $c_{i,j}$ of a ray-like triangulation, it suffices to specify shifts of the planar variables $X_{a,b}$ in the triangulation. This is because we can solve for all the remaining $X$ variables in terms of the $c_{i,j}$'s and $X_{a,b}$. While any shifts of $n{-}3$ initial $X_{a,b}$ preserve $c_{i,j}$ appearing for this triangulation, asking for the shift to preserve \textbf{all} the $c_{i,j}$, so that we have \textbf{all} the zeros and factorizations is much more constraining. Let us choose a specific triangulation, say the ray-like one with $(1,3), (1,4), \cdots, (1, n{-}1)$, and specify the initial shifts as \begin{equation}
X_{1,i} \to X_{1,i} + \delta_{1,i},\quad{\rm for}~i=3,4, \cdots, n{-}1.    
\end{equation}  

In order to preserve $c_{i,j}$ of this triangulation (with $1\leq i<j<n$) we must shift $X_{a,b}$ according to the solution of \eqref{eq:ceq}; in particular, we find the following shifts for $X_{i,n}$ (the variables in the opposite edge of the triangular region):
\begin{equation}
X_{i,n} \to X_{i,n} -\delta_{1, i{+}1},  \quad{\rm for}~i=2,3, \cdots, n{-}2. 
\end{equation}

In other words, $X_{2,n}$ must be shifted by the same amount but in the opposite direction as $X_{1,3}$, same for $X_{3,n}$ and $X_{1,4}$, and so on. Now how can we preserve the remaining $c_{i,n}$ for $i=2,3,\cdots, n{-}2$? Note $c_{2,n}=X_{2,n}+X_{1,3}-X_{3,n}$ and the shifts of $X_{2,n}$ and $X_{1,3}$ cancel on the RHS, thus to preserve $c_{2,n}$ we must have the shift of $X_{3,n}$ vanishes, $\delta_{1,4}=0$. To preserve $c_{3,n}=X_{3,n}+X_{1,4}-X_{1,3}-X_{4,n}$, we must have $\delta_{1,5}=\delta_{1,3}$. Similarly we find $\delta_{1,6}=0$, $\delta_{1,7}=\delta_{1,5}$ and so on. Thus the only $c$-preserving shifts correspond to $\delta_{1, e}=0$ and all $\delta_{1,o}$ equal. However, for odd $n$ we further conclude that $\delta_{1, o}=0$ as well, and only for even $n$ we can have non-vanishing and equal $\delta_{1,o}$, which we call $-\delta$. This leads to $\delta X_{e,e}=-\delta X_{o,o}=\delta$, and $\delta X_{o,e}=0$, which are the only possible $c$-preserving shifts!

\subsection{Realization of the shift on momenta}
We have expressed our shift directly in terms of its action on our basis of invariants $X_{i,j}$, but we can of course also describe it explicitly in terms of a shift on the momenta directly. It is actually slightly more convenient to describe the shift in terms of the vertices $x_i^\mu$ of the momentum polygon, from which the shift on the momenta $p_i^\mu = (x_{i+1}^\mu - x_i^\mu)$ can be inferred. To realize the shift for an $2n$ particle process, we imagine adding $2n$ dimensions of spacetime, orthogonal to the ones the original momentum polygon lives in, which are grouped into $n$ pairs of different timelike and spacelike directions $t_a^\mu, s_a^\mu$ for $a=1, \cdots, n$.  So we have $s_a \cdot x_j = t_a \cdot x_j = 0$ for all $a, j$, and also $t_a \cdot t_b,\, t_a \cdot s_b,\, s_a \cdot t_b,\, s_a \cdot s_b = 0$ for $a \neq b$. Finally we normalize $t_a^2 =\frac{\delta}{2}, s_a^2 = -\frac{\delta}{2}$. 
Then we can define the shift 
\begin{equation}
x^\mu_{2k} \to x^\mu_{2k} + t^\mu_k, \, \, \, x^\mu_{2k+1} \to x^\mu_{2k+1} + s^\mu_k,
\end{equation}
sending 
\begin{eqnarray}
X_{2k, 2l} = (x_{2k} - x_{2l})^2 &\to& (x_{2k} - x_{2l} + t_{k} - t_{l})^2 = X_{2k,2l} + \delta ,\nonumber \\ X_{2k+1,2l+1} = (x_{2k+1} - x_{2l +1})^2 &\to& (x_{2k+1} - x_{2l + 1} + s_k - s_l)^2 = X_{2k+1,2l+1} - \delta ,\nonumber \\ X_{2k, 2l+1} = (x_{2k} - x_{2l+1})^2 &\to& (x_{2k} - x_{2l+1} +t_k - s_l)^2 = X_{2k,2l+1},
\end{eqnarray}
which is just our kinematic shift.

\subsection{4-point amplitudes}
\label{sec:4ptdeformed}

In order to gain more intuition for the physics of our shifts, it is useful to study the shifted four-particle amplitude in some detail. Recall that with our shifts we have 
\begin{equation}
{\cal I}^\delta_4 = \frac{\Gamma[\alpha^\prime (X_{1,3} - \delta)] \Gamma[\alpha^\prime(X_{2,4} + \delta)]}{\Gamma[\alpha^\prime(X_{1,3} + X_{2,4})]}.
\end{equation}

Of course for $\delta = 0$, the low-energy amplitude is $\frac{1}{ \alpha^\prime X_{1,3}} + \frac{1}{\alpha^\prime X_{2,4}}$ of the massless Tr$(\phi^3)$ theory. We can get an idea of the massive spectrum of states in the UV completion by looking at the residue on the first massive pole at e.g. $\alpha^\prime X_{1,3} = -1$. This residue is $(1 - \alpha^\prime X_{2,4})$, and using the familiar translation from $s,t$ variables to center-of-mass energies and angles, $t = ({\rm cos} \theta - 1)s/2 $, the residue at $\alpha^\prime X_{1,3} = \alpha^\prime s = -1$ becomes $\frac{({\rm cos} \theta + 1)}{2}$. The angular dependence on ${\rm cos} \theta$ allows to read off that at this mass level we have the exchange of a particle of spin 0 and one of spin 1. 
\begin{figure}[t]
    \centering
    \includegraphics[width=\textwidth]{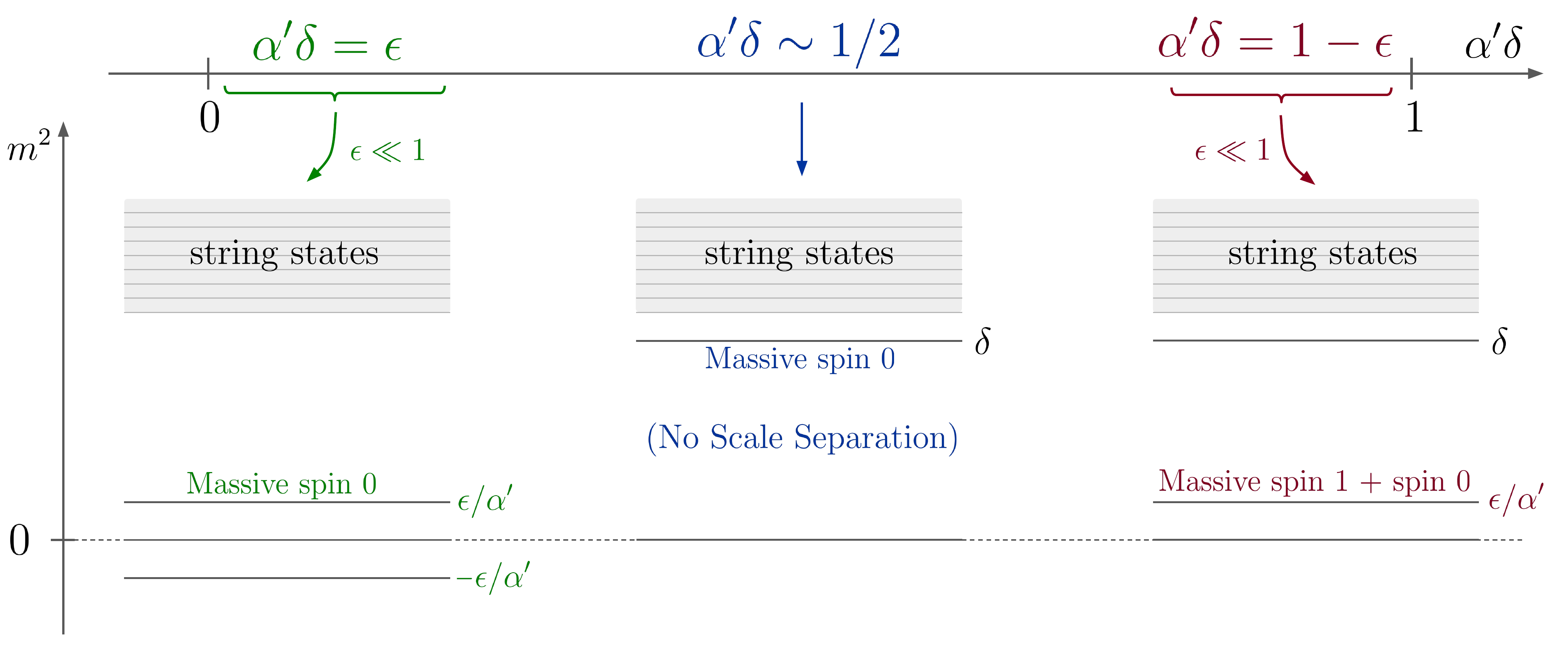}
    \caption{UV completion for different values of $\alpha^\prime \delta$.}
    \label{fig:spectrum}
\end{figure}
Let us now turn on $\delta$ and see what happens as we vary $\alpha ^\prime \delta$ from very small values near $0$, to intermediate fractional values, and then near $\alpha^\prime \delta \to 1$ (see figure \ref{fig:spectrum}). At very low-energies for $\alpha^\prime X_{1,3}, \alpha^\prime X_{2,4} \ll 1$, we have 
\begin{equation}
{\cal I}^\delta_4 \to \alpha^\prime\left(\Gamma[-\alpha^\prime \delta] \Gamma[+\alpha^\prime \delta] \right) \times (X_{1,3} + X_{2,4}),
\end{equation}
giving us the amplitude for NLSM with $\lambda^2 = \alpha^\prime \Gamma[-\alpha^\prime \delta] \Gamma[+ \alpha^\prime \delta]$. 
We can again get an idea of the spectrum of massive states, by looking at the first massive level; this is actually a shifted version of the massless pole we had at $\delta = 0$; the residue at $X_{1,3} = \delta$ is simply 1! Thus we learn that at the first massive level where $X_{1,3} = \delta$, we are exchanging a massive spin-0 particle. If $\alpha ^\prime \delta \ll 1$, there is a separation of scales between this state and the rest of the string states. At very low energies where $X \ll \delta$, we have the amplitude for pions. At $X_{1,3} = \delta$ we encounter an extra massive spin-0 particle. The amplitudes in the intermediate region $\delta \ll X_{1,3},X_{2,4} \ll \frac{1}{\alpha^\prime}$ is simply that of the Tr$(\phi^3)$ theory, which softens the UV power-law growth of the low-energy NLSM amplitudes into falling $1/X$ behavior. And ultimately for energies above the string scale $\alpha^\prime X_{1,3},\alpha^\prime X_{2,4} \gg 1$ we see the tower of stringy excitations and the softest UV behavior characteristic of string theory. 

When $\alpha^\prime \delta$ is no longer small, but say near $\sim 1/2$, there is no separation between the first massive scalar and the rest of the string states, and so we get a purely stringy UV completion with no intermediate Tr$(\phi^3)$ regime. 

The situation becomes more interesting as we approach $\alpha^\prime \delta \to 1$. Let us put $\alpha^\prime \delta = 1 - \epsilon$. Then we have a light state $X_{1,3} = - \epsilon$. The residue on this pole is $\Gamma[\alpha^\prime X_{2,4} + 1 - \epsilon]/\Gamma[X_{2,4} - \epsilon] = (X_{2,4} - \epsilon)$. Again translating to energies and angles, we see that at this small mass we are exchanging both a spin-0 and a spin-1 particle. Thus, for $
\alpha^\prime \delta$ very close to 1, we see pion amplitude at low energies, but encounter, instead of a massive spin-0 particle as we did for $\alpha^\prime \delta$ close to 0, a massive spin-1 particle (and a scalar), once again softening the UV behavior of the low-energy theory. This is familiar physics from real-world QCD, where the $\rho$ meson plays the dominant role in the unitarization of pion scattering. 

This makes it clear what we must expect at $\alpha^\prime \delta \to 1$. The light massive spin-1 particle becomes massless in this limit, and the only consistent theory we could be describing is Yang-Mills theory! Of course we have an amplitude for external scalars, and so we are describing colored scalars coupled to gluons. Exactly at $\alpha^\prime \delta = 1$, the amplitude is 
\begin{equation} 
{\cal I}^{{\rm \delta = 1}}_4 = \frac{\Gamma[\alpha^\prime X_{1,3} - 1] \Gamma[\alpha^\prime X_{2,4} + 1]}{\Gamma[\alpha^\prime(X_{1,3} + X_{2,4})]}.
\end{equation}

Note that these shifts keep a massless gluon pole at $X_{1,3} = 0$, but remove it at $X_{2,4} = 0$. Thus we must interpret this amplitude as that of two \textbf{different} colored scalars $A,B$, in the configuration $1^A 2^A 3^B 4^B$, so that the $X_{1,3}$ channel gluon exchange is allowed by the $X_{2,4}$ channel exchange is forbidden. This interpretation is easily confirmed by looking at the low-energy limit, where the amplitude becomes $X_{2,4}/X_{1,3}+1$, precisely corresponding to $X_{1,3}$ channel gluon exchange for $1^A 2^A 3^B 4^B$ scattering plus the four-vertex contact diagram. 

As we will now see, this story generalizes for all shifted $2n$ particle amplitudes. For $\delta =0$ we have the amplitudes for Tr$(\phi^3)$ theory at low energies. Instead for general fractional $\delta$, the low-energy amplitudes are those of NLSM. When $\alpha^\prime \delta$ is small, the NLSM amplitudes are first UV softened into those of Tr$(\phi^3)$ at energies above $\sim \delta$, before being further UV softened into string amplitudes above the scale $1/\alpha^\prime$. For $\alpha^\prime \delta$ of order one, the UV softening of the low-energy NLSM amplitudes is purely stringy. But as $\alpha^\prime \delta$ increases further and approaches one, colored massive spin-1 particles descend from the string states, and at $\alpha^\prime \delta = 1$, we are describing the amplitudes for pairs of $n$ distinct colored scalars, ordered as  $M_{1,\cdots, n} ( 1^{A_1} 2^{A_1} 3^{A_2} 4^{A_2} \cdots (2n - 1)^{A_n} (2n)^{A_n})$. 
As we will discuss further below and explore at much greater length in \cite{Gluons}, these amplitudes give us, inter-alia, direct access to $n$-gluon amplitudes, by factorizing $2n$-scalar amplitudes on gluon poles where $(p_{2k - 1} + p_{2k})^2 \to 0$. 

\subsection{$\alpha^\prime \delta \in (0,1)$ and the Non-linear Sigma Model}
For $\alpha^\prime \delta$ non-integer we claim that
\begin{equation}
\mathcal{I}_{2n}^{\delta}=\int_{\mathbb{R}_{>0}^{2n{-}3}} \prod_{I=1}^{2n{-}3} \frac{\diff y_I}{y_I}~\prod_{(e, e)} u_{e,e}^{\alpha' (X_{e,e}+\delta)} \times \prod_{(o, o)} u_{o,o}^{\alpha' (X_{o,o}-\delta)} \times \prod_{(o, e)} u_{o,e}^{\alpha'X_{o,e}} ,
\label{eq:stringyNLSM}
\end{equation}
yields NLSM tree-level amplitudes at low energies, $i.e.$ in the $\alpha^\prime \rightarrow 0$ limit, thus explaining all the mysterious zeros and factorizations observed for these amplitudes.
 
To show that this is the case let us understand the consequences of this shift. For $\delta=0$, the fact that the string amplitude has poles associated with massless resonances for $X_{i,j}\rightarrow0$, is because $u_{i,j}^{\alpha^\prime X_{i,j}} \rightarrow 1$, and thus the integral develops a singularity in one of the integration boundaries. This is particularly easy to see for the $X_{i,j} \in \mathcal{T}$: in this case, the integral becomes singular for $X_{i,j}=0$ because it diverges near $y_{i,j}=0$ as we see in \eqref{eq:stringyphi3}. For $\delta \neq 0$, we lose the poles corresponding to $X_{e,e},X_{o,o}$=0, since the divergences of the integral are still regulated by $\delta$. Note that poles corresponding to  $X_{e,e},X_{o,o}$ chords are associated with propagators enclosing \textbf{odd-point} interactions. Therefore, at leading order in the low energy expansion, we only have poles when $X_{o,e}=0$, which are precisely the propagators appearing in diagrams built only out of even-point interactions. This is exactly the pole structure that we expect for NLSM amplitudes, and thus the first hint that we are going in the correct direction. 
 
Now provided we have the correct pole structure, $i.e.$ we don't have any poles for $X_{e,e},X_{o,o}$=0, then the fact that the amplitude vanishes in the skinny-rectangle type zero (which it does because the undeformed amplitude does) implies that this amplitude has vanishing soft limit, $i.e.$ it has the \textbf{Adler zero}. This is exactly what we concluded before just for the case of field theory NLSM, that our zero \textbf{implies} the Adler zero for these amplitudes. 
 
Finally, just from $u$-equations, we have that on the $X_{o,e} \rightarrow 0$, the amplitude \eqref{eq:stringyNLSM} factorizes into the product of the respective lower-point amplitudes. At tree-level, the Adler zero together with factorization ensure that the low energy limit corresponds to NLSM amplitude~\cite{Cheung:2014dqa,Cheung:2015ota}. Alternatively, we could make the same conclusion via the uniqueness theorem~\cite{Arkani-Hamed:2016rak,Rodina:2016jyz} even without the knowledge of factorization.
 
For this note, this is enough since we are interested in understanding the zeros and factorizations of tree-level amplitudes. However, as it is explained in \cite{NLSM} this shift also allows us to obtain NLSM loop-level amplitudes. 
 
This way we have proved that \eqref{eq:stringyNLSM} defines a new completion of the NLSM amplitudes. Different stringy completions of the NLSM have been proposed \cite{Carrasco:2016ldy,Bianchi:2020cfc}, but none make the zeros and factorizations manifest. In both cases, the stringy completion is \textbf{manifestly} cyclic, as opposed to our stringy completion, which is manifestly not cyclically symmetric but in which the leading order at low energies restores the cyclic symmetry expected for NLSM amplitudes. 
 
We thus see that the UV completion provided by this stringy formulation is not a familiar one. To understand this better let us look directly at what happens in the field-theory limit. 

\subsubsection{NLSM from field theory Tr$(\phi^3)$}
To extract the field theory limit, let us assume that we also have $\alpha^\prime \delta \ll 1$. Therefore we obtain:
\begin{equation}
\begin{aligned}
\mathcal{I}_{2n}^{\delta}=&\int_{\mathbb{R}_{>0}^{2n{-}3}} \prod_{I=1}^{2n{-}3} \frac{\diff y_I}{y_I}~\prod_{(e, e)} u_{e,e}^{\alpha' (X_{e,e}+\delta)} \times \prod_{(o, o)} u_{o,o}^{\alpha' (X_{o,o}-\delta)} \times \prod_{(o, e)} u_{o,e}^{\alpha'X_{o,e}} \\
\rightarrow& \, \mathcal{A}_{2n}^{\text{Tr}(\phi^3)}( X_{e,e} \to X_{e,e} + \delta, X_{o,o}\to  X_{o,o}-\delta),
\end{aligned}
\end{equation}
where $\mathcal{A}_{2n}^{\text{Tr}(\phi^3)}$ stands for the field theory amplitude in Tr$(\phi^3)$ theory. Finally to get the real low energy behavior we need to further expand in $X\ll \delta$ or, equivalently, $\delta \rightarrow \infty$. From our previous argument, we have that the leading non-vanishing order in this expansion is the NLSM amplitude. 
 
This means that we can get the NLSM amplitude directly from the Tr$(\phi^3)$ field theory amplitude: 
\begin{equation}
A_{2n}^{\rm NLSM}=\lim_{\delta\to \infty} \delta^{2n-2} A_{2n}^{\text{Tr}(\phi^3)}( X_{e,e} \to X_{e,e} + \delta,  X_{o,o}\to  X_{o,o}-\delta), 
\end{equation}
where the prefactor $\delta^{2n-2}$ is there ensure the correct units. 
 
So we have that the UV completion provided by this stringy integral is one in which the NLSM is given as the low energy limit of a theory of colored massive scalars. These scalars can be regular scalars with mass$^2=\delta$ or tachyons with mass$^2=-\delta$. This is certainly an unfamiliar UV completion of the NLSM. Apart from the unusual presence of positive and negative mass$^2$ particles with precisely equal magnitudes, the UV amplitude is not cyclically invariant, while the NLSM amplitudes certainly are. Indeed the full UV amplitude does have a cyclic symmetry under $i \to i+1$ but only if we also flip the sign $\delta \to -\delta$. This implies that in the $1/\delta$ expansion, all terms with even powers of $\delta$ will be cyclically invariant while those with odd powers of $\delta$ will pick up a minus sign under cyclic shift. Quite beautifully, for the shifted $2n$ particle amplitude, after the naively leading powers of $1/\delta$ all cancel, we are left with an amplitude that begins with an even power $1/\delta^{2(n-1)}$ and hence is cyclically invariant as desired. A simple Lagrangian that generates the shifted Tr$(\phi^3)$ amplitudes and explains the non-cyclic nature of the UV completion will be presented in \cite{NLSM}. 

\paragraph{4-point} Let us now look at the 4-point amplitude. Starting from the Tr$(\phi^3)$ and performing the shift we get:
\begin{equation}
\mathcal{ A}_4^{\text{Tr}(\phi^3)}( X_{1,3} \to X_{1,3} - \delta,  X_{2,4}\to  X_{2,4}+\delta) = \frac{1}{X_{1,3}-\delta} + \frac{1}{X_{2,4}+\delta},
\end{equation}
now expanding in $\delta \gg 1$ yields:
\begin{equation}
  \mathcal{ A}_4^{\text{Tr}(\phi^3)}( X_{1,3} - \delta,  X_{2,4}+\delta) \,\xrightarrow{\delta \rightarrow \infty}\, \frac{1}{\delta} (1 -1) - \frac{1}{\delta^2} \underbrace{\left(X_{1,3}+X_{2,4}\right)}_{ \mathcal{ A}_4^{\text{NLSM}}} + \mathcal{O}(1/\delta^3),
\end{equation}
where indeed the first order cancels and the leading non-vanishing order gives the pion amplitude. Already in this small $4$-point problem, we can appreciate how important it is that the mass of $X_{e,e}$ is minus that of the mass of $X_{o,o}$. If this was not the case the leading order would be non-vanishing and we would \textbf{not} get the NLSM, instead, it would be closer to $\phi^4$ theory. 

\paragraph{6-point} At 6-point it is convenient to start by writing the Tr$(\phi^3)$ amplitude in a way that makes manifest the $X_{\text{o,e}}$ poles, as follows:

\begin{equation}
\begin{aligned}
    \mathcal{ A}_6^{\text{Tr}(\phi^3)}( X \to X \pm \delta) =& \frac{1}{X_{1,4}} \left( \frac{1}{X_{1,3}} + \frac{1}{X_{2,4}}\right)\left(\frac{1}{X_{4,6}} + \frac{1}{X_{1,5}} \right) + (\text{cyclic}, i\rightarrow i+2) \\ 
    & + \frac{1}{X_{1,3}X_{3,5}X_{1,5}}  + \frac{1}{X_{2,4}X_{4,6}X_{2,6}}.
\end{aligned}
\end{equation}

Performing the shift on the 6-point Tr$(\phi^3)$ amplitude yields:
\begin{equation}
\begin{aligned}
    \mathcal{ A}_6^{\text{Tr}(\phi^3)}( X \to X \pm \delta) =& \frac{1}{X_{1,4}} \left( \frac{1}{X_{1,3}-\delta} + \frac{1}{X_{2,4}+\delta}\right)\left(\frac{1}{X_{4,6}+\delta } + \frac{1}{X_{1,5}-\delta} \right) + (\text{cyclic}, i\rightarrow i+2)  \\ 
    & + \frac{1}{(X_{1,3}-\delta)(X_{3,5}-\delta)(X_{1,5}-\delta)}  + \frac{1}{(X_{2,4}+\delta)(X_{4,6}+\delta)(X_{2,6}+\delta)},
\end{aligned}
\end{equation}
gathering and expanding in $\delta \gg 1$ we get:
\begin{equation}
\begin{aligned}   \mathcal{A}_6^{\text{Tr}(\phi^3)}( X \to X \pm \delta) \,\xrightarrow{\delta \rightarrow \infty}\, &-\frac{1}{\delta^4} \left(- \frac{(X_{1,3}+X_{2,4})(X_{1,5}+X_{4,6})}{X_{1,4}} + (\text{cyclic},i\rightarrow i+2) 
       \right. \\
     & \left.+ X_{1,3}+X_{3,5}+X_{1,5}+X_{2,4}+X_{4,6}+X_{2,6}
     \right) +  \mathcal{O}(1/\delta^5),
\end{aligned}
\end{equation}
which we can identify with the 6-point NLSM amplitude, $\mathcal{A}_6^{\text{NLSM}}$.

\subsubsection{Factorizations near zeros}

The identifications of NLSM amplitudes with those of Tr$(\phi^3)$ theory with the simple shifted kinematics $X_{ee} \to X_{ee} + \delta, X_{oo} \to X_{oo} - \delta, X_{e o} \to X_{e o}$ allows to also very simply understand the pattern of factorization near zeroes we had observed experimentally in section \ref{sec:FTNLSM}. Precisely because these shifts are $c$-preserving, at the level of the Tr$(\phi^3)$ amplitudes, the factorization patterns are precisely the same before and after the shifts.

To begin with, we consider the near-zero factorizations for $2n$ particle amplitudes into ``even points $\times$ even points'', even when taking into account the necessary kinematic shifts, we still end up with the same $\delta$ shifts for the $X_{ee},X_{oo},X_{eo}$ for each of the lower point amplitudes. This proves that the ``even $\times$ even'' near-zero factorizations for NLSM amplitudes simply factor into the product of NLSM amplitudes, just as we saw in section \ref{sec:FTNLSM}. 

\begin{figure}[t]
    \centering
\includegraphics[width=0.6\textwidth]{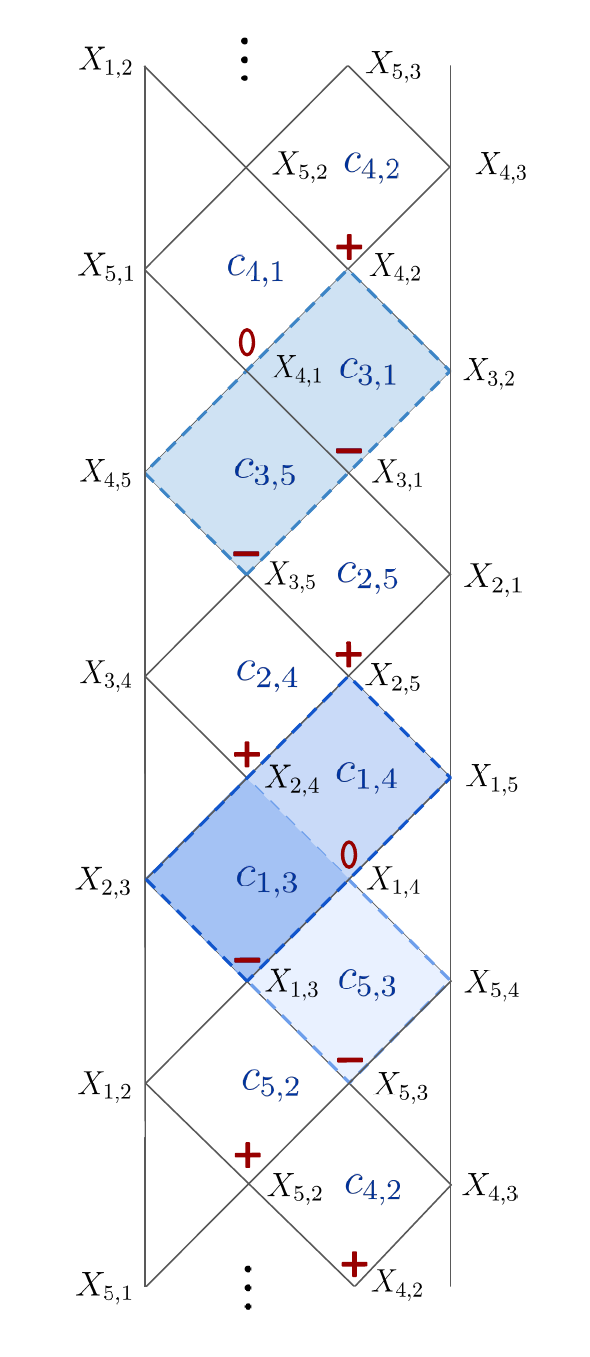}
    \caption{Shifts of kinematic variables in the 5-point factor in the factorization near zero of 6-point NLSM.}
    \label{fig:MixedAmp}
\end{figure}

The case of near-zero factorization to ``odd $\times$ odd'' amplitudes is somewhat more interesting, since as we observed experimentally in section \ref{sec:FTNLSM}, we encounter the mixed amplitudes for cubic scalars $\phi$ and pions $\pi$. We can now easily understand the reason for this, as well as the interesting rule for ``who is a $\phi$ and who is a ``$\pi$'' we delineated in section \ref{sec:FTNLSM}. As an example let us consider the $n=6$ 
factorization associated the upper skinny rectangle in figure \ref{fig:mixedFact}, where we turn on $c_2$. Let us focus on the $5$-point factor, with the appropriate kinematic replacements. For clarity, we will denote the kinematics of the $n=5$ problem by $Y$ variables $Y_{ij}$. So all the $Y_{ij} = X_{ij}$ except of course $Y_{15} = 0$, and we have the kinematic replacement $Y_{25} = X_{26}$.  We can now perform the $\delta$ shift on all the variables: $Y_{1,3} \to Y_{1,3} - \delta, Y_{1,4} \to Y_{1,4}, Y_{2,4} \to Y_{2,4} + \delta, Y_{2,5} = X_{2,6} \to X_{2,6} + \delta = Y_{2,5} + \delta, Y_{3,5} \to Y_{3,5} - \delta$. We denote the effect of this on the 5-point mesh by attaching a ``+/-'' to each variable as denoted in figure \ref{fig:MixedAmp} presented at the end of the paper.

We can now see that the shifts in the $n=5$ mesh do {\it not} preserve all the $c$'s. Nonetheless, some of the $c$'s are preserved, as represented in the shaded meshes in the figure. This picture enables us to see that this $n=5$ factor does {\it not} have all of our zeros; not all the skinny rectangles remain unshifted. However, some of the zeros do survive: the ones naturally associated with soft limits for particles $1,3,5$ are still clearly present in the picture. Now for particles $1,3$, we can see that the collinear poles $(p_1 + p_2)^2 = Y_{1,3} \to Y_{1,3} - \delta$, $(p_5 + p_1)^2 = Y_{2,5} \to Y_{2,5} + \delta$ are shifted, so these massless poles are absent, and hence the skinny rectangle zero implied an Adler zero when particle 1 becomes soft. The same holds for particle 3. However, for particle 5, we have that collinear pole $(p_4 + p_5)^2 = Y_{1,4}$ is unshifted, and hence the skinny rectangle zero does not imply a soft zero for particle 5. It is easy to see that factorization of the Tr$(\phi^3)$ amplitude after the shifts implies that the remaining particles must be interpreted as scalars with a Tr$(\phi^3)$ coupling. So we have learned that the $5$-pt factor in the near zero factorization associated with turning on $c_2$ gives us a mixed amplitude ${\cal A}^{{\rm NLSM} + \phi^3}(\pi, \phi, \pi, \phi, \phi)$~\cite{Cachazo:2016njl}. This argument extends to the general pattern of ``odd $\times$ odd'' factorizations of NLSM amplitudes explained in section \ref{sec:FTNLSM}. 

\subsection{$\alpha^\prime \delta =1$ and scaffolded gluons} 

Now let us explore what happens when the deformation becomes one in string units, $i.e.$ $\alpha^\prime \delta =1$. Then the stringy integral becomes:

\begin{equation}  \label{eq:stringyYMS}
\mathcal{I}_{2n}^{\delta}=\int_{\mathbb{R}_{>0}^{2n{-}3}} \prod_{I=1}^{2n{-}3} \frac{\diff y_I}{y_I}~\prod_{(a, b)} u_{a,b}^{\alpha' X_{a,b}} \frac{\prod_{(e,e)}u_{e,e}}{\prod_{(o,o)}u_{o,o}},
\end{equation}
as explained in \ref{sec:4ptdeformed}, at low energies, we expect to get a theory of colored scalars and gluons, just like that given by Lagrangian \eqref{eq:LYMS}.
 
At tree-level, we can see that \eqref{eq:stringyYMS} is what we get in the bosonic string amplitude, describing the scattering of $2n$ gluons, by choosing a particular kinematic configuration. This connection will make clear the origin of the scalars.  

\subsubsection{Bosonic string connection}
The open bosonic string amplitude for $2n$ gauge bosons with polarizations $\epsilon_i$ is given by:
\begin{equation}	\mathcal{A}^{\text{tree}}_n(1,2,\dots,2n) = \int \frac{\diff^{2n} z_i}{\text{SL(2,}\mathbb{R})} \,  \left( \prod_{i<j} z_{i,j}^{2\alpha^{\prime} p_i \cdot p_j}\right) \, \left. \exp\left( \sum_{i\neq j} 2 \frac{\epsilon_i \cdot \epsilon_j }{z_{i,j}^2} - \frac{\sqrt{\alpha^{\prime}}\epsilon_i \cdot p_j}{z_{i,j}}\right) \right\vert_{\text{multi-linear in }\epsilon_i},
\label{eq:BosStringTree}
\end{equation}
where $z_{i,j}=z_i-z_j$ are the usual worldsheet coordinates ({\it c.f.}~\cite{Green:1987sp}). Let us now assume that the space is sufficiently high-dimensional, with $\tilde{D}$ dimensions, so that we can achieve the following kinematical configuration: 

\begin{equation}
\begin{aligned}
	p_i \cdot \epsilon_j &= 0, \quad \forall \, \, (i,j) \in (1,...,2n) , \\
	\epsilon_i \cdot \epsilon_j &= \begin{cases}
		1 \quad \text{if } (i,j) \in \{(1,2);(3,4);(5,6);...;(2n-1,2n)\} ,\\
		0 \quad \text{otherwise}.
	\end{cases}
\end{aligned}
\label{eq:kinConfig}	
\end{equation}

This kinematic configuration can be easily achieved by considering the momentum to live in the first $D$ dimensions (the ones corresponding to the usual dimensionality of space), and the polarizations to live in the extra dimensions. 
With this kinematical choice, all the polarizations are fixed, and the remaining degrees of freedom are the $2n$  $D$-dimensional momenta -- exactly that of a $2n$-scalar problem. The bosonic string integral \eqref{eq:BosStringTree} simplifies enormously and we get the following single term:
\begin{figure}[t]
    \centering
\includegraphics[width=\linewidth]{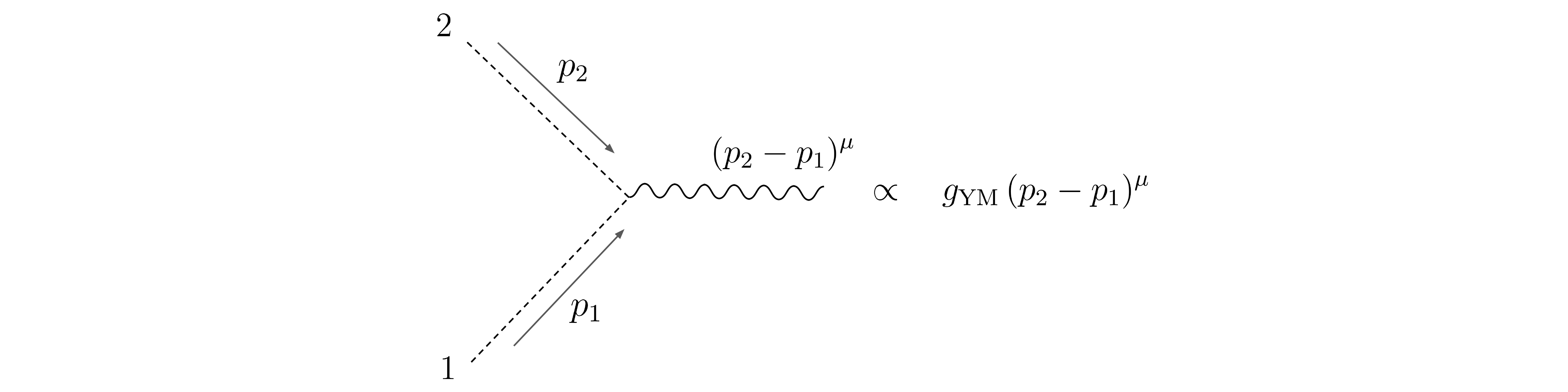}
    \caption{Two scalar - one gluon interaction}
    \label{fig:ScalarGluonInt}
\end{figure}
\begin{equation}
\begin{aligned}
	\mathcal{A}_{2n}(1,2,...,2n) 
&\xrightarrow{\eqref{eq:kinConfig}
\text{ kinematics}}\int \frac{\diff^{2n} z_i}{\text{SL(2,}\mathbb{R})} \,   \prod_{i<j} z_{i,j}^{2\alpha^{\prime} p_i \cdot p_j} \, \frac{1}{z_{1,2}^2z_{3,4}^2z_{5,6}^2...z_{2n-1,2n}^2} \\
 =&\int \underbrace{\frac{\diff^{2n} z_i}{\text{SL(2,}\mathbb{R})} \,  \frac{1}{z_{1,2}z_{2,3}z_{3,4}\dots z_{2n,1} }\prod_{i<j} z_{i,j}^{2\alpha^{\prime} p_i \cdot p_j}}_{\text{Stringy Tr$(\phi^3)$}} \, \frac{z_{2,3}z_{4,5}z_{6,7}\dots z_{2n,1}}{z_{1,2}z_{3,4}z_{5,6}\dots z_{2n-1,2n}}
	\label{eq:2nScalarTree}
\end{aligned}
\end{equation}

Now the $u$-variables are defined in terms of the worldsheet coordinates, as the following SL($2,\mathbb{R}$)-invariant cross-ratios:
\begin{equation}
    u_{i,j} = \frac{z_{i,j-1} \, z_{i-1,j}}{z_{i,j} \, z_{i-1,j-1}}.
	\label{eq:us}
\end{equation}

Using this definition, one can see that the extra ratio in the integrand is exactly equal to $\left(\prod_{(e,e)}u_{e,e}/\prod_{(o,o)}u_{o,o}\right)$, giving us back \eqref{eq:stringyYMS}. 
 
Therefore, we now understand that the scalars scattering in \eqref{eq:stringyYMS} are secretly gluons in higher dimensions. Moreover, we see that they only interact with their immediate neighbors, since only $\epsilon_{2i}\cdot\epsilon_{2i-1}$ are non-zero -- which can be interpreted as there being $n$ different species of such scalars that do not mix. Finally, the original cubic gluon interaction gives rise to a cubic interaction between a pair of scalars, $(2i,2i-1)$, and a gluon with the corresponding Feynman rule (see figure \ref{fig:ScalarGluonInt}). 

\begin{figure}[t]
    \centering
    \includegraphics[width=0.8\linewidth]{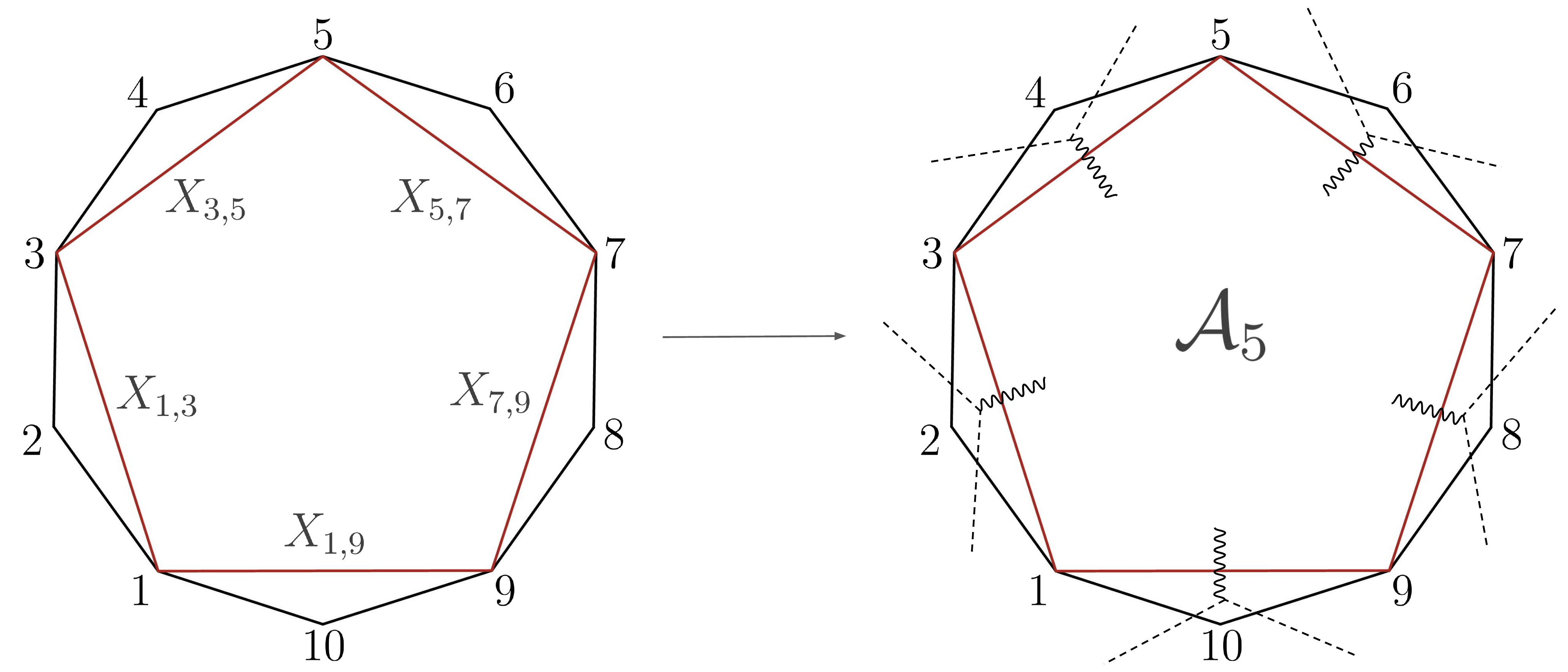}
    \caption{From 10-point scalar amplitude to 5-point gluon amplitude after taking the scaffolding residue.}
    \label{fig:5ScaffRes}
\end{figure}

Therefore, starting from the $2n$-scalar scattering, to access the $n$-point gluon amplitude, we need to take $n$ residues, that put the gluons on-shell, $i.e.$ take residues corresponding to $X_{1,3} = X_{3,5} = \dots = X_{1,2n-1}=0$. So from this perspective, we think of each gluon in the scattering process as coming from a pair of scalars -- the gluons are scaffolded by scalars. This allows us to talk about spin-1 particles in a purely scalar way, which ultimately allows the connection to the simple theory of colored scalars that we started with. See figure \ref{fig:5ScaffRes} for the case where, starting with a 10-point scalar, we can access the 5-point gluon amplitude, after taking the scaffolding residue. Such $2n$-scalar bosonic string amplitudes as well as YMS amplitudes~\cite{Cachazo:2014xea} in the field-theory limit, were studied in~\cite{He:2018pol, He:2019drm}.  

As we explain in \cite{Gluons}, the momentum and polarization of the gluons can be determined in terms of the momentum of the external scalars:
\begin{equation}
    \begin{cases}
        q_i^\mu &= (p_{2i}+p_{2i-1})^\mu \\
        \epsilon_i^\mu &\propto (p_{2i}-p_{2i-1})^\mu
    \end{cases},
\end{equation}
where the momentum of the gluon can be read off directly by momentum conservation and the polarization through the vertex Feynman rule in figure \ref{fig:ScalarGluonInt}. In this scalar language, gauge invariance and linearity in the polarizations have their own avatar that is explained in detail in \cite{Gluons}.

\subsubsection{Scaffolding residue}

In order to extract the gluon amplitude we need to take the residues $X_{2i+1,2i-1}=0$. To easily access this residue it is useful to pick a positive parametrization $\{y_\mathcal{T}\}$ corresponding to a triangulation \textbf{including} chords $X_{2i+1,2i-1}$. This way the singularity associated with $X_{2i+1,2i-1}=0$ comes from the divergence of the integral near 
$y_{2i+1,2i-1}=0$, and the residue of the amplitude turns into the residue of the integrand at $y_{2i+1,2i-1}=0$.
 
As it is explained in \cite{Gluons}, for such a triangulation, the factor $\left(\prod_{(e,e)}u_{e,e}/\prod_{(o,o)}u_{o,o}\right)$ simplifies to:
\begin{equation}
    \frac{\prod_{(e,e)}u_{e,e}}{\prod_{(o,o)}u_{o,o}}\longrightarrow \frac{1}{y_{1,3}y_{3,5}\dots y_{1,2n-1}} \times \frac{1}{\prod_{(k,m) \in \mathcal{T}^\prime} y_{k,m}},
\end{equation}
where $\mathcal{T}^\prime$ stands for the triangulation of the inner $n$-gon, with vertices $\{1,3,5,\dots,2n-1\}$, corresponding to the gluon amplitude. And thus the $2n$-scalar amplitude becomes:
\begin{equation}
\mathcal{I}_{2n}^{\delta}=\int_{\mathbb{R}_{>0}^{2n{-}3}} \underbrace{\prod_{i=1}^{n} \frac{\diff y_{2i-1,2i+1}}{y_{2i-1,2i+1}^2}\prod_{I\in \mathcal{T}^\prime} \frac{\diff y_I}{y_I^2}~\prod_{(a, b)} u_{a,b}^{\alpha' X_{a,b}}}_{\Omega_{2n}} ,
\end{equation}
which is exactly the stringy Tr$(\phi^3)$ integral where instead of a d$\log$ form, we have $\diff y/y^2$. So n-point gluon amplitude is then given by the low energy of:
\begin{equation}
    \mathcal{I}_{n}^{\text{gluon}} = \int_{\mathbb{R}_{>0}^{n{-}3}}\text{Res}_{y_{1,3}=0}\left( \text{Res}_{y_{3,5}=0} \left( \dots \left(\text{Res}_{y_{1,2n-1}=0} \left( \Omega_{2n} \right)\right) \dots \right) \right) \bigg \vert_{X_{2i-1,2i+1}=0}.
\end{equation}

\subsubsection{Zeros}

The stringy $2n$-scalar has all the same zeros and factorizations as stringy Tr$(\phi^3)$ theory, but in order to understand the zeros/factorizations for gluon amplitudes,  we need to study which of these survive after the scaffolding residue. 
 
\begin{figure}
    \centering
\includegraphics[width=0.85\linewidth]{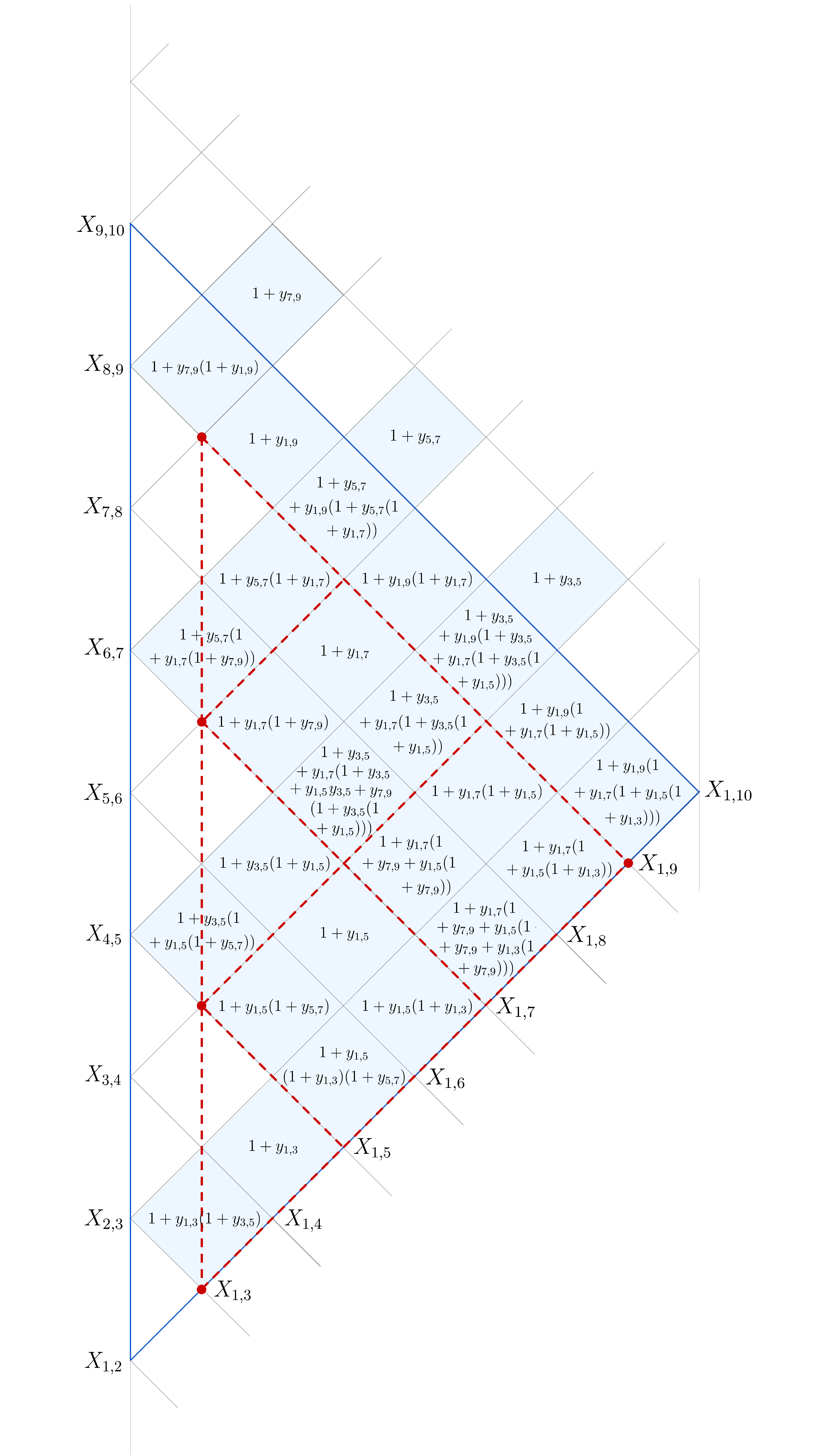}
    \caption{10-point kinematic mesh and $F$ polynomials for underlying scaffolding triangulation. Highlighted in red, 5-point effective gluon kinematic mesh.}
    \label{fig:10zeros}
\end{figure}

As pointed out previously, to access the scaffolding residue it is useful to pick the underlying triangulation to contain $\{X_{2i-1,2i+1}\}$. To talk about the zeros we will pick the underlying triangulation to be as close as possible to the usual ray-like one by choosing the triangulation of the $n$-gon to be ray-like. Still, the fact that we have chords $\{X_{2i-1,2i+1}\}$ means that the $c_{i,j}$ appearing in the stringy integral are not exactly those of the usual triangle (see figure \ref{fig:10zeros} at the end of the paper). 
 
We are now going to study the zeros of the gluon amplitude for the case of $10$ scalars $\rightarrow$ $5$ gluons. This example is big enough to illustrate the non-trivial features and understand how it generalizes to higher points. At 10 points, let us consider the triangulation of the 10-gon: $\{X_{1,3},X_{3,5},X_{5,7},X_{7,9},X_{1,9},X_{1,5},X_{1,7}\}$, $i.e.$ we have the scaffolding chords and a ray-like triangulation for the inner pentagon. For this choice of triangulation, the resulting region of the mesh is represented in figure \ref{fig:10zeros} as the shaded region. We see that there are only $3$ meshes from the usual triangle that are now missing and instead are replaced by 3 meshes on the top. 
 
In figure \ref{fig:10zeros} we represent the $F$-polynomials entering the string integral inside the mesh, $c_{i,j}$, corresponding to their exponents. We further mark with red dots the scaffolding poles, $X_{i,j}$, that we need to take the residue on to localize on the gluon problem. The claim is that effectively, to read off the zeros/factorizations, we should think of the gluon mesh as being the one highlighted in red, as this is the one in which each mesh point is associated with one of the $X_{\text{odd},\text{odd}}$ entering the gluon amplitude. In this case, we should have a 5-point mesh, so we see that each usual square in a scalar mesh gets replaced with 4 squares, in the gluon mesh. In this new picture, to get a zero we need to see exactly the same patterns of meshes to zero, of course, now each individual mesh gets subdivided into four smaller meshes. Thus at 5-point, our usually expected codimension-2 zeros are mapped to codimension $2 \times 4 = 8$ zeros. This is exactly the codimension we obtained when we phrase them in terms of $p_i \cdot p_j, \epsilon_i \cdot p_j, \epsilon_j \cdot p_i, \epsilon_i \cdot \epsilon_j$. It is worth noting that this is \textbf{not} a zero of the full $2n$-scalar problem, but instead a zero only after taking the scaffolding residue to land on the $n$-gluon amplitude.  
 
Looking back at figure \ref{fig:10zeros}, one zero would correspond to setting $c_{i,7}=c_{i,8}=0$ for $i\in \{1,\dots,4\}$, or, in the full stringy case, to a negative integer. The reason why the answer vanishes in this limit is because, after the scaffolding residue, it reduces to a sum of integrals of the form:
\begin{equation}
    \int_{0}^{\infty} \frac{\diff y_{1,7}}{y_{1,7}} y_{1,7}^{X_{1,7}+ n} \times \left( \text{remaining integrations} \right) \quad , \quad \text{with } n\in \mathbb{N}_0,
    \label{eq:resy17}
\end{equation}
and so, for the usual reason, we get zero from the integration in $y_{1,7}$. It is easy to understand why this happens. Note that all the $F_{i,j}$ inside the zero causal diamond depend on $y_{1,7}$, however, there are still other $F$-polynomials, \textbf{outside} this causal diamond, depending on $y_{1,7}$, such as $F_{i,9}$ for $i\in\{1,\dots,5\}$. But in all these cases $y_{1,7}$ always appears multiplied by some $y_{i,j}$ associated with a scaffolding variable. Therefore, after taking the scaffolding residue, this dependence will either vanish since we are evaluating at $y_{\text{scaff}}=0$, or shift $y_{1,7}^{X_{1,7}}$. In either case, we will have the form \eqref{eq:resy17} after the scaffolding residue. Instead, the $F$-polynomials that are inside the zero causal diamond are those in which the dependence on $y_{1,7}$ is of the form: $1 + y_{1,7} + \dots$ which are unaffected by the scaffolding residue and thus would not lead to \eqref{eq:resy17}. 
 
Another possible zero is obtained by setting $c_{1,j}=c_{2,j}=0$ for $j\in \{5,\dots,8\}$, or, in the full stringy case, to a negative integer. In this case, the claim is that, after the scaffolding residue, the amplitude reduces to sums of integrals of the form of \eqref{eq:resy17} but where $y_{1,7}$ and $X_{1,7}$ get replaced for $y_{1,5}$ and $X_{1,5}$, and thus the zero comes from the integration in $y_{1,5}$. As for the case in $y_{1,7}$, all $F$-polynomials depending on $y_{1,5}$ \textbf{outside} this causal diamond are such that $y_{1,5}$ always appears multiplied by some $y_{2i-1,2i+1}$. Whereas, for all the $F$-polynomials inside the zero causal diamond, the dependence in $y_{1,5}$ is either of the form $1+y_{1,5} + \dots$, or $y_{1,5}$ appears multiplying one of the remaining $y$'s of the inner $n$-gon, which in the 5-pt case is only $y_{1,7}$. The presence of these last terms would also not lead to \eqref{eq:resy17}, and thus why these need to be inside the zero causal diamond.
 
We can translate the locus of zeros we have just phrased in terms of vanishing $c$'s in the more familiar language of dot products between polarization vectors and momenta. For instance consider our first set of zeros, where $c_{i,7} = c_{i,8} = 0$ for $i=1, \cdots, 4$. So e.g. for $i=1,2$ we have the four constraints $p_1 \cdot p_7, p_1 \cdot p_8, p_2 \cdot p_7, p_2 \cdot p_8 = 0$. Taking linear combinations of these relations is equivalent to the four statements $(p_1 \pm p_2) \cdot (p_7 \pm p_8) = 0$, and given the map between polarization vectors and momenta where e.g. $q_j = p_{2j - 1} + p_{2 j}, \epsilon_j = (p_{2j} - p_{2j-1})$, this turns into the statements that $q_1 \cdot q_4 = 0, q_1 \cdot \epsilon_4 = 0, \epsilon_1 \cdot q_4 = 0, \epsilon_1 \cdot \epsilon_4 = 0$. This generalizes in the obvious way: every ``big'' mesh $(i,j)$is divided into four ``small'' meshes that are set to zero, and this is equivalent to the statements $q_i \cdot q_j=q_i \cdot \epsilon_j = \epsilon_i \cdot q_j = \epsilon_i \cdot \epsilon_j = 0$, just as we observed experimentally in section \ref{sec:FTYM}.

For a general $2n$-scalar to $n$-gluon amplitude, by drawing the effective gluon mesh (represented in red for the 5-point gluon problem in \ref{fig:10zeros}) the zero causal diamonds are exactly those identified in the usual scalar mesh, where now a square gets replaced by a set of 4 squares. Let us say the zero for one such causal diamond comes from the integration in $y_I$ of the internal $n$-gon, then the claim is that all the $F$-polynomials lying inside this causal diamond contain all the $F$-polynomials in which $y_I$ does \textbf{not} appear multiplying one of the scaffolding $y$'s.

\subsubsection{Factorizations}

Once more, since the $2n$-scalar amplitude is just a $c$-preserving deformation of stringy Tr$(\phi^3)$, all the factorizations of the latter are also true for the former. So, to understand which factorizations of the $2n$-scalar turn into gluon factorizations, we need to study which ones survive after the scaffolding residue. The idea is then to find the factorizations in which the scaffolding variables appear in lower point amplitudes so that the residue is non-vanishing. There are two different such cases - either the lower point amplitudes are both odd points or even points. 
 
To explain the different possible factorization patterns we can find we will be studying the 10-point scalar problem, as this example is big enough to include all the subtleties we find for general $2n$ scalars. 
 
Let us consider the even-point factorizations of the $10$-point scalar amplitude. For even-point factorizations, as long as we ensure that all the scaffolding variables appear on the lower problems, then we have two $2n$-scalar lower point amplitudes. In addition, since for even-point factorizations the $X_{\text{B}}$ and $X_{\text{T}}$ associated with the causal diamond we're probing are both $X_{\text{o,e}}$, the prefactor is the usual 4-point amplitude of Tr$(\phi^3)$ theory:
\begin{equation}
    \mathcal{A}_{2n}^{\alpha^\prime \delta =1 }( c_\star \neq 0 ) \rightarrow \frac{\Gamma(\alpha^\prime X_{\text{B}})\Gamma(\alpha^\prime X_{\text{T}})}{\Gamma(\alpha^\prime(X_{\text{B}}+X_{\text{T}}))} \times \mathcal{A}_{2n_1}^{\alpha^\prime \delta =1, \text{up} } \times \mathcal{A}_{2n_2}^{\alpha^\prime \delta =1, \text{down} },
\end{equation}
where $c_\star$ is the mesh we are turning on inside the zero causal diamond, and $2n_1$ and $2n_2$ are the lower even-point amplitudes, such that $n_1 +n_2 = n+1$.
 
\begin{figure}[t]
    \centering
    \includegraphics[width=\linewidth]{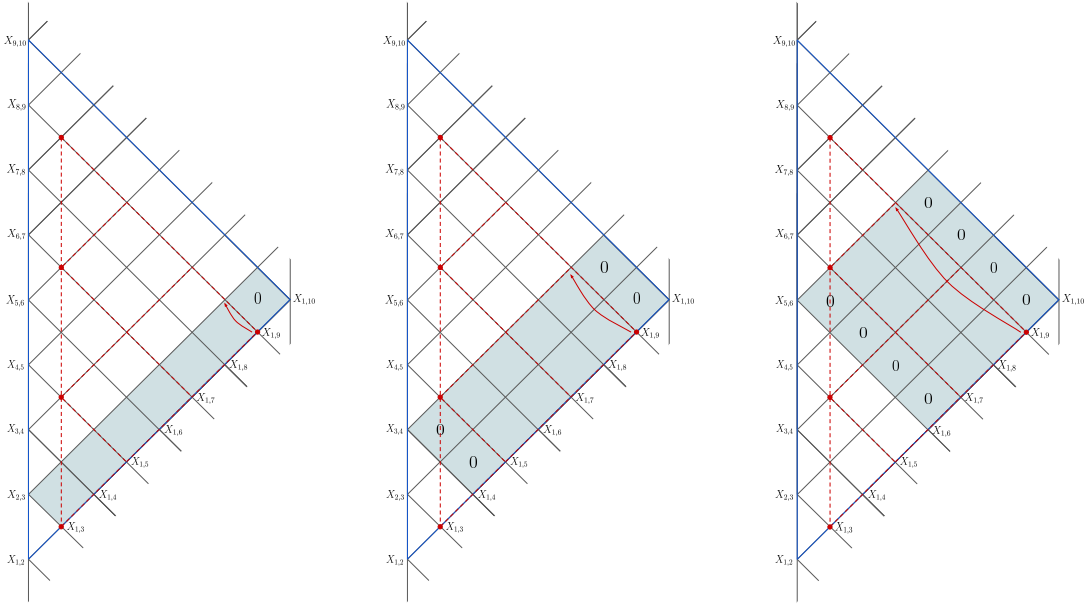}
    \caption{Factorization of the 10-point scalars and the scaffolded 5-point gluons.}
    \label{fig:10Fact}
\end{figure}

Let us go back to the 10-point example. In this case, there are two possible ways of factorizing into even-point amplitudes, presented in figure \ref{fig:10Fact} (center and right mesh). Let us start by looking at the right case corresponding to a square -- in this case, the lower point amplitudes are:
\begin{equation}
    \mathcal{A}_{10} (c_\star \neq 0) \rightarrow \frac{\Gamma(\alpha^\prime X_{1,6})\Gamma(\alpha^\prime X_{5,10})}{\Gamma(\alpha^\prime(X_{1,6}+X_{5,10}))}\times \mathcal{A}^{\text{down}}_6\times \mathcal{A}^{\text{up}}_6.
\end{equation}

Let us focus on those factorizations that can also be accessed in the field theory limit for gluons, by considering setting all the $c$'s inside the square to zero but for one $c_\star$ inside this square to be not zero. Asking for all the scaffolding variables to appear in the appropriately kinematic shifted lower amplitudes forces two columns of the square vanish, and therefore the $c$'s we can choose to turn on are exactly those inside the gluon mesh (see figure \ref{fig:10Fact}, right). Now note that while in the ``up'' amplitude taking the scaffolding residue in the original problem, corresponds to taking three residues on this 6-point amplitude and thus exactly producing a \textbf{3-point gluon} amplitude, the same is not true for the down amplitude. In the 6-point amplitude at the bottom, we are only taking residues in two variables, $X_{1,3}$ and $X_{3,5}$, not an $X_{1,5}$. Therefore we are only producing two gluons and the remaining pair of scalars is left untouched, and so after taking the scaffolding residue, the down amplitude is that of \textbf{two gluons + two scalars}. Of course, if we further take a residue in $X_{1,5}$, then we further turn the down amplitude into a 3pt gluon amplitude, and this becomes a pure gluon factorization. The last piece we need to understand is the prefactor, which in the field theory limit becomes:
\begin{equation}
     \mathcal{A}_{10} (c_\star \neq 0) \rightarrow \left(\frac{1}{X_{1,6}}+\frac{1}{X_{5,10}} \right)\times \mathcal{A}^{\text{down}}_6\times \mathcal{A}^{\text{up}}_6,
\end{equation}
so taking the scaffolding residue in each term to turn the scalar amplitudes into gluon amplitudes will not affect the prefactor. This seems then to say that in the end we are left with an answer that has poles when $X_{\text{o,e}} \rightarrow 0$, which is incompatible with the fact that after scaffolding residue we should be left with only poles of $X_{\text{o,o}} \rightarrow 0$, corresponding to chords of the gluon problem. However, note that, crucially, since we are also setting the rectangles, $c_{i,6}=c_{i,9}=0$ with $i=1,\dots,4$, we have that $X_{1,6} = X_{1,7}$ and $X_{5,10} = X_{5,9}$, allowing us to conclude that after taking the scaffolding residue we get:
\begin{equation}
    \mathcal{A}_{5}^{\text{gluons}} (c_\star \neq 0) \rightarrow \left(\frac{1}{X_{1,7}}+\frac{1}{X_{5,9}} \right)\times \mathcal{A}^{\text{down, gluons + $\phi$}}_4\times \mathcal{A}^{\text{up, gluons}}_3,
\end{equation}
where this is now an honest factorization of the 5-point gluon amplitude. 
 
Another possible zero causal diamond that leads to even lower point amplitudes is the one represented in the central mesh of figure \ref{fig:10Fact}, where the lower point amplitudes are now 4-point and 8-point. Once more, asking for the scaffolding variables to be present in the lower point problems, forces us to set $c_{i,4}=c_{i,9}=0$ for $i=1,2$, and thus the $c_{i,j}$ we choose to turn on lives inside the effective gluon mesh. So the factorization pattern is now in the field theory limit:
\begin{equation}
 \mathcal{A}_{10} (c_\star \neq 0) \rightarrow \left(\frac{1}{X_{1,4}}+\frac{1}{X_{3,10}}\right)
  \times \mathcal{A}^{\text{down}}_4\times \mathcal{A}^{\text{up}}_8.
\end{equation}

As opposed to the $6\times6$ factorization, in this case, by taking the scaffolding residue both the $4$-point and the $8$-point scalar amplitudes turn into pure gluon amplitudes, and similarly, the prefactors change appropriately so that we get an honest 5-point gluon factorization:

\begin{equation}
    \mathcal{A}_{5}^{\text{gluon}} (c_\star \neq 0) \rightarrow \left(\frac{1}{X_{1,5}}+\frac{1}{X_{3,9}} \right)\times \mathcal{A}^{\text{down, gluons}}_2\times \mathcal{A}^{\text{up, gluons}}_4.
\end{equation}

Finally, let us look at the case of an odd-point factorization. At $10$-point, one example of this is by considering the skinny rectangle (figure \ref{fig:10Fact}, left). In this case, the down amplitude is a $3$-point amplitude which is trivial, and the up amplitude is $9$-point. Asking for the scaffolding variables to be present in the lower point problems, forces $c_{1,9}=0$, and by turning on any of the remaining $c$'s inside the skinny rectangle we obtain:
\begin{equation}
    \mathcal{A}_{10} (c_\star \neq 0) \rightarrow \underbrace{\int_{0}^\infty \frac{\diff y_{1,3}}{y_{1,3}^2} y_{1,3}^{\alpha^\prime X_{1,3}}(1+y_{1,3})^{-\alpha^\prime c_\star}}_{\mathcal{A}_4^{\alpha^\prime \delta =1}} \times \mathcal{A}^{\text{up}}_9,
\end{equation}
so we see that in this case the $4$-point pre-factor becomes the $4$-point $2n$-scalar amplitude, while for the lower point amplitude, it's just the $9$-point amplitude with the extra $u_{\text{e,e}}/u_{\text{o,o}}$. So in the low energy limit, and further taking the scaffolding residue we get:
\begin{equation}
     \mathcal{A}_{10} (c_\star \neq 0) \xrightarrow{\alpha^\prime X \ll 1} \frac{X_{2,10}}{X_{1,3}} \times \mathcal{A}^{\text{up}}_9 ; \quad   \mathcal{A}_{5}^{\text{gluon}} (c_\star \neq 0) 
     \rightarrow  X_{2,10} \times \mathcal{A}^{\text{up,scaffolded}}_9.
\end{equation}

Similarly to what we saw in the NLSM, we predict that lower odd-point amplitudes generated in these factorizations could be some mixed amplitudes of gluons and scalars. Regardless this is certainly another honest factorization of the gluon 5-point amplitude. The other reason to expect the lower-point object to be a mixed theory amplitude is exactly because the skinny rectangle is closely connected to soft limits, exactly the case in which it was first observed the appearance of these extended theory amplitudes.

\section{Outlook}

There are many obvious avenues for exploration following from the observations in this paper, so instead of attempting an exhaustive accounting of all of them, we will highlight a few that seem especially ripe for immediate development. 

Already at tree-level, it is interesting to ask whether the pattern of zeros suffices to completely determine the amplitude. For both the Tr$(\phi^3)$ theory and the NLSM, there is an obvious ansatz to make for the $n$-particle tree amplitude. Combining all the poles into a common denominator, we can assume that the numerator is a polynomial of correct units for each theory, and add the further crucial assumption that this numerator is at most linear in each $X_{ij}$ variable; this last requirement enforces good behavior in the Regge limit. This still leaves an enormous number of free parameters in the ansatz, but we can further impose our hidden zeroes. Quite remarkably we have found that experimentally for Tr$(\phi^3)$ amplitudes with $n=5,6,7$ points and for NLSM amplitudes with $n=6,8$, imposing the zeros in this way does fully fix the amplitude! For $n$-point Tr$(\phi^3)$ amplitude, the number of parameters is $\frac{n(n{-}3)}{2}$ choose $(n{-}3)$, and there are $51$ and $6700$ parameters for $n=6,8$ ansatz of NLSM amplitude respectively. Indeed imposing the simplest (cyclic class of) ``skinny rectangle'' zeros is enough to fully determine the amplitude in all these cases. It would be fascinating to prove this fact in general since this provides an entirely new way to uniquely determine scattering amplitudes, complementary to the traditional picture relying on poles and factorization. It would also be very interesting to ask how does the string amplitude is constrained by the zeros. It has been demonstrated in~\cite{Boels:2014dka} that by combining the monodromy relations and the behaviour under the (generalised) Britto-Cachazo-Feng-Witten shifts~\cite{Britto:2005fq}, the zeros~\cite{DAdda:1971wcy} of the residues of amplitudes at the kinematic poles are suffcient to dertermine certain string amplitudes.

Our focus on the numerator structure of the amplitude dovetails with an approach to determining the canonical form of positive geometries~\cite{Arkani-Hamed:2017tmz}, first seen in the context of the amplituhedron for ${\cal N}=4$ SYM,  by looking at pattern of zeros for the numerator demanded by killing illegal singularities/enforcing the ``dlog form'' structure~\cite{Arkani-Hamed:2014dca}. In this setting, the variety associated with the vanishing of the numerator is called the ``adjoint'' of the positive geometry. This set of zeros is more closely associated with familiar facts about singularities of amplitudes, while our new zeros reflect something entirely different, the collapsing of the geometry as kinematics are varied. It would be interesting to study the analog of these ``collapsing geometry'' zeros for the amplituhedron, to begin with even for the simplest case of the best-understood $m=2$ amplituhedron. 

It is interesting to phrase the existence of our pattern of zeros in an algebraic-geometric language. If we combine all the diagrams into a common denominator consisting of the product of all the $X_{i,j}$, the amplitude is ${\cal A} = {\cal N}(X)/{\cal D}(X)$, where the numerator ${\cal N}(X)$ is a degree $\frac{(n-2)(n-3)}{2}$ polynomial in the $X_{i,j}$. The complete locus of zeros of the amplitude is then a complicated variety in the $X$ space defined by ${\cal N}(X) = 0$. From this perspective, the hidden zeros tell us something striking about this complicated variety: it contains a large number of {\it linear subspaces} of various dimensionalities. This is certainly not a generic property for high-dimensional varieties! It would be fascinating if this ``numerator variety'' had further special properties. A speculative but intriguing thought is that this variety should be in some sense ``maximally nice''. For instance one might hope that the variety is ``determinantal'', with ${\cal N}(X)$ expressible as the determinant of a predictable matrix, in a way that would make all our hidden zeros manifest. 

In this paper, we have focused on zeroes and factorization at tree level, where at least for the Tr$(\phi^3)$ theory the amplitude is given by the canonical form for the associahedron. We now know that the integrand for the Tr$(\phi^3)$ theory at 1-loop is also given by the canonical form of a cousin of the associahedron, also naturally presented as a Minkowski sum of simplices~\cite{Salvatori:2018aha, Arkani-Hamed:2019vag}. Thus we expect a similar locus in kinematic space where the one-loop integrand has zero/factorization patterns, which would be interesting to flesh out. Of course this locus will in general involve sending kinematic variables involving the loop momenta to zero, and so don't immediately imply zeros for integrated amplitudes. It would be interesting to see if any trace of these zeros survives post-loop integration. Beyond one loop and to all orders in the topological expansion, there are various notions of a loop integrand naturally associated with surfaces, with the simplest ``infinite integrand'' reflecting the action of the mapping class group and the concomitant infinite repetition of Feynman diagrams/triangulations of the surface. This infinite integrand is also the canonical form of associated polytopes--``surfacehedra''--having infinitely many facets with a fractal structure~\cite{surfacehedron}. Surfacehedra are also Minkowski sums, so at least the infinite integrand should also have patterns of zeros and factorization, though the implications of this fact both for naturally finite integrands obtained by truncating the surfacehedra or modding out by the mapping class group must be properly understood. 

The phenomenon of factorization near zeros is clearly striking, and it would be fascinating to understand if and how it generalizes. We have seen an especially simple geometric understanding of this factorization, by understanding how the associahedron degenerates as Minkowski summands are shut off, so that at the penultimate step before it collapses entirely to lower dimensions, it simplifies drastically to what we described in the introduction as the ``sandwich'', with an interval separating two opposite facets. It is natural to wonder whether there are other predictable patterns for the degenerating associahedron as we turn on further $c$'s, that might also have interpretations in terms of factorization. We are aware of one such pattern: if instead of turning on a single $c_{i_*,j_*}$ in our maximal causal diamond, we turn on an entire strip i.e. $c_{i_*, k}$ for all $k$ inside the diamond, then the amplitude {\it also} factorizes. This is an interpretation in our mesh picture of the ``3-splits'' for Tr$(\phi^3)$ amplitudes discovered in \cite{Cachazo:2021wsz}. It would be interesting to try and interpret this in the language of the associahedron and examine how it might extend to full string amplitudes. More generally it would be interesting to classify the general set of factorization properties for string amplitudes associated with shutting off various patterns of $c$'s. 

In another vein, it is interesting that starting purely from the NLSM, we are naturally led to discover the mixed $\pi/\phi$ amplitudes from the near-zero factorizations. The particular mixed amplitudes we discover in this way are clearly only a tiny subset of all possible mixed amplitudes--for instance, they all contain only three $\phi$'s. In \cite{NLSM} we will show how general NLSM + $\phi^3$
amplitudes can be obtained by a simple set of kinematic shifts of the Tr$(\phi^3)$ amplitudes. Of course, the mixed amplitudes do not have all of our zeros, and so the shifts are not the $c$-preserving shifts featured in this paper. 

Nonetheless, this general phenomenon of kinematic shifts generating non-trivial theories from simple ones is a fascinating one, and it would be interesting to see how far it extends. One especially simple example is worth mentioning as an interesting contrast to the shift we have highlighted in this paper. Suppose we shift the $g$ Tr$(\phi^3)$ amplitudes by $X_{e,e; o,o} \to X_{e,e;o,o} + \delta, X_{e,o} \to X_{e,o}$, i.e. no $\pm \delta$ difference. This still removes all the massless poles $X_{ee},X_{oo}$ and leaves us only with poles associated with even-particle scattering amplitudes. It is trivial to see that the low-energy amplitudes for $X \ll \delta$ with the deformation are nothing but that of $\lambda$ Tr$(\phi^4)$ theory with quartic coupling $\lambda = g^2/\delta$, augmented with a further tower of higher-dimension operators. It is striking that this seemingly minor but unusual change--the sign difference between even-even and odd-odd shifts--makes all the difference in the world in going from generating the relatively boring Tr$(\phi^4)$ theory to the much richer and more intricate amplitudes for pions and gluons! 

As another amusing example, suppose we take the $g$ Tr$(\phi^3)$ theory but this time shift {\it all} $1/X \to 1/X + \kappa$. Clearly, the new functions we obtain in this way will still factorize onto themselves on the massless poles, and so still define a consistent set of amplitudes. But for which theory? We can actually determine a Lagrangian for the amplitudes of this theory in a very simple way. At any $n$, there is the part of the amplitude with no poles at all--purely a contact interaction. At $n$ points, this is given by simply replacing every propagator by $\kappa$; since there are Catalan$_{n-3}$ many diagrams at $n$ points, this contact interaction is $C_{n-3} \kappa^{n-3} g^{n-2}$. Hence we can identify an interesting non-linear Lagrangian we can call the ``Catalan Lagrangian'' that gives rise to the amplitudes associated with this shift: ${\cal L}^{\rm Catalan} = \sum_{n=3}^\infty C_{n-3} g^{n-2} \kappa^{n-3} {\rm Tr}(\phi^n) = g {\rm Tr}\left( (\sqrt{1 - 4 g \kappa \phi} -1)/(2 \kappa)\right)$. This is again an interesting cousin of the more interesting and surprising linear shift in our paper, which starts from the polynomial Tr$(\phi^3)$ theory and generates the amplitudes for the non-polynomial Lagrangian describing pion scattering. It would be interesting to map out the space of these possible deformations more systematically. 

Finally, it is clearly interesting to study the widest class of theories connected by sharing hidden zeros and factorization. The most obvious place to begin is to simply consider our shifted theories with general values for $\delta$. As we have explained for generic fractional $\delta$, the low-energy amplitudes are always those of the NLSM. But for $\alpha^\prime \delta$ being integers, something more interesting happens. As we have seen for $\alpha^\prime \delta = 1$ we have a theory of massless gluons interacting with the ``scaffolding'' of the external scalars. It is easy to see that for e.g. $\alpha^\prime \delta = 2$, while the leading interactions at low-energies are those of gluons coupled to the scaffolding scalars, there are {\it also} interactions that must be interpreted as arising from a theory of massless colored particles of spin 2. In general for $\alpha^\prime \delta = J$, at least kinematically we are describing gluons coupled to a tower of massless colored particles of spin up to $J$. Of course there are famous theorems~\cite{Weinberg:1980kq} about the impossibility of consistent theories for massless colored particles of high spin, so at first blush these theories for $J>1$ should be discredited on physical grounds. But the way they naturally connect to Tr$(\phi^3)$, NLSM, and Yang-Mills, simply via further continuing the $\delta$ deformation, suggests these theories may somehow have a purpose in life, perhaps especially in the limit as $\delta \to \infty$, where an infinite tower of higher-spin colored particles become massless. 

\acknowledgments

It is our pleasure to thank Alfredo Guevara, Daniel Longenecker, Giulio Salvatori, Yichao Tang, Jaroslav Trnka, Ellis Ye Yuan, Yaoqi Zhang for inspiring discussions. The work of N.A.H. is supported by 
the DOE (Grant No. DE-SC0009988), by the Simons Collaboration on Celestial Holography, and further support was made possible by the Carl B. Feinberg cross-disciplinary program in innovation at the IAS. The work of Q.C. is supported by the National Natural Science Foundation of China under Grant No. 123B2075. The work of C.F. is supported by FCT/Portugal
(Grant No. 2023.01221.BD). The work of S.H. has been supported by the National Natural Science Foundation of China under Grant No. 12225510, 11935013, 12047503, 12247103, and by the New Cornerstone Science Foundation through the XPLORER PRIZE.

\bibliographystyle{JHEP}\bibliography{Refs}
\newpage
\newpage
\newpage

\end{document}